%% file: thesis.tex
\documentclass{nuthesis}
\usepackage[utf8]{inputenc}
\usepackage[T1]{fontenc}
\usepackage{fixltx2e}
\usepackage{graphicx}
\usepackage{grffile}
\usepackage{longtable}
\usepackage{wrapfig}
\usepackage{rotating}
\usepackage[normalem]{ulem}
\usepackage{amsmath}
\usepackage{textcomp}
\usepackage{amssymb}
\usepackage{capt-of}
\usepackage{hyperref}
\adviser{Professor Vladimir Itskov}
\adviserAbstract{Vladimir Itskov}
\major{Mathematics}
\degreemonth{December}
\degreeyear{2015}

\input{header.tex}
\allowdisplaybreaks

\numberwithin{theorem}{chapter}
\usepackage[sorting=none,backend=bibtex]{biblatex}
\newcounter{TodoList} 
\newcounter{NoteList} 
\reversemarginpar

\author{Zachary J. Roth}
\date{August 2015}
\title{Analysis of neuronal sequences using pairwise biases}
\hypersetup{
 pdfauthor={Zachary J. Roth},
 pdftitle={Analysis of neuronal sequences using pairwise biases},
 pdfkeywords={},
 pdfsubject={},
 pdfcreator={Emacs 24.3.1 (Org mode 8.3.2)},
 pdflang={English}}

\begin{document}
\maketitle
\frontmatter

\begin{abstract}
Sequences of neuronal activation have long been implicated in a variety of brain
functions. In particular, these sequences have been tied to memory formation and
spatial navigation in the hippocampus, a region of mammalian brains.
Traditionally, neuronal sequences have been interpreted as noisy manifestations
of neuronal templates (i.e., orderings), ignoring much richer structure
contained in the sequences. This paper introduces a new tool for understanding
neuronal sequences: the bias matrix. The bias matrix captures the probabilistic
tendency of each neuron to fire before or after each other neuron. Despite
considering only pairs of neurons, the bias matrix captures the best total
ordering of neurons for a sequence (\Cref{lem:skbias-gamma}) and, thus,
generalizes the concept of a neuronal template.

We establish basic mathematical properties of bias matrices, in particular
describing the fundamental polytope in which all biases reside
(\Cref{thm:fundamental-polytope}). We show how the underlying simple digraph of
a bias matrix, which we term the bias network, serves as a combinatorial
generalization of neuronal templates. Surprisingly, every simple digraph is
realizable as the bias network of some sequence (\Cref{thm:bias-network}). The
bias-matrix representation leads us to a natural method for sequence
correlation, which then leads to a probabilistic framework for determining the
similarity of one set of sequences to another. Using data from rat hippocampus,
we describe events of interest and also sequence-detection techniques. Finally,
the bias matrix and the similarity measure are applied to this real-world data
using code developed by us.
\end{abstract}

\begin{copyrightpage}
\end{copyrightpage}

\begin{dedication}
To my family, thank you. My mother and father both worked hard and without
complaint in raising me despite not receiving anything in return, and they each
strived to live well as examples for me. Mom, much of my personality and humor
stems from you; that has been key to so much of what I have done and how I have
lived. Dad, I find more and more that I share many natural similarities to you;
and I find more and more that this is a good thing. Justin, whether you know it
or not, you are much of the reason that I ever cared about doing well in school.
Though I know you would contest it, I am convinced to this day that you are far
more intelligent than I. Micah, Noah, and Samuel, you influenced me less in
these developmental ways since you're younger than I; but you deserve
recognition for putting up with me all these years. Rachel, that you are
choosing to become part of my family is astounding to me. I look forward to our
life together.
\end{dedication}

\begin{acknowledgments}
As with any accomplishment, there are many to be acknowledged for their
contributions (more or less directly) to this work. I would be remiss if I did
not first acknowledge God (1) for providing me with any abilities I do have and
(2) for filling in the gaps with support from friends, family, and advisors. His
grace and generosity to me are unfathomable.

During my somewhat extended tenure as a graduate student, many people served as
advisors to me. Judy Walker provided me with much encouragement, direction, and
patience as my first advisor. Her support during the early years of my graduate
school experience was invaluable, and the time that we spent laughing during
meetings helped to make the experience enjoyable. I am also grateful to Vladimir
Itskov and Carina Curto who took something of a risk in allowing me to work with
them later on. Vladimir taught me that I need to be willing to fight for any
worthwhile idea or approach, and Carina kept me encouraged as I fought some of
these fights. Vladimir pushed me to try to keep up as he leaped ahead in his
thoughts, and Carina demonstrated that even complex ideas can be crafted and
organized in such a way as to make them seem straightforward. Lastly, in
inviting me to visit Janelia Research Campus, Eva Pastalkova assumed the role of
an advisor of sorts. Although Eva never served as a mathematics advisor, she
spent much time advising me as I found myself floundering to stay afloat in the
foreign world of neuroscience. It was encouraging to work with such a principled
scientist as I found myself questioning so much of what science seems to have
become, and it was good to share with Eva in conversations of deeper importance
than science.

In addition to Vladimir, Carina, and Judy, I would also like to explicitly thank
the remainder of my supervisory committee: Jamie Radcliffe and Khalid Sayood.
Jamie unhesitatingly agreed to serve as a reader on this committee at the last
minute, and he did this despite having had to read my combinatorics "proofs"
during my first semester of graduate school. Khalid always offered a
much-appreciated perspective on work and life. Though not part of my committee,
the math department staff made my life easier during this time. Marilyn Johnson
and Liz Youroukos, in particular, helped me with various things throughout the
course of graduate school; and each slipped me lots of, uh, brain food to help
me along.

Those who have known me beyond the walls of academia during graduate school have
also contributed much toward this accomplishment. During this time, these
friendships have kept me sane and grounded. It is, perhaps, unreasonable to try
to mention all of the people who have aided me with their friendship over these
years; but that will not stop me from trying.

To those who started school with me in the Fall of 2007, I appreciated our time
together learning of analysis, algebra, combinatorics, and the such; and, yes, I
even appreciated your jabs at me for being mostly indistinguishable from Orlando
Bloom. To my many officemates throughout the years, I appreciated our time
together and the many conversations that you each allowed me to thrust upon you
while I was avoiding real work. I wouldn't have expected my life in a math
department to contain so many thoughtful conversations about theology.

Mark DeBoer, moving to Lincoln was easier because I knew we would be living
together at first; I appreciated our years living together. David Lessor and Joe
LaChance, eating pizza, playing Halo, and climbing with you guys kept me feeling
mostly normal as I fell deeper into the world of math. Mike Janssen and Tanner
Auch, it's somewhat shocking how much time you were willing to spend with me
outside of the time spent together in both school and church; shared meals,
board games, \emph{King of the Hill}, and vidya games with you were staples of my
time in Lincoln. Laura Janssen and Amy Auch, you facilitated this time together
and became great friends in the process. Mike Callen, thank you for opening up
your house to me while I was still mostly a stranger; no other close friendship
of mine has ever formed quite so quickly as ours.

My time in Lincoln was truly special in that my school and church communities
overlapped heavily. To my Grace Chapel community, thank you for all of your
generosity. During these years in Lincoln, my close friends the Janssens, Auchs,
and Rawsons probably fed me more than I fed myself. The opportunity to share in
life with the same people in school, in church, and beyond was a blessing that I
didn't fully appreciate at the time. Grace Fellowship Church, you made being in
State College during my last year of graduate school not just bearable but
worthwhile.

So many other people should be mentioned for some reason or another: Ben
Nolting, Brian Kloppenborg, David and Jenny Rawson, Doug Dailey, Warren Wright,
Jon Walker, David Beevers, Brian and Gretchen Bredemeier, Yingxue Wang, and
Brian Lustig, to name a few. To those I have not mentioned, I appreciated our
friendships, too. From others in my joint math/church community to landlords to
friendships continued from times past and to many others, thank you.
\end{acknowledgments}

\setcounter{tocdepth}{2}
\tableofcontents*

\mainmatter

\chapter{Sequential information in the brain}
\label{sec:orgheadline4}
\label{ch:intro}

\section{Phenomenology}
\label{sec:orgheadline1}
\label{sec:phenomenology}

The inner workings of the brain are still poorly understood. Study it though we
may, the deep truths of its inner workings---the ways that it both processes and
stores information---elude us. Even in organisms far simpler than humans, this
problem remains. In 1986, White et al.~\cite{White1986} revealed the the
entire connectome---the neurons and their synaptic connections---for the
hermaphroditic roundworm C. elegans. Yet with this map of all 302 neurons and
around 7000 connections, we still do not fully understand how the brains of C.
elegans work. Comparing this to the estimated 86 billion neurons in the average
adult human nervous system \cite{Azevedo2009}, the task of understanding our own
brains is daunting to say the least. As such, it is not surprising that the body
of literature describing and modeling the observed phenomena of the brain keeps
growing. Here, we develop tools for use in the analysis of neuronal data that
accompanies these phenomena. Though our tools are of general purpose, we focus
on specific phenomena observed in a particular area of the mammalian brain: the
hippocampus.

The hippocampus is widely believed to aid in two main functions: episodic memory
and spatial navigation. Episodic memories are, broadly speaking, memories that
are associated with events in one's life (times, places, emotions, etc.); this
type of memory is in contrast with other types of memory such as semantic memory
(knowledge of facts) and implicit memory (e.g., how to ride a bike). Evidence
for the participation of the hippocampus in the formation of new episodic
memories was first reported in \cite{Scoville1957} in 1957. A man referred to as
Patient HM underwent a surgery to remove large portions of his hippocampal
formation in an attempt to be cured of severe epileptic seizures. Although the
surgery was successful in regards to reducing the frequency and severity of the
seizures, there was an unexpected side effect: Patient HM no longer seemed to be
capable of forming new memories about his experiences. Research on Patient HM
and his condition continued until his death in 2008.

\begin{figure}
  \centering
  \includegraphics[width=6in]{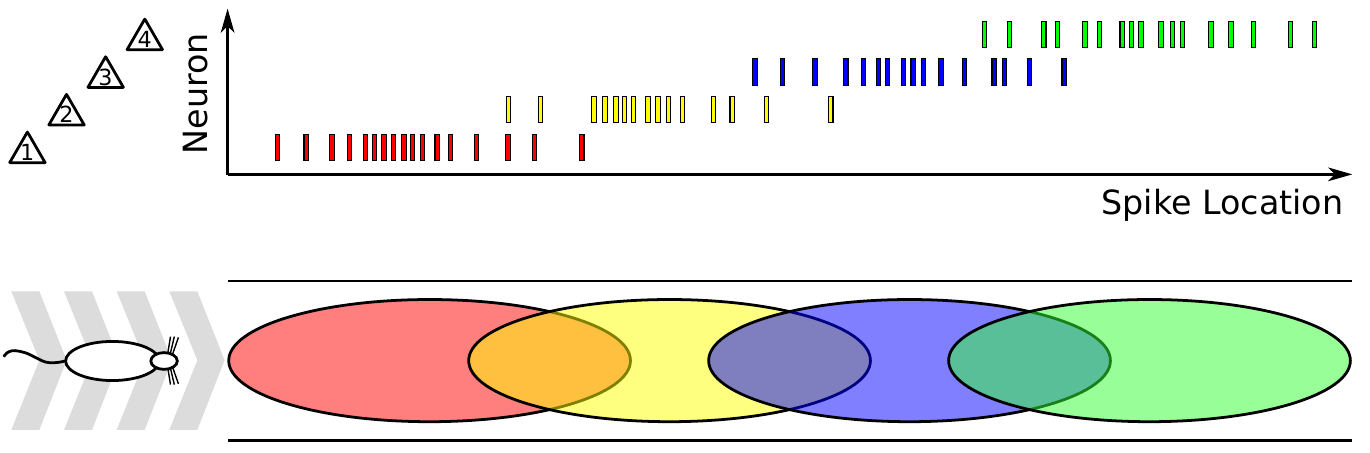}
  \caption{\label{fig:place-field-cartoon}Place fields and neuronal spiking. The
    animal here has four place cells (not pictured) each with a preference for
    spiking while the animal is at a particular location within a straight
    track. The four place fields are shown as colored regions on the track
    (bottom). The traversal of such a track results in the generation of a
    neuronal sequence (top) called a place-field sequence, which reflects the
    direction of traversal through the track.}
\end{figure}

The clearest evidence of the participation of the hippocampus in spatial
navigation is the existence of place cells, first observed in 1971 in
\cite{OKeefe1971}. These neurons have a preference for spiking when the animal
is occupying a certain location in space, as illustrated in
\Cref{fig:place-field-cartoon}. The spatial area in which a place cell has a
tendency to spike is called its place field. Traversal of a series of place
fields results in the generation of a neuronal sequence called a place-field
sequence, such as pictured in \Cref{fig:place-field-cartoon}. It is believed
(e.g., \cite{Skaggs1996}, \cite{Dragoi2006}, \cite{Dragoi2010}) that generation
of place-field sequences is intimately related to, and perhaps dependent upon, a
hippocampal oscillation of the local-field potential (LFP) called the theta
oscillation, which exists whenever an animal is in locomotion.

\begin{figure}
  \centering
  \includegraphics[width=6in]{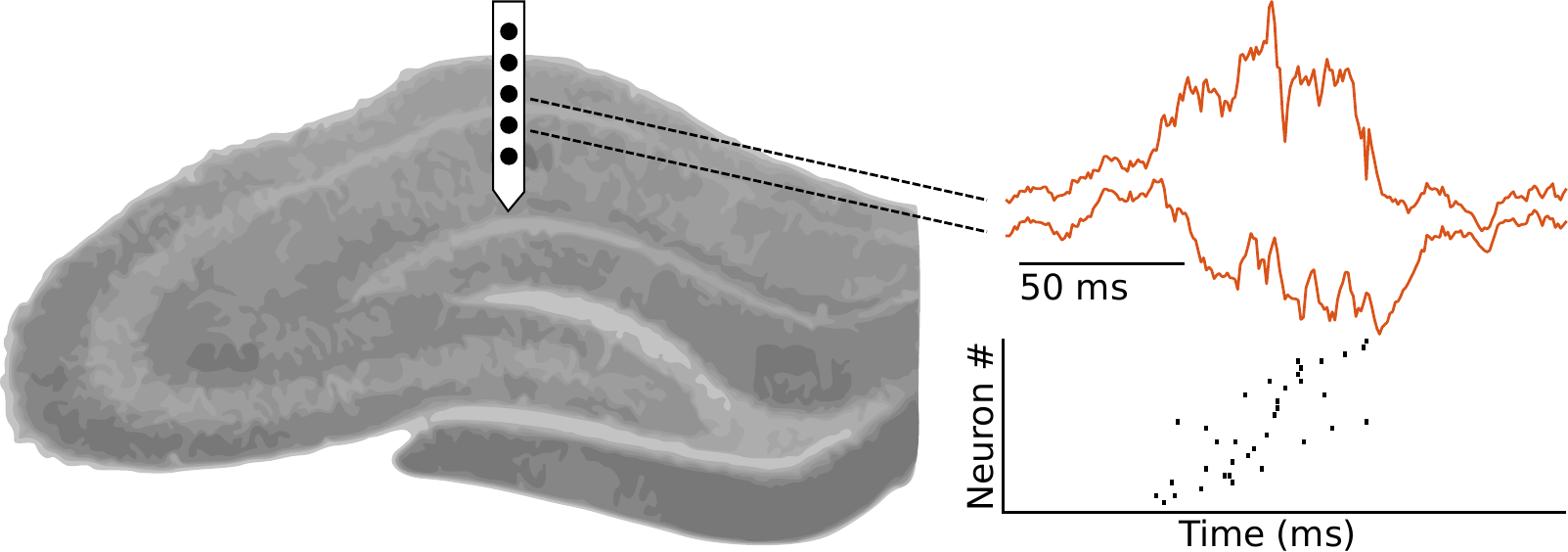}
  \caption{\label{fig:ripple-probe-envelope}A sharp-wave ripple in the
    hippocampus. On the left, we see an image of a rat hippocampus with a probe
    across the pyramidal layer of the CA1 region of the hippocampus. In the
    upper-right section of this figure, we see a real example of local-field
    potentials from above and below the pyramidal layer. This separation of the
    local-field potential is called a sharp wave. Below the sharp wave, the
    increase in spiking that accompanies the SWR is showin in a spike raster
    plot.}
\end{figure}

A hippocampal event called a sharp-wave ripple (SWR) has also been tied to the
formation of new episodic memories. Such an event is illustrated in
\Cref{fig:ripple-probe-envelope}. As pictured in this figure, a SWR is
characterized by two main features: (1) a sharp wave, which is a separation in
the local-field potential (LFP) across the pyramidal layer of the hippocampus,
and (2) a burst in neuronal activity, often observed as a high-frequency ripple
in the LFP within the pyramidal layer. When SWRs were selectively eliminated
from neural activity during post-training memory-consolidation periods in
\cite{Girardeau2009}, animals exhibited reduced performance in spatial memory
tasks, indicating that SWRs are necessary for the consolidation of episodic
memories.

\section{Neuronal sequences}
\label{sec:orgheadline2}
\label{sec:neuronal-sequences}

Given that neuronal sequences are important for storage and processing of
information in the brain, it is crucial that we possess a reasonable model of
sequences. Similarities between place-field sequences and SWR sequences
\cite{Diba2007} give reason to believe that properties such as sequence duration
and total number of spikes (and even the precise timing of spikes) are perhaps
less important (or, rather, manifestations of) some other properties that are
less immediately obvious when considering and comparing sequences. An
appropriate sequence model should ideally capture these intrinsic qualities and,
in so doing, address the fundamental question of what a sequence actually is.

\begin{figure}
  \centering
  \begin{subfigure}[b]{0.76in}
    \includegraphics[height=1.25in]{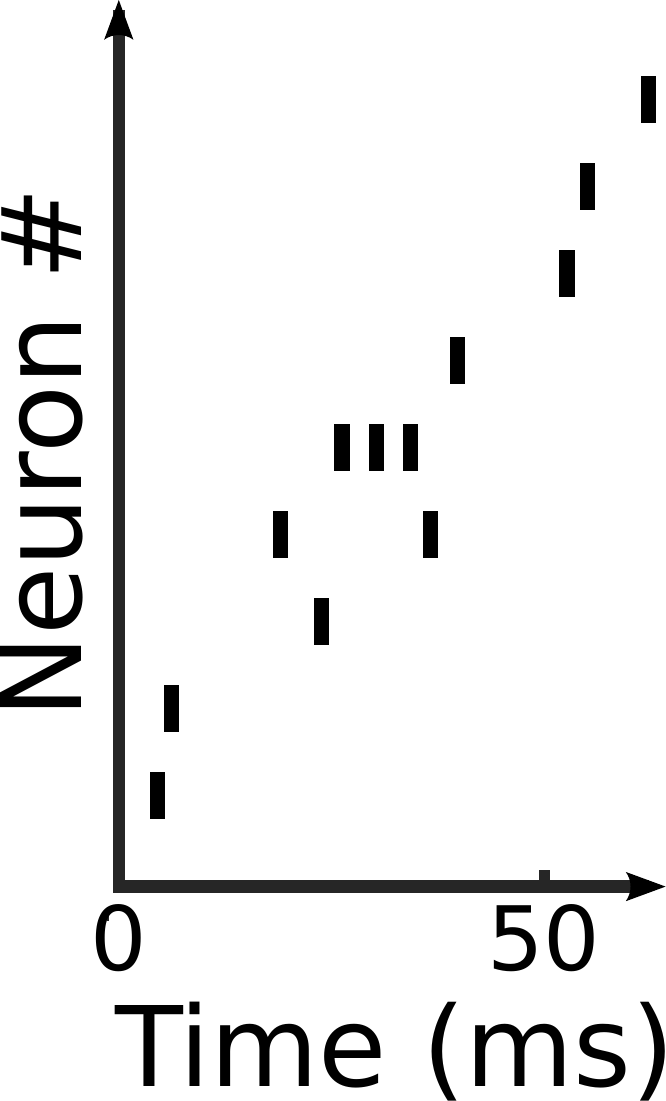}
    \caption{SWR}
  \end{subfigure}
  \hspace{1.5em}
  \begin{subfigure}[b]{3.76in}
    \includegraphics[height=1.25in]{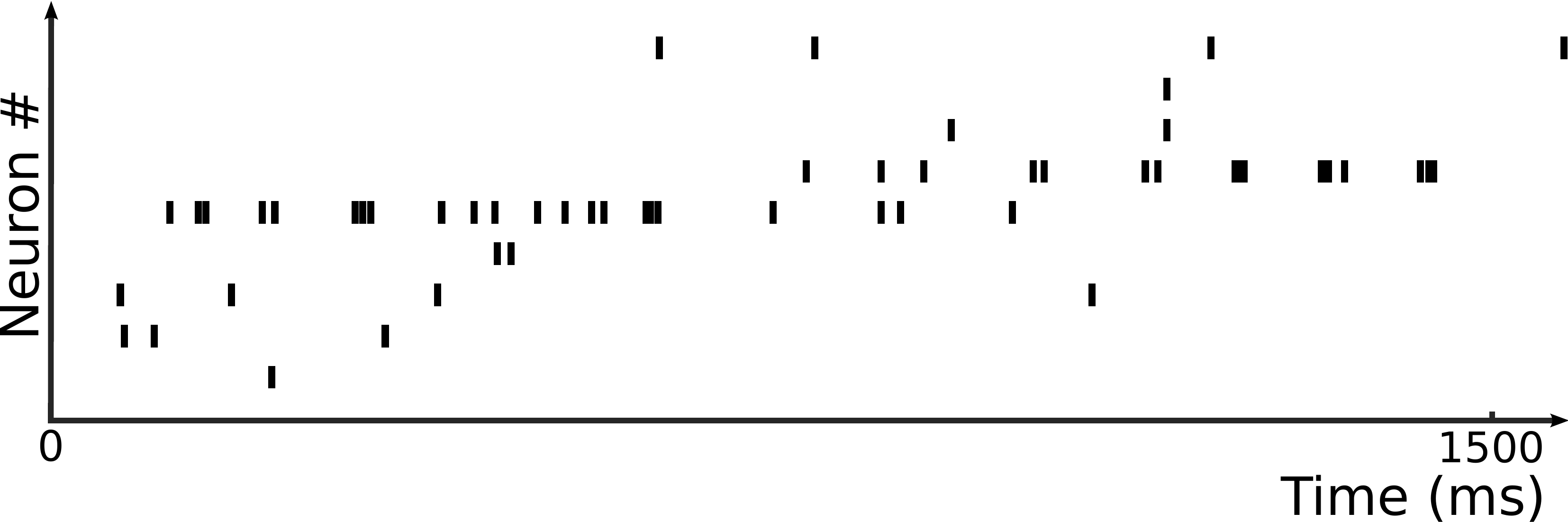}
    \caption{Place-field}
  \end{subfigure}
  \caption{\label{fig:noisy-sequences}Sequences from rat hippocampus. The SWR
    sequence in panel~(a) and the place-field sequence in panel~(b) are very
    different from each other in many ways including duration and number of
    spikes. Still, these sequences share a similar trend in the general order in
    which neurons seem to spike.}
\end{figure}

With clean illustrations such as that given in \Cref{fig:place-field-cartoon},
the concept of sequence seems simple: A sequence is intuitively, more or less,
just an ordering of neurons. Though other characteristics stand out, the fact
that there seems to be a clear ordering is a large part of what makes a sequence
seem to be structured to us. This intuition, however, breaks down as we start to
look noisy, real-world data such as that in \Cref{fig:noisy-sequences}. How do
the two sequences in this figure compare to each other? These sequences appear
to the eye to have similar trends in their neuronal firing patterns. How can we
characterize and quantify this relationship? This example immediately reveals
that our intuitive notion of a sequence is not broad enough to encapsulate the
sequences observed in real data. The situation becomes even hairier when we
think about asking a computer to tell us which sequences are similar.

In the literature that deals with neuronal sequences, oversimiplified
mathematical models have arisen to describe the concept of a sequence. These
models have served a purpose and pushed the field forward, but they are limited
in their representation of such a broad class of events.
\Cref{ch:characterizations} introduces a generalized model that more robustly
represents these sequences.

\section{Summary of thesis}
\label{sec:orgheadline3}

In the first half of this thesis
(\Crefrange{ch:characterizations}{ch:bias-matrices}), we introduce a novel tool
for characterizing and comparing neuronal sequences: the bias matrix.
\Cref{ch:characterizations} provides characterizations of neuronal sequences,
including the introduction of the bias matrix (\Cref{sec:biases}). In
\Cref{ch:bias-matrices}, we investigate mathematical properties of bias
matrices, including a description of their relationship to the previously used
center-of-mass representation (\Cref{sec:com-relationship}), a proof that all
bias matrices lie in the fundamental polytope (\Cref{thm:fundamental-polytope}
in \Cref{sec:bias-space}), and a proof that the adjacency matrix of any simple
digraph can be realized as the sign matrix of some bias matrix
(\Cref{thm:bias-network} in \Cref{sec:bias-network}).

The second part of this thesis (\Crefrange{ch:comparing}{ch:experiment})
describes how we use these tools to analyze neuronal sequences from the
hippocampus of rats. \Cref{ch:comparing} uses the bias matrix to introduce a
statistical tool for detecting non-random correlations among sequences of
neuronal activation. \Cref{ch:experiment} introduces the experiment in which the
data was recorded and describes the various types of sequences that are of
interest. In \Cref{ch:analysis}, the techniques developed in \Cref{ch:comparing}
are used to investigate the similarities between the aforementioned sequence
types. \Cref{ch:readme} serves as a README file for the code base that was
developed for these analyses. Most of the code used for this analysis was
written from scratch for this analysis.

\chapter{Characterizations of sequences}
\label{sec:orgheadline9}
\label{ch:characterizations}

As discussed in \Cref{sec:neuronal-sequences}, representation of neuronal
sequences as orderings of neurons is appealing because it makes sequences easy
to describe and easy to interpret; but it falls short in that (1) it does not
capture the complex relationships that exist between neurons, especially as
observed in noisy data, and (2) it has no simple, realistic biological
interpretation. In this chapter, we describe a new representation of sequences
that addresses these issues and, as we will see in \Cref{ch:bias-matrices},
generalizes the concept of a neuronal ordering.

\section{A combinatorial characterization}
\label{sec:orgheadline5}

Throughout the remainder of this document, we denote the collection of neurons
by \(\neurons := [n]\), where \(n \in \N\). The raw form of a sequence with which we
start is called a spike train:
\begin{definition}
A \emph{spike train} is a set of pairs of firing times and neurons, that is, a set of
the form \(\set{(t, i) \mid t \in \R,\ i \in \neurons}\).
\end{definition}
We have already seen graphical examples of spike trains in
\Cref{fig:ripple-probe-envelope,fig:noisy-sequences} in the form of spike raster
plots. As seen in the real-world sequences introduced in \Cref{ch:intro},
neuronal sequences occur on various timescales; but their sequential nature is
independent of the timescales on which they occur. As such, an appropriate
sequence representation will be timescale invariant while still maintaining
complex relationships between the spikes of neurons. Although spike trains
maintain complex relationships between the spikes of neurons, this
representation is not timescale-invariant since each spike has an associated
time. If we instead discard this time information and maintain only the order in
which spikes occur, the resultant representation will be timescale-invariant and
also represent complex inter-spike relationships. This brings us to our formal
definition of a sequence.

\begin{figure}
  \centering
  \includegraphics[height=0.75in]{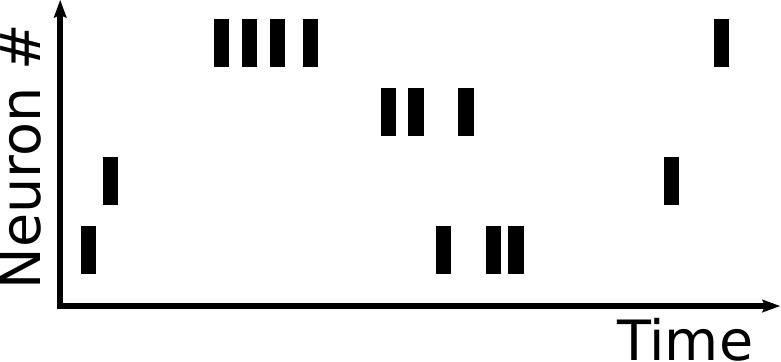}
  \caption{\label{fig:running-example}Spike raster plot. To get a sequence from
    a spike train pictured as a spike-raster plot (as shown here), we simply
    list the neuron for each spike in order of occurrence (left to right).
    Hence, the corresponding sequence is $\b{s} = (1, 2, 4, 4, 4, 4, 3, 3, 1, 3,
    1, 1, 2, 4) = 12444433131124$.}
\end{figure}

\begin{definition}
A \emph{sequence} is a finite list of neurons.
\end{definition}
If \(\b{s} = (s_1, s_2, \dots, s_\l)\) is a sequence, we will often write \(\b{s}\)
as \(\b{s} = s_1 s_2 \cdots s_\l\). Given a spike train \(T = \set{(t_1, i_1),
\dots, (t_\l, i_\l)}\) with \(t_1 \leq \cdots \leq t_\l\) and \(i_k < i_{k + 1}\) if
\(t_k = t_{k + 1}\), the corresponding sequence is \((i_1, \dots, i_\l)\). An
example of this process is illustrated in \Cref{fig:running-example}.

The set of all sequences on the neuron set \(\neurons\) is represented by
\(\sequences(\neurons)\) or, if \(\neurons\) is understood, by \(\sequences\).
\Cref{tbl:terminology} contains some terminology and notation for working with a
sequence \(\b{s} = (s_1, \dots, s_\l)\) and a neuron \(i\).

\begin{table}[htb]
\caption{\label{tbl:terminology}
The terminology of sequences}
\centering
\begin{tabular}{l|l|l}
Notation & Terminology & Meaning\\
\hline
 & a \emph{spike} of \(\b{s}\) & an entry of a sequence \(\b{s}\)\\
 & \(i\) \emph{spikes} in \(\b{s}\) & \(i = s_k\) for some \(k \in [\l]\)\\
\(\len(\b{s})\) & \emph{length} of \(\b{s}\) & the number of spikes in \(\b{s}\)\\
\(\supp(\b{s})\) & \emph{support} of \(\b{s}\) & collection of neurons that spike in \(\b{s}\)\\
\(L_i(\b{s})\) & \emph{spike set} of \(i\) & collection of indices where \(i\) spikes\\
\(c_i(\b{s})\) & \emph{spike count} of \(i\) & number of times that \(i\) spikes\\
\(\rho_i(\b{s})\) & \emph{firing rate} of \(i\) & proportion of spikes that are equal to \(i\)\\
\end{tabular}
\end{table}

\begin{example}
\label{ex:spike-count}
With \(\b{s}\) from the previous example, we see that neuron \(1\) spikes four times
and has spike set \(L_1(\b{s}) = \set{1, 9, 11, 12}\); hence, the spike count of
\(1\) in \(\b{s}\) is \(c_1(\b{s}) = 4\). Similarly, \(c_2(\b{s}) = 2\), \(c_3(\b{s}) =
3\), \(c_4(\b{s}) = 5\), and \(c_i(\b{s}) = 0\) for any other neuron \(i \in
\neurons\). Since \(\len(\b{s}) = 14\), the firing rate of each neuron in \(\b{s}\)
is \(\rho_1(\b{s}) = \frac{4}{14}\), \(\rho_2(\b{s}) = \frac{2}{14}\),
\(\rho_3(\b{s}) = \frac{3}{14}\), and \(\rho_4(\b{s}) = \frac{5}{14}\).
\end{example}

\begin{definition}
Given a sequence \(\b{s} = (s_1, \dots, s_\l)\) and an index set \(K = \set{k_1,
\dots, k_{\l'}} \sset [\l]\) with \(k_1 < \cdots < k_{\l'}\), we define the
\emph{subsequence} of \(\b{s}\) corresponding to \(K\) to be \(\b{s}_K := (s_{k_1}, \dots,
s_{k_{\l'}})\). We say that a sequence \(\b{s}'\) is a subsequence of \(\b{s}\) if
\(\b{s}' = \b{s}_K\) for some index set \(K \sset [\l]\).
\end{definition}

\begin{example}
Let \(\b{s} = (1, 2, 4, 4, 4, 4, 3, 3, 1, 3, 1, 1, 2, 4)\) be the sequence from
\Cref{fig:running-example}, and let \(K_1 = [6]\) and \(K_2 = \set{1, 2, 9, 11, 12,
13}\). Then the subsequences of \(\b{s}\) corresponding to \(K_1\) and \(K_2\) are
\(\b{s}_{K_1} = (1, 2, 4, 4, 4, 4)\) and \(\b{s}_{K_2} = (1, 2, 1, 1, 1, 2)\),
respectively.
\end{example}

The following operations will be useful in our discussion of sequences.

\begin{definition}
Given any two sequences \(\b{u} = (u_1, \dots, u_\l)\) and \(\b{v} = (v_1, \dots,
v_m)\), define the \emph{concatenation} of \(\b{u}\) with \(\b{v}\) to be the sequence
\(\b{u} \cdot \b{v} = \b{u} \b{v} = (u_1, \dots, u_\l, v_1, \dots, v_m)\).
\end{definition}

\begin{notation}
For any neuron \(i \in \neurons\) and any positive integer \(m \in \N\), let \(i^m\)
denote the sequence \((s_1, \dots, s_m)\) with \(s_k = i\) for all \(k \in [m]\).
\end{notation}

\section{Center-of-mass sequences}
\label{sec:orgheadline6}

A sequence can induce an ordering by many means. One sandard way of performing
this reduction is by considering the average location in which a neuron fires.

\begin{definition}
For each neuron \(i \in \neurons\) and sequence \(\b{s} = (s_1, \dots, s_\l)\), we
define the \emph{(unnormalized) center of mass} of neuron \(i\) in sequence \(\b{s}\) to
be
\[
  \com_i(\b{s}) :=
  \begin{cases}
    \frac{1}{c_i(\b{s})} \sum_{k \in L_i(\b{s})} k, & c_i(\b{s}) > 0; \\
    \frac{\len(\b{s}) + 1}{2}, & c_i(\b{s}) = 0.
  \end{cases}
\]
\end{definition}
Note that \(\com_i(\b{s}) \in [1, \len(\b{s})]\) for all \(i \in \neurons\).
Moreover, when a neuron does not fire in a sequence, the corresponding
center-of-mass value is chosen so that it falls in the middle of the range \([1,
\l]\). Thus, \(\com_i(\b{s})\) reduces all spikes from neuron \(i\) in sequence
\(\b{s}\) to a single value; but this value is dependent on the length of \(\b{s}\).
To allow for comparison of sequences of different lengths, we shift and
normalize \(\com_i(\b{s})\) so that the resultant value lies in the interval \((-1,
1)\) regardless of the length of \(\b{s}\):
\begin{definition}
For each neuron \(i \in \neurons\) and sequence \(\b{s} = (s_1, \dots, s_\l)\), we
define the \emph{(normalized) center of mass} of neuron \(i\) in sequence \(\b{s}\) to be
\[ \gamma_i(\b{s}) := \frac{2\com_i(\b{s}) - 1}{\len(\b{s})} - 1. \]
\end{definition}
Note that with this definition, neurons that do not fire in the sequence have
\(\gamma_i(\b{s}) = 0\). If the sequence consists of just one spike, then the
corresponding neuron has \(\com_i(\b{s})=1\) and hence \(\gamma_i(\b{s})=0\). On the
other hand, if a neuron fires only once in the sequence and this corresponds to
the first or last spike, then \(\gamma_i(\b{s}) = -1+\frac{1}{\len(\b{s})}\) or
\(\gamma_i(\b{s}) = 1-\frac{1}{\len(\b{s})}\), respectively, so that the endpoints
of the \([-1,1]\) range are only attained in the limit as \(\len(\b{s}) \to
\infty\). These centers of mass can now be used to induce an ordering on the
underlying neurons.

\begin{definition}
We define the \emph{center-of-mass ordering} of \(\b{s}\) to be \(\s(\b{s}) := (i_1,
\dots, i_n)\) where \(\set{i_1, \dots, i_n} = [n]\) are such that
\(\gamma_{i_1}(\b{s}) \leq \cdots \leq \gamma_{i_n}(\b{s})\) and, in the case that
\(\gamma_{i_k}(\b{s}) = \gamma_{i_{k + 1}}(\b{s})\), such that \(i_k < i_{k + 1}\).
\end{definition}

\begin{example}
Let \(\b{s} = 12444433131124\). Then, as we saw in \Cref{ex:spike-count},
\(L_1(\b{s}) = \set{1, 9, 11, 12}\) and \(c_1(\b{s}) = 4\). Now, \(\com_1(\b{s}) =
\frac{1}{c_1(\b{s})} \sum L_1(\b{s}) = \frac{33}{4}\) and \(\gamma_1(\b{s}) =
\frac{2 (\nicefrac{33}{4}) - 1}{14} - 1 = \frac{3}{28}\). Similarly, we find that
\(\com(\b{s}) = (\frac{33}{4}, \frac{15}{2}, \frac{25}{3}, \frac{32}{5})\) and
\(\gamma(\b{s}) = (\frac{3}{28}, 0, \frac{5}{42}, -\frac{11}{70})\). Further,
\(\gamma_4(\b{s}) < \gamma_2(\b{s}) < \gamma_1(\b{s}) < \gamma_3(\b{s})\) and,
hence, \(\s(\b{s}) = (4, 2, 1, 3)\).
\end{example}

It is worth mentioning that when the concept of centers of mass is used in the
literature, the starting point is usually spike trains, not our concept of a
neuronal sequence. When starting with spike trains, the spike times of a neuron
are averaged instead of its indices in the sequence; these average firing times
then induce an ordering in the same way that the centers of mass do.

\section{Pairwise biases}
\label{sec:orgheadline7}
\label{sec:biases}

If each neuron active in a sequence spikes only once, then the sequence itself
is an ordering of its active neurons. In such a case, the center-of-mass
ordering equals the original sequence. A broader class of sequences induce an
ordering in a similarly unambiguous way. For instance, in the sequence
\(3311222\), every spike of neuron \(3\) comes before every spike of neuron \(1\);
further, every spike of neuron \(1\) occurs before every spike of neuron \(2\).
Hence, \(3311222\) induces the unambiguous order \(312\). Although this ordering can
be computed with centers of mass, the ordering is also immediately obvious upon
inspection of the sequence since all of the spikes from each neuron are grouped
together. We give a name to sequences such as this.

\begin{definition}
Disjoint sets \(I, J \sset \N\) are \emph{interlaced} if there exist \(i, i' \in I\) and
\(j, j' \in J\) such that \(i < j\) and \(i' > j'\).
\end{definition}

\begin{definition}
A sequence \(\b{s}\) is \emph{pure} if the spike sets of the active neurons are not
interlaced.
\end{definition}

In a pure sequence, all of the spikes from one neuron must follow (or precede)
all of the spikes from any other neuron. In particular, the indices of the
spikes of each individual neuron form a contiguous set of natural numbers. To
get a sequence that is not pure from one that is pure, we could transpose
adjacent spikes from different neurons. For example, the pure sequence
\(111122223333\) can be transformed into the impure sequence \(111212223333\) by
transposing the fourth and fifth spikes in the pure sequence. In a sense, this
new sequence is minimally impure: The spikes from neuron \(3\) are still
contiguous, and only one pair of spikes from neurons \(1\) and \(2\) are "out of
order." This perspective suggests that the purity of a sequence (1) is somehow
related to the pairwise relationships between the spikes of different neurons
and (2) is dependent on the collection of all such pairs of neurons. In other
words, the purity of a sequence depends on the bias of each pair of neurons to
fire in a particular order. We begin to quantify these biases with the
bias-count matrix below.

\begin{definition}
The \emph{bias-count matrix} of the sequence \(\b{s}\) is the matrix \(C(\b{s}) :=
[c_{ij}(\b{s})]_{i, j \in \neurons}\) where \(c_{ij}(\b{s}) :=
\abs{L_{ij}(\b{s})}\) and \(L_{ij}(\b{s}) := \set{ (k, k') \mid k < k',\ s_k = i,\
\text{and}\ s_{k'} = j}\). In words, \(c_{ij}(\b{s})\) is the number of length-2
subsequences of \(\b{s}\) that are equal to \((i, j)\).
\end{definition}

\begin{example}
Consider the sequence \(\b{s} = (1, 2, 4, 4, 4, 4, 3, 3, 1, 3, 1, 1, 2, 4)\) from
\Cref{fig:running-example}. Note that \(L_1(\b{s}) = \set{1, 9, 11, 12}\) and
\(L_2(\b{s}) = \set{2, 13}\). Hence, we see that \[ L_{12}(\b{s}) = \set{(1, 2),
(1, 13), (9, 13), (11, 13), (12, 13)} \] and \[ L_{21}(\b{s}) = \set{(2, 9), (2,
11), (2, 12)}; \] so \(c_{12}(\b{s}) = \abs{L_{12}(\b{s})} = 5\) and
\(c_{21}(\b{s}) = \abs{L_{21}(\b{s})} = 3\). In a similar way for the other
entries \(c_{ij}(\b{s})\), we find that the bias-count matrix for \(\b{s}\) is
\[
C(\b{s}) =\begin{bmatrix}
 6 & 5 & 4 & 8 \\
 3 & 1 & 3 & 6 \\
 8 & 3 & 3 & 3 \\
 12 & 4 & 12 & 10 \\
\end{bmatrix}.
\]\end{example}

A few simple facts about the bias-count matrix are immediate. First, since the
definition of the bias-count matrix arose out of a discussion of pure sequences,
it is worth noting that the bias-count matrix completely captures whether or not
a sequence is pure.

\begin{lemma}
A sequence \(\b{s}\) with \(\supp(\b{s}) = \neurons\) has at least
\(\binom{\nneurons}{2}\) non-zero, off-diagonal entries, with equality if and only
if \(\b{s}\) is pure.

\begin{proof}
Any \(C(\b{s})\) must obviously have at least \(\binom{\nneurons}{2}\) non-zero,
off-diagonal entries since each of the \(\binom{\nneurons}{2}\) pairs \(i, j \in
\neurons\) of active neurons must fire in at least one of the two possible
orderings \((i, j)\) or \((j, i)\). If \(\b{s}\) is pure, it must be the case that
exactly one of the orderings \((i, j)\) or \((j, i)\) appears in \(\b{s}\) for each
\(i, j \in \neurons\) with \(i \neq j\), of which there are exactly
\(\binom{\nneurons}{2}\). If \(C(\b{s})\) has more than \(\binom{\nneurons}{2}\)
non-zero, off-diagonal entries, then, by the pigeonhole principle, there exist
distinct neurons \(i, j \in \neurons\) such that \(c_{ij}(\b{s}) > 0\) and
\(c_{ji}(\b{s}) > 0\); so we must have that the spikes of \(i\) and \(j\) are
interlaced, meaning that \(\b{s}\) is not pure. This finishes the proof.
\end{proof}
\end{lemma}

Next, one cannot always recover a sequence from its bias-count matrix. Consider,
for example, \(\b{s}_1 = 1221\) and \(\b{s}_2 = 2112\). Then \(C(\b{s}_1) =
C(\b{s}_2) = \sbmtx{1 & 2 \\ 2 & 1}\). Hence, thought of as a map, \(C: \sequences
\to \N^{\nneurons \x \nneurons}\) is not injective. Though we cannot recover
\(\b{s}\) from \(C(\b{s})\), we can recover the spike counts \(c_i(\b{s})\). To see
this, first note that \[ c_{ii}(\b{s}) = \binom{c_i(\b{s})}{2}. \] Since \(f : \N
\to \N\) by \(f(n) = \binom{n}{2}\) is injective, it has an inverse; therefore, \[
c_i(\b{s}) =f^{-1}(c_{ii}(\b{s})) = \half + \half \sqrt{1 + 8 c_{ii}(\b{s})}. \]

Last, note that \(c_{ij} + c_{ji} = c_i c_j\). In words, \(c_{ij} + c_{ji}\) is
equal to the product of the number of spikes of \(i\) and of \(j\). This is easy to
see because we just choose one index for \(i\) and one index for \(j\) (which can be
done in \(c_i c_j\) ways); in each such choice, either \(i\) follows \(j\) (counted by
\(c_{ji}\)) or \(j\) follows \(i\) (counted by \(c_{ji}\)).

Though \(C(\b{s})\) encapsulates the idea of purity, it does so in a somewhat
unsatisfactory way. The sequences \(123\) and \(112233\), for instance, are both
pure sequences with the same neuronal ordering, yet their bias-count matrices
\[
C(123) =\begin{bmatrix}
 0 & 1 & 1 \\
 0 & 0 & 1 \\
 0 & 0 & 0 \\
\end{bmatrix}
\qquad\text{and}\qquad C(112233) =\begin{bmatrix}
 1 & 4 & 4 \\
 0 & 1 & 4 \\
 0 & 0 & 1 \\
\end{bmatrix}
\]are quite different. The issue is that \(112233\) has more pairs of spikes than
\(123\) does. We can correct for this by normalizing each entry by an appropriate
value. How likely is neuron \(j\) to follow neuron \(i\)? There are \(c_{ij}\) index
pairs in which \(j\) follows \(i\) out of a total of \(c_i c_j\) possible index pairs.
This leads us to the definition of a second type of bias matrix.

\begin{definition}
The \emph{probability-bias matrix} of a sequence \(\b{s}\) is the matrix
\(\bias(\b{s}) := [b_{ij}(\b{s})]_{i, j \in \neurons}\) where \(b_{ij}(\b{s}) :=
\frac{c_{ij}(\b{s})}{c_i(\b{s}) c_j(\b{s})}\).
\end{definition}

\begin{example}
The probability-bias matrix for the sequence \(\b{s} = 12444433131124\) from
\Cref{fig:running-example} is
\[
\bias(\b{s}) =\begin{bmatrix}
 \nicefrac{6}{16} & \nicefrac{5}{8} & \nicefrac{4}{12} & \nicefrac{8}{20} \\
 \nicefrac{3}{8} & \nicefrac{1}{4} & \nicefrac{3}{6} & \nicefrac{6}{10} \\
 \nicefrac{8}{12} & \nicefrac{3}{6} & \nicefrac{3}{9} & \nicefrac{3}{15} \\
 \nicefrac{12}{20} & \nicefrac{4}{10} & \nicefrac{12}{15} & \nicefrac{10}{25} \\
\end{bmatrix}
=\begin{bmatrix}
 \nicefrac{3}{8} & \nicefrac{5}{8} & \nicefrac{1}{3} & \nicefrac{2}{5} \\
 \nicefrac{3}{8} & \nicefrac{1}{4} & \nicefrac{1}{2} & \nicefrac{3}{5} \\
 \nicefrac{2}{3} & \nicefrac{1}{2} & \nicefrac{1}{3} & \nicefrac{1}{5} \\
 \nicefrac{3}{5} & \nicefrac{2}{5} & \nicefrac{4}{5} & \nicefrac{2}{5} \\
\end{bmatrix}.
\]\end{example}

The probability-bias matrix maps the set \(\sequences\) of sequences into the
matrices \([0, 1]^{\nneurons \x \nneurons}\), and it does so in a way that each
off-diagonal entry of the matrix has a natural interpretation in terms of how
likely it is to see a pair of spikes in a given order. As such, it should hold
that \(b_{ij} + b_{ji} = 1\). Recalling that \(c_{ij} + c_{ji} = c_i c_j\), we see
that this indeed holds: \[ b_{ij} + b_{ji} = \frac{c_{ij}}{c_i c_j} +
\frac{c_{ji}}{c_j c_i} = \frac{c_{ij} + c_{ji}}{c_i c_j} = \frac{c_i c_j}{c_i
c_j} = 1. \] Stated simply, when \(i\) and \(j\) both spike in a sequence, each pair
of spikes must occur in one order or the other. As defined, this likelihood is
global in terms of the sequence, which gives it a sort of skew-symmetric nature:
\(b_{ij} = 1 - b_{ji}\). As such, it is clear that not every point in the
hypercube \([0, 1]^{\nneurons \x \nneurons}\) can be obtained as a
probability-bias matrix. Though that much is known simply because no irrational
coordinate can ever be hit by the map, the above relation shows that at least
one further relation is imposed on this image. In a sense, this relation
suggests that the bias matrix is still not quite in the right form: Since
\(b_{ij} + b_{ji} = 1\), we see that \(b_{ij} - \half = \half - b_{ji} = -(b_{ji} -
\half)\). Because the relation comes from the fact that the probability biases
are probabilities, the entries are centered at \(\half\), whereas we might more
naturally think of a non-bias (i.e., a maximally unbiased bias) as being equal
to \(0\). Similarly, we might want the sign of a bias to convey the tendency of
neurons to spike in the given order (a positive bias) or the opposite order (a
negative bias). Perhaps we then care more about the difference between \(b_{ij}\)
and \(b_{ji}\) than we do about each of them individually. This leads us to define
our final version of the bias matrix.

\begin{definition}
The \emph{skew-bias matrix} of the sequence \(\b{s}\) is the matrix \(\skbias(\b{s}) :=
\bias(\b{s}) = \bias(\b{s})^T = [\beta_{ij}(\b{s})]_{i, j \in \neurons}\) with
entries \(\beta_{ij}(\b{s}) = b_{ij}(\b{s}) - b_{ji}(\b{s}) =
\frac{c_{ij}(\b{s}) - c_{ji}(\b{s})}{c_i(\b{s}) c_j(\b{s})}\). The skew-bias
matrix \(\skbias(\b{s})\) is a scaling of the skew-symmetric part of the
probability-bias matrix \(\bias(\b{s})\).
\end{definition}

Note that when \(i \neq j\), \(\beta_{ij}(\b{s}) = b_{ij} - b_{ji} = b_{ij} - (1 -
b_{ij}) = 2b_{ij} - 1\). The skew biases are now centered at zero and have a sign
that indicates the direction of the bias. Further, the skew-bias matrix is
skew-symmetric as a matrix since \(\beta_{ij} = b_{ij} - b_{ji} = -(b_{ji} -
b_{ij}) = -\beta_{ji}\).

\begin{example}
\label{ex:skew-bias}
The skew-bias matrix for the sequence \(\b{s} = (1, 2, 4, 4, 4, 4, 3, 3, 1, 3, 1,
1, 2, 4)\) from \Cref{fig:running-example} is
\[
\skbias(\b{s}) =\begin{bmatrix}
 0 & \nicefrac{1}{4} & -\nicefrac{1}{3} & -\nicefrac{1}{5} \\
 -\nicefrac{1}{4} & 0 & 0 & -\nicefrac{1}{5} \\
 \nicefrac{1}{3} & 0 & 0 & -\nicefrac{3}{5} \\
 \nicefrac{1}{5} & \nicefrac{1}{5} & \nicefrac{3}{5} & 0 \\
\end{bmatrix}.
\]Here, it is obvious at a glance that, for instance, neuron \(2\) tends to follow
neuron \(1\) and that neurons \(2\) and \(3\) are maximally unbiased toward each
other.
\end{example}

\section{Higher-order biases}
\label{sec:orgheadline8}

When creating a mathematical model, it is important to question what information
the model discards. The pairwise biases from the previous section attempt to
capture the tendency of pairs of neurons to fire in a particular order. While
this certainly preserves some information about complex relationships between
neurons/spikes, it may at times discard more information than is desired.

\begin{example}
Consider the sequences \(\b{s}_1 = 123321\) and \(\b{s}_2 = 321123\). These
sequences are indistinguishable by pairwise biases:
\[
C(\b{s}_1) = C(\b{s}_2) =\begin{bmatrix}
 1 & 2 & 2 \\
 2 & 1 & 2 \\
 2 & 2 & 1 \\
\end{bmatrix}.
\]Note that, since the probability- and skew-bias matrices are computed from the
bias-count matrix, these matrices also cannot distinguish between \(\b{s}_1\) and
\(\b{s}_2\).
\end{example}

We might want, however, for the sequences from the above example to be
distinguishable using some bias-like method. This may or may not be possible for
all pairs of sequences, but we can extend the idea behind the pairwise-bias
matrices to consider more neurons within each individual bias. For instance, we
can count the subsequences that are equal to each permutation of \(123\) (where,
before, we counted the subsequences equal to each permutation of \(12\)). We call
these higher-order biases.

\begin{definition}
For a list \(\b{v} = (v_1, \dots, v_k)\) of \(k\) distinct neurons, we define
\[
  L_\b{v}(\b{s}) := \{ (i_1, \dots, i_k) \mid i_1 < \dots < i_k \text{ and }
                        s_{i_j} = v_j \text{ for all } j \in [k] \}.
\]
to be the collection of subsequences of \(\b{s}\) that are equal to \(\b{v}\) and
define \[ c_\b{v}(\b{s}) := \abs{L_\b{v}(\b{s})} \] to be the count of those
subsequences; we call \(c_\b{v}(\b{s})\) an \emph{order-\(k\) bias count}. Moreover,
analogously to the bias-count matrix, we define the \emph{order-\(k\) bias-count
vector} \[ C_k(\b{s}) := \sbmtx{c_{\b{v}_1}(\b{s}) \\ \vdots
\\ c_{\b{v}_{k'}}(\b{s})} \] where \(\b{v}_1, \dots, \b{v}_{k'}\) is the list of
length-\(k\) biases in lexicographical order and \(k' = \binom{\nneurons}{k} k! =
\frac{\nneurons!}{(\nneurons - k)!} = \nneurons_{(k)}\).
\end{definition}

\begin{example}
Continuing the previous example, let \(\b{s}_1 = 123321\) and \(\b{s}_2 = 321123\).
Note that \(L_{231}(\b{s}_1) = \set{(2, 3, 6), (2, 4, 6)}\) and \(L_{231}(\b{s}_2)
= \emptyset\). In particular, \(c_{231}(\b{s}_1) = 2\) but \(c_{231}(\b{s}_2) = 0\).
Hence, the order-3 biases can distinguish between the sequences \(\b{s}_1\) and
\(\b{s}_2\).
\end{example}

Note that \(c_\b{v}\) is a generalization of \(c_{ij}\) and that \(c_{ij} = c_{(i,
j)} = c_\b{v}\) for \(\b{v} = (1, 2)\). This also unifies the definitions of the
spike-count \(c_i\) and the bias-count \(c_{ij}\); that is, \(c_i\) and \(c_{ij}\) are
order-\(1\) and order-\(2\) biases, respectively. But this added information comes
at a cost: The number of order-\(k\) biases is \(O(\nneurons^k)\). With this
increase in the number of biases, it is valuable to see that a higher-order bias
vector contains at least as much information about a sequence as a lower-order
conterpart.

\begin{lemma}
The collection of order-\((k - 1)\) biases can be computed from the collection of
order-\(k\) biases if the spike counts are also known.

\begin{proof}
Let \(\b{s} \in \sequences\) be given, and let \(\b{v} = (v_1, \dots, v_{k - 1})
\in \neurons^{k - 1}\) be a list of distinct neurons. We want to compute the
order-\((k - 1)\) bias \(c_\b{v}(\b{s})\) from the collection of order-\(k\) biases.
Recall that \(c_\b{v}(\b{s})\) is the number of times that \(\b{v}\) appears as a
subsequence of \(\b{s}\). Let \(U\) be the collection of length-\(k\) sequences of
distinct neurons and of which \(\b{v}\) is a subsequence. To make this more
precise, define the sets \(V = \set{v_1, \dots, v_{k - 1}}\) and \(I = \neurons
\setminus V\). We can now write \(U\) as \[ U = \set{\b{v}_{[m]} \cdot v \cdot
\b{v}_{\set{m + 1, \dots, k}} \mid v \in N,\ m \in [k]}. \] We want to
double-count the number \(d\) of subsequences of \(\b{s}\) that are equal to some
element of \(U\). First, by definition of \(d\) and the bias count \(c_\b{u}(\b{s})\),
we have that \(d = \sum_{\b{u} \in U} c_\b{u}(\b{s})\).

To count in another way, let \(B = \bigcup_{v \in N} L_v(\b{s})\) be the
collection of indices of \(\b{s}\) that are not spikes from the set \(V\). Now, each
subsequence of \(\b{s}\) that is equal to an element of \(U\) has the form \(\b{s}_{J
\cup \set{m}}\) for some \(m \in B\) and some \(J \sset [\l]\) with \(\b{s}_J =
\b{v}\); moreover, each such \(J\) and \(m\) yields such a subsequence. As there are
\(c_\b{v}(\b{s})\) such \(J\) and \(\sum_{v \in N} c_v(\b{s})\) such \(m\), we find that
\(d = c_\b{v}(\b{s}) \sum_{v \in N} c_v(\b{s})\). Finally, putting together the
two expressions for \(d\), we find that \[ c_\b{v}(\b{s}) = \parens{\sum_{\b{u}
\in U} c_\b{u}(\b{s})} \Big/ \parens{\sum_{v \in N} c_v(\b{s})} .\] This
completes the proof.
\end{proof}
\end{lemma}

Similarly to the pairwise biases, we may not care about subsequence counts
\(c_\b{v}\) so much as the likelihood that the neurons \(v_1, \dots, v_k\) will
appear in the order \(\b{v} = (v_1, \dots, v_k)\) provided that they all appear.
\begin{definition}
Given a list \(\b{v} = (v_1, \dots, v_k)\) of distinct neurons, define the
\emph{order-\(k\) probability-bias} to be the number \(b_\b{v}(\b{s}) :=
\frac{c_\b{v}(\b{s})}{c_{v_1}(\b{s}) \cdots c_{v_k}(\b{s})}\). The \emph{order-\(k\)
probability-bias vector} is the vector \[ B_k(\b{s}) :=
\sbmtx{b_{\b{v}_1}(\b{s}) \\ \vdots \\ b_{\b{v}_{k'}}(\b{s})}, \] where
\(\b{v}_1, \dots, \b{v}_{k'}\) is the list of length-\(k\) biases in lexicographical
order.
\end{definition}
Note that the denominator of \(b_\b{v}(\b{s})\) is the total number of length-\(k\)
subsequences of \(\b{s}\) that contain all neurons \(v_1, \dots, v_k\); together
with the fact that \(c_\b{v}(\b{s})\) is a count of such sequences, we see that
\(b_\b{v}(\b{s})\) is really the proportion of all length-\(k\) subsequences of
\(\b{s}\) containing \(v_1, \dots, v_k\) that are equal to \(\b{v}\).

\begin{example}
Let \(\b{s} = 12444433131124\). Then \(L_1(\b{s}) = \set{1, 9, 11, 12}\),
\(L_2(\b{s}) = \set{2, 13}\), and \(L_3(\b{s}) = \set{7, 8, 10}\). So we see that
\(L_{123}(\b{s}) = \set{(1, 2, 7), (1, 2, 8), (1, 2, 10)}\) and \(L_{132}(\b{s}) =
\set{(1, 7, 13), (1, 8, 13), (1, 10, 13), (9, 10, 13)}\); hence,
\[
  c_{123}(\b{s}) = \abs{L_{123}(\b{s})} = 3
  \qquad \text{and} \qquad
  c_{132}(\b{s}) = \abs{L_{132}(\b{s})} = 4.
\]
Since \(c_1(\b{s}) = 4\), \(c_2(\b{s}) = 2\), and \(c_3(\b{s}) = 3\), we also have
that
\[
  b_{123}(\b{s}) = \frac{c_{123}(\b{s})}{c_1(\b{s}) c_2(\b{s}) c_3(\b{s})} = \frac{1}{8}
  \qquad \text{and} \qquad
  b_{123}(\b{s}) = \frac{c_{123}(\b{s})}{c_1(\b{s}) c_2(\b{s}) c_3(\b{s})} = \frac{1}{6}.
\]
Similarly, we can compute the remainder of the order-3 bias-count and
probability-bias vectors. The vectors are shown as matrices here strictly for
formatting purposes:
\[
C_3(\b{s}) =\begin{bmatrix}
 c_{123}(\b{s}) & c_{213}(\b{s}) & c_{312}(\b{s}) & c_{412}(\b{s}) \\
 c_{124}(\b{s}) & c_{214}(\b{s}) & c_{314}(\b{s}) & c_{413}(\b{s}) \\
 c_{132}(\b{s}) & c_{231}(\b{s}) & c_{321}(\b{s}) & c_{421}(\b{s}) \\
 c_{134}(\b{s}) & c_{234}(\b{s}) & c_{324}(\b{s}) & c_{423}(\b{s}) \\
 c_{142}(\b{s}) & c_{241}(\b{s}) & c_{341}(\b{s}) & c_{431}(\b{s}) \\
 c_{143}(\b{s}) & c_{243}(\b{s}) & c_{342}(\b{s}) & c_{432}(\b{s}) \\
\end{bmatrix}=
\begin{bmatrix}
 3 & 1 & 8 & 12 \\
 9 & 3 & 8 & 4 \\
 4 & 8 & 0 & 0 \\
 4 & 3 & 3 & 0 \\
 4 & 12 & 0 & 32 \\
 12 & 12 & 0 & 12 \\
\end{bmatrix}
\]and
\[
B_3(\b{s}) =\begin{bmatrix}
 b_{123}(\b{s}) & b_{213}(\b{s}) & b_{312}(\b{s}) & b_{412}(\b{s}) \\
 b_{124}(\b{s}) & b_{214}(\b{s}) & b_{314}(\b{s}) & b_{413}(\b{s}) \\
 b_{132}(\b{s}) & b_{231}(\b{s}) & b_{321}(\b{s}) & b_{421}(\b{s}) \\
 b_{134}(\b{s}) & b_{234}(\b{s}) & b_{324}(\b{s}) & b_{423}(\b{s}) \\
 b_{142}(\b{s}) & b_{241}(\b{s}) & b_{341}(\b{s}) & b_{431}(\b{s}) \\
 b_{143}(\b{s}) & b_{243}(\b{s}) & b_{342}(\b{s}) & b_{432}(\b{s}) \\
\end{bmatrix}=
\begin{bmatrix}
 \nicefrac{1}{8} & \nicefrac{1}{24} & \nicefrac{1}{3} & \nicefrac{3}{10} \\
 \nicefrac{9}{40} & \nicefrac{3}{40} & \nicefrac{2}{15} & \nicefrac{1}{15} \\
 \nicefrac{1}{6} & \nicefrac{1}{3} & 0 & 0 \\
 \nicefrac{1}{15} & \nicefrac{1}{10} & \nicefrac{1}{10} & 0 \\
 \nicefrac{1}{10} & \nicefrac{3}{10} & 0 & \nicefrac{8}{15} \\
 \nicefrac{1}{5} & \nicefrac{2}{5} & 0 & \nicefrac{2}{5} \\
\end{bmatrix}.
\]\end{example}

We define no higher-order version of the skew-bias matrix \(\skbias\) because the
skew-nature of \(\skbias\) arose from the fact that only two probabilities were
involved for each pair of neurons, that is, from the fact that \(b_{ij} + b_{ji}
= 1\). In the higher-order version, a more complicated relationship exists,
namely \[ \sum_{\s \in \perms{\set{n_1, \dots, n_k}}} b_{(n_{\s(i)}, \dots,
n_{\s(n)})} = 1 \] for each set of distinct neurons \(\set{n_1, \dots, n_k} \sset
\neurons\), where \(\perms(\Omega)\) is the collection of permutations of the set
\(\Omega\).

\chapter{Bias Matrices}
\label{sec:orgheadline14}
\label{ch:bias-matrices}

In the previous chapter, we formalized the concept of a neuronal sequence and
introduced the pairwise biases associated with those sequences. These pairwise
biases---of which we have three separate formulations---capture the tendency of
pairs of neurons to fire in a particular order. As the main drive of the
discussion in the previous chapter was to motivate and define these biases, it
is worth taking more space here to investigate further properties of these
matrices. Recall that we defined the bias-count matrix \(C(\b{s}) = [c_{ij}]_{i,
j \in \neurons}\), the probability-bias matrix \(\bias(\b{s}) = [b_{ij}]_{i, j \in
\neurons}\), and the skew-bias matrix \(\skbias(\b{s}) = [\beta_{ij}]_{i, j \in
\neurons}\).

\begin{lemma}
Let \(\b{s}\) and \(\b{s}'\) be sequences, and consider the following statements:
\begin{enumerate}
\item \(C(\b{s}) = C(\b{s}')\)
\item \(\bias(\b{s}) = \bias(\b{s}')\)
\item \(\skbias(\b{s}) = \skbias(\b{s}')\)
\end{enumerate}
Then statements (1) and (2) are equivalent. Further, (1) and (2) imply (3).

\begin{proof}
First, note that that (1) implies (2) since \(\bias\) was defined solely in terms
of \(C\). Similarly, (2) implies (3) since \(\skbias\) was defined solely in terms
of \(\bias\). All that remains to be shown is that (2) implies (1). To see this,
note that the spike count \(c_i(\b{s})\) can be recovered from the diagonal entry
\(b_{ii}(\b{s})\) of \(\bias(\b{s})\): \[ b_{ii}(\b{s}) =
\frac{c_{ii}(\b{s})}{c_i(\b{s}) c_i(\b{s})} =
\frac{\binom{c_i(\b{s})}{2}}{c_i(\b{s}) c_i(\b{s})} = \half \parens{1 -
\frac{1}{c_i(\b{s})}} \implies c_i(\b{s}) = \frac{1}{1 - 2b_{ii}(\b{s})} .\]
Knowing each spike count now allows us to recover any entry of \(C(\b{s})\): \[
b_{ij}(\b{s}) = \frac{c_{ij}(\b{s})}{c_i(\b{s}) c_j(\b{s})} \implies
c_{ij}(\b{s}) = b_{ij}(\b{s}) c_i(\b{s}) c_j(\b{s}) .\] Hence, \(C(\b{s}) =
C(\b{s}')\) if \(\bias(\b{s}) = \bias(\b{s}')\). Therefore, (2) implies (1), as
claimed. This completes the proof.
\end{proof}
\end{lemma}

Note that statement~(3) from the above lemma is not equivalent to
statements~(1) and~(2). This is easily seen, for example, with the pure
sequences \(12\) and \(1122\): \[ \skbias(12) = \skbias(1122) = \bmtx{0 & 1 \\ -1 &
0} \] but \[ C(12) = \bmtx{0 & 1 \\ 0 & 0} \neq \bmtx{1 & 4 \\ 0 & 1} = C(1122).
\] The reason for this is that \(\skbias(\b{s})\) does not contain sufficient
information to recover the spike counts since \(\beta_{ii}(\b{s}) = 0\) for all
neurons \(i \in \neurons\) and all sequences \(\b{s}\).

\section{Relationship to center-of-mass vectors}
\label{sec:orgheadline10}
\label{sec:com-relationship}

Bias matrices were introduced in an attempt to deal with issues associated with
center-of-mass vectors, but these two concepts were reached through very
different approaches. Still, a surprising relationship exists between them: The
center-of-mass vector \(\com(\b{s})\) can be computed from the bias-count matrix
\(C(\b{s})\).

\begin{lemma}
\label{lem:center-from-bias-count}
If \(\b{s} \in \sequences\) is a sequence and \(i \in \supp(\b{s})\) is an active
neuron, then \(\com_i(\b{s}) = 1 + \frac{1}{c_i(\b{s})} \sum_{j=1}^n
c_{ji}(\b{s})\).

\begin{proof}
Let \(\l = \len(\b{s})\). Also, let \(c_{ji}^{(k)}(\b{s})\) denote the number of
times neuron \(j\) spikes before the \$k\$-th spike in the sequence \(\b{s}\) provided
that \(s_k = i\): \[ c_{ji}^{(k)}(\b{s}) := |\{k' < k \mid s_{k'} = j \text{ and }
s_k = i\}|. \] Observe that
\[
   \sum_{k=1}^{\l} c_{ji}^{(k)}(\b{s}) = c_{ji}(\b{s})
   \qquad \text{and} \qquad
   \sum_{j=1}^n c_{ji}^{(k)}(\b{s}) =
   \begin{cases}
     k - 1, & s_{k} = i; \\
     0, & s_k \neq i.
   \end{cases}
\]
Further, since we know \(C(\b{s})\), we know \(c_{ii}(\b{s})\) for each \(i \in
\neurons\). Recalling that \(c_{ii}(\b{s}) = \binom{c_i(\b{s})}{2}\), we note that
\(c_i(\b{s}) = \half + \half \sqrt{1 + 8 c_{ii}(\b{s})}\). Moreover, knowing
\(c_{ii}(\b{s})\) for each \(i \in \neurons\) allows us to find \(\l\) as \(\l =
\sum_{i \in \neurons} c_i(\b{s})\). Putting these facts together, we see that
\begin{align*}
  \sum_{j=1}^n c_{ji}(\b{s})
    &= \sum_{j=1}^n \sum_{k=1}^{\l} c_{ji}^{(k)}(\b{s}) \\
    &= \sum_{k=1}^{\l} \sum_{j=1}^n c_{ji}^{(k)}(\b{s}) \\
    &= \sum_{k \in L_i(\b{s})} (k - 1) \\
    &= \sum L_i(\b{s}) - c_i(\b{s}) \\
    &= c_i(\b{s}) [\com_i(\b{s}) - 1].
\end{align*}
Because \(c_i(\b{s}) \neq 0\) by assumption, we can solve for \(\com_i(\b{s})\) to
obtain the desired result.
\end{proof}
\end{lemma}

In particular, the above \namecref{lem:center-from-bias-count} says that
\(\com(\b{s}_1) = \com(\b{s}_2)\) if \(C(\b{s}_1) = C(\b{s}_2)\). However, the
converse to this statement is not true. That is, the center-of-mass vector
\(\com(\b{s})\) does not contain enough information to compute the bias-count
matrix. For example, consider the sequences \(\b{s}_1 = 123123\) and \(\b{s}_2 =
211332\). Then
\[
\com(\b{s}_1) = \com(\b{s}_2) =\begin{bmatrix}
 \nicefrac{5}{2} \\
 \nicefrac{7}{2} \\
 \nicefrac{9}{2} \\
\end{bmatrix}
\]but
\[
C(\b{s}_1) =\begin{bmatrix}
 1 & 3 & 3 \\
 1 & 1 & 3 \\
 1 & 1 & 1 \\
\end{bmatrix}\neq
\begin{bmatrix}
 1 & 2 & 4 \\
 2 & 1 & 2 \\
 0 & 2 & 1 \\
\end{bmatrix}= C(\b{s}_2).
\]As such, the bias-count matrix \(C(\b{s})\) is in fact a generalization of the
center-of-mass vector \(\com(\b{s})\).

One might also suspect that the center-of-mass vector \(\gamma(\b{s})\) can be
recovered from the skew-bias matrix \(\skbias(\b{s})\). However, upon closer
inspection of the previous \namecref{lem:center-from-bias-count}, we notice that
the bias-count matrix was used to recover the spike-count vector, which cannot
be done when starting with the skew-bias matrix. As the next
\namecref{lem:skbias-gamma} shows, however, we can reconstruct the
center-of-mass vector \(\gamma(\b{s})\) from the skew-bias \(\skbias(\b{s})\) if we
also know the sequence's firing-rate vector \(\rho(\b{s})\).

\begin{proposition}
\label{lem:skbias-gamma}
If \(\b{s} \in \sequences\) is a sequence, then \[ \gamma(\b{s}) =
\skbias(\b{s})^T \rho(\b{s}) . \]

\begin{proof}
Let \(\l = \len(\b{s})\), and assume that \(i \in \supp(\b{s})\). Recall that
\(\gamma_i(\b{s}) = \frac{2\com_i(\b{s}) - 1}{\l} - 1\), from which we see that \[
\com_i(\b{s}) = \frac{\l}{2} [\gamma_i(\b{s}) + 1] + \frac{1}{2}. \] We now want
to rewrite the previous \namecref{lem:center-from-bias-count} to give us an
expression for \(\gamma_i(\b{s})\) in terms of the skew-biases
\(\beta_{ij}(\b{s})\). Dropping the \(\b{s}\) dependence from the notation, the
above equation and the previous \namecref{lem:center-from-bias-count} yield the
following:
\begin{align*}
  \frac{\l}{2}(\gamma_i + 1) + \frac{1}{2}
    &= \com_i \\
    &= 1 + \frac{1}{c_i} \sum_{j = 1}^n c_{ji} \\
    &= 1 + \frac{1}{c_i} \parens[\Big]{c_{ii} + \sum_{j \neq i} c_{ji}} \\
    &= 1 + \frac{1}{c_i} \parens[\Big]{\frac{c_i(c_i - 1)}{2} + \sum_{j\neq i} \frac{c_i c_j}{2}(\beta_{ji} + 1)} \\
    &= 1 + \frac{1}{2} \parens[\Big]{c_i - 1 + \sum_{j\neq i} c_j(\beta_{ji} + 1)} \\
    &= 1 + \frac{1}{2} \parens[\Big]{\brkts[\big]{c_i + \sum_{j \neq i} c_j} - 1 + \sum_{j\neq i} c_j\beta_{ji}} \\
    &= 1 + \frac{1}{2} \parens[\Big]{\l - 1 + \sum_{j = 1}^n c_j\beta_{ji}} ,
\end{align*}
where we have recalled in the last line that \(\beta_{ii} = 0\) and \(\sum_{j =
1}^n c_j = \l\). Solving for \(\gamma_i\) and including the \(\b{s}\) dependence into
our notation, we obtain
\[ \gamma_i(\b{s}) = \sum_{j=1}^n \beta_{ji}(\b{s}) \rho_j(\b{s}) \]
since \(\rho_i(\b{s}) = \frac{c_i(\b{s})}{\l}\). Note that although we derived
this equation using the assumption that \(c_i(\b{s}) > 0\), it still holds if
\(c_i(\b{s}) = 0\) because in this case we have \(\gamma_i(\b{s}) = 0\) and
\(\beta_{ij}(\b{s}) = 0\) for all \(j\). Since \(\sum_{j=1}^n \beta_{ji}(\b{s})
\rho_j(\b{s})\) is the entry in row \(i\) of \(\skbias(\b{s})^T \rho(\b{s})\), this
completes the proof.
\end{proof}
\end{proposition}

If we substitute the firing-rate vector \(\rho(\b{s})\) in \Cref{lem:skbias-gamma}
for an arbitrary firing-rate vector, then the skew-bias matrix can be used to
generate other center-of-mass vectors and orderings.

\begin{definition}
Given a sequence \(\b{s}\) and a firing-rate column vector \(\rho = (\rho_1, \dots,
\rho_n)\), define the \emph{center of mass of \(\b{s}\) induced by \(\rho\)} to be
\(\g_\b{s}(\rho) := \skbias(\b{s})^T \rho\). Further, define the corresponding
\emph{center-of-mass ordering of \(\b{s}\) induced by \(\rho\)}, denoted
\(\s_\b{s}(\rho)\), just as with the center-of-mass ordering of a sequence: If
\(\g_\b{s}(\rho) = (g_1, \dots, g_n)\), then \(\s_\b{s}(\rho) = (i_1, \dots, i_n)\)
where \(\set{i_1, \dots, i_n} = [n]\) are such that \(g_{i_1}(\b{s}) \leq \cdots
\leq g_{i_n}(\b{s})\) and, in the case that \(g_{i_k}(\b{s}) = g_{i_{k +
1}}(\b{s})\), such that \(i_k < i_{k + 1}\).
\end{definition}

\begin{example}
Recall that the sequence \(\b{s} = 12444433131124\) has skew-bias matrix
\[
\skbias(\b{s}) =\begin{bmatrix}
 0 & \nicefrac{1}{4} & -\nicefrac{1}{3} & -\nicefrac{1}{5} \\
 -\nicefrac{1}{4} & 0 & 0 & \nicefrac{1}{5} \\
 \nicefrac{1}{3} & 0 & 0 & -\nicefrac{3}{5} \\
 \nicefrac{1}{5} & \nicefrac{1}{5} & \nicefrac{3}{5} & 0 \\
\end{bmatrix}.
\]The firing-rate vector of \(\b{s}\) is \(\rho(\b{s}) = \frac{1}{14} [4, 2, 3, 5]\)
and \(\s(\b{s}) = (4, 2, 1, 3)\). With firing-rate vectors \(\rho_1 = \frac{1}{4}
[1, 1, 1, 1]\) and \(\rho_2 = \frac{1}{3} [1, 1, 1, 0]\), we find that
\[
  \g_\b{s}(\rho_1) =
  \bmtx{\nicefrac{17}{240} \\ \nicefrac{1}{80} \\
    \nicefrac{1}{15} \\ -\nicefrac{1}{4}}
  \longrightarrow
  \s_\b{s}(\rho_1) = \bmtx{4 \\ 2 \\ 3 \\ 1}
  \qquad
  \text{and}
  \qquad
  \g_\b{s}(\rho_2) =
  \bmtx{\nicefrac{1}{36} \\ \nicefrac{1}{12} \\
    -\nicefrac{1}{9} \\ -\nicefrac{1}{3}}
  \longrightarrow
  \s_\b{s}(\rho_2) = \bmtx{4 \\ 3 \\ 1 \\ 2}.
\]
So, \(\rho_1\) and \(\rho_2\) induce the orderings \((4, 2, 3, 1)\) and \((4, 3, 1,
2)\), respectively. In particular, notice that the activation of neuron \(4\)
changed the induced ordering of \(\set{1, 2, 3}\).
\end{example}

In the case of a pure sequence, this process can result in only one neuronal
ordering, as the following lemma shows.

\begin{lemma}
\label{lem:recover-pure-order}
If \(\b{s}\) is a pure sequence, then \(\s_\b{s}(\rho) = \s(\b{s})\) for each
firing-rate vector \(\rho = (\rho_1, \dots, \rho_n)\) satisfying \(\rho_i > 0\) for
each \(i \in \neurons\).

\begin{proof}
Without loss of generality, we can assume that \(\b{s} = \s(\b{s})\) since
\(\skbias(\b{s}) = \skbias(\s(\b{s}))\) whenever \(\b{s}\) is pure. For each neuron
\(i \in [n - 1]\), we know that \(\beta_{j s_i}(\b{s}) = \beta_{j s_{i +
1}}(\b{s})\) for each \(j \in \neurons \setminus \set{s_i, s_{i + 1}}\) since
\(\b{s}\) is pure and since \(s_i\) and \(s_{i + 1}\) are adjacent spikes (not equal
to \(j\)); additionally, we know that \(\beta_{s_i s_i}(\b{s}) = \beta_{s_{i + 1}
s_{i + 1}}(\b{s}) = 0\) and that \(\beta_{s_i s_{i + 1}}(\b{s}) = 1\) and
\(\beta_{s_{i + 1} s_i}(\b{s}) = -1\). Letting \(\g_\b{s}(\rho) = (\g_1, \dots,
\g_n)\), these facts tell us that \[ \g_{s_i} = \sum_{j = 1}^n \beta_{j
s_i}(\b{s}) \rho_j < \sum_{j = 1}^n \beta_{j s_{i + 1}}(\b{s}) \rho_j =
\g_{s_{i + 1}} .\] Hence, regardless of the choice of \(\rho\), we always have
that \(\g_{s_1} < \cdots < \g_{s_n}\). In particular, letting \(\rho =
\rho(\b{s})\), we find that \(\s_\b{s}(\rho(\b{s})) = \s(\b{s})\), thus completing
the proof.
\end{proof}
\end{lemma}

\section{Permutations of sequences}
\label{sec:orgheadline11}

To understand a bias matrix \(A\), it would be useful to know which sequences give
rise to that bias matrix; that is, we would like to be able to describe the set
\(\skbias^{-1}(A) \sset \sequences\). We can "fill in" part of this set if we know
a sequence in that set, that is, a sequence \(\b{s}\) such that \(\skbias(\b{s}) =
A\).

Consider the sequence \(\b{s} = 12321\) and its bias-count matrix
\[
C(\b{s}) =\begin{bmatrix}
 1 & 2 & 1 \\
 2 & 1 & 1 \\
 1 & 1 & 0 \\
\end{bmatrix}.
\]If we flip any two adjacent spikes, the result in the bias-count matrix will be
that exactly two entries are modified. For instance, if we flip the first two
spikes, the bias-count matrix becomes
\[
C(21321) =\begin{bmatrix}
 1 & 1 & 1 \\
 3 & 1 & 1 \\
 1 & 1 & 0 \\
\end{bmatrix}.
\]Since we want the bias-count matrix to remain unchanged if we are to find
another sequence in \(\skbias^{-1}(\skbias(\b{s}))\), we can try to undo the
effect of the first flip by flipping another pair of spikes; that is, since we
made a subsequence of the form \(12\) into one of the form \(21\), we can look for a
subsequence of the form \(21\) to change into \(12\). Here, the last two spikes are
\(21\); changing these and the first two, we get the sequence \(\b{s}' = 21312\).
Now, \(C(\b{s}') = C(\b{s})\). So, indeed, it is sometimes possible to find
another sequence with the same biases by simply permuting the spikes of the
original sequence. It will be convenient to have some notation and terminology
for such an operation:
\begin{definition}
For a sequence \(\b{s} = (s_1, \dots, s_\l)\) and a permutation \(\pi \in
\perms{[\l]}\), define the \emph{shuffling of \(\b{s}\) by \(\pi\)} to be \(\b{s}^\pi :=
(s_{\pi(1)}, \dots, s_{\pi(\l)})\).
\end{definition}

\begin{example}
Let \(\b{s} = 1212333\). The following table shows permutations of \([\len(\b{s})]
= [7]\) along with the corresponding shuffled sequence.
\begin{center}
\begin{tabular}{l|l}
Permutation & Shuffled sequence\\
\hline
\(\pi_1 = (1\ 7)(2\ 6)(3\ 5)\) & \(\b{s}^{\pi_1} = 3332121\)\\
\(\pi_2 = (2\ 3)\) & \(\b{s}^{\pi_2} = 1122333\)\\
\(\pi_3 = (1\ 2)(3\ 4)\) & \(\b{s}^{\pi_3} = 2121333\)\\
\(\pi_4 = (1\ 3)(5\ 6\ 7)\) & \(\b{s}^{\pi_4} = 1212333\)\\
\end{tabular}
\end{center}
Note that the non-trivial permutation \(\pi_4\) induces a shuffled sequence
\(\b{s}^{\pi_4}\) that is equal to the original sequence \(\b{s}\).
\end{example}

\begin{definition}
For a sequence \(\b{s} = (s_1, \dots, s_\l)\) and a collection \(\w\) of neurons,
define the \emph{restriction of \(\b{s}\) to \(\w\)} to be \(\b{s}|_\w := s_{I_\w}\) where
\(I_\w := \cup_{i \in \w} L_i(\b{s})\).
\end{definition}

\begin{example}
Let \(\b{s} = 12444433131124\), and let \(\w_1 = \set{1, 2}\) and \(\w_2 = \set{3,
4}\). Then \(\b{s}_{\w_1} = 121112\) and \(\b{s}_{\w_2} = 44443334\).
\end{example}

\begin{lemma}
Let \(\b{s} = (s_1, \dots, s_\l)\) be a sequence, and let \(\pi = (k_1\ k_1+1)(k_2\
k_2+1)\) be a permutation such that \(s_{k_1} = s_{k_2 + 1}\) and \(s_{k_2} =
s_{k_1 + 1}\). Then \(C(\b{s}) = C(\b{s}^\pi)\).

\begin{proof}
Let \(\b{s} = (s_1, \dots, s_\l)\) be a sequence, and let \(\pi = (k_1\ k_1+1)(k_2\
k_2+1)\) be a permutation of \([\l]\) such that \(s_{k_1} = s_{k_2 + 1}\) and
\(s_{k_2} = s_{k_1 + 1}\). Let \(i_1 = s_{k_1} \in \neurons\) and \(i_2 = s_{k_2} \in
\neurons\), and let \(N = \neurons \setminus \set{i_1, i_2}\). Observe that
\(\b{s}|_N = \b{s}^\pi|_N\) since \(\pi\) fixes the indices of the spikes from
neurons in \(N\); in particular, this tells us that \(c_{j_1 j_2}(\b{s}) = c_{j_1
j_2}(\b{s}|_N) = c_{j_1 j_2}(\b{s}^\pi|_N) = c_{j_1 j_2}(\b{s}^\pi)\).

We now want to show that \(c_{ij}(\b{s}) = c_{ij}(\b{s}^\pi)\) and \(c_{ji}(\b{s})
= c_{ji}(\b{s}^\pi)\) if \(\set{i, j} \cap \set{i_1, i_2} \neq \emptyset\).
Recalling that \(c_{ij} = |L_{ij}|\), we proceed by listing the relationships
between index pairs in the sets \(L_{ij}\) and \(L_{ji}\). Suppose that \(m \in [\l]
\setminus \set{k_1, k_1 + 1, k_2, k_2 + 1}\) is a spike index and that \(j := s_m
\in N\). Then the following relationships hold:
\begin{align*}
  (k_1, m) \in L_{i_1 j}(\b{s})                             &\iff (k_1 + 1, m) \in L_{i_1 j}(\b{s}^\pi) \\
  (k_2 + 1, m) \in L_{i_1 j}(\b{s})                         &\iff (k_2, m) \in L_{i_1 j}(\b{s}^\pi) \\
  (k_1 + 1, m) \in L_{i_2 j}(\b{s})                         &\iff (k_1, m) \in L_{i_2 j}(\b{s}^\pi) \\
  (k_2, m) \in L_{i_2 j}(\b{s})                             &\iff (k_2 + 1, m) \in L_{i_2 j}(\b{s}^\pi) \\
  (m, k_1) \in L_{j i_1}(\b{s})                             &\iff (m, k_1 + 1) \in L_{j i_1}(\b{s}^\pi) \\
  (m, k_2 + 1) \in L_{j i_1}(\b{s})                         &\iff (m, k_2) \in L_{j i_1}(\b{s}^\pi) \\
  (m, k_1 + 1) \in L_{j i_2}(\b{s})                         &\iff (m, k_1) \in L_{j i_2}(\b{s}^\pi) \\
  (m, k_2) \in L_{j i_2}(\b{s})                             &\iff (m, k_2 + 1) \in L_{j i_2}(\b{s}^\pi) \\
  (k_1, k_1 + 1), (k_1, k_2) \in L_{i_1 i_2}(\b{s})         &\iff (k_1 + 1, k_2 + 1), (k_2, k_2 + 1) \in L_{i_1 i_2}(\b{s}^\pi) \\
  (k_1 + 1, k_2 + 1), (k_2, k_2 + 1) \in L_{i_2 i_1}(\b{s}) &\iff (k_1, k_1 + 1), (k_1, k_2) \in L_{i_2 i_1}(\b{s}^\pi)
\end{align*}
This completes the proof since each line above indicates that the corresponding
index-pair sets are of the same size for both \(\b{s}\) and \(\b{s}^\pi\).
\end{proof}
\end{lemma}

Since a sequence of such permuations can be applied recursively, one might
wonder whether all sequences with equal bias-count matrices can be obtained from
a single sequence by repeated application of such permutations. This is not the
case: For example, if \(\b{s} = 123231\) and \(\b{s}' = 231123\), then \(C(\b{s}) =
C(\b{s}')\) but no such permutation \(\pi = (i\ i+1)(j\ j+1)\) exists such that
\(\b{s}' = \b{s}^\pi\) since there is no pair \(\{i, j\}\) with \(i \neq j\) such that
\(s_i = s_{j+1}\) and \(s_{i+1} = s_j\). More succinctly, if \(\bias(\b{s}) =
\bias(\b{s}')\), it is not necessarily the case that \(\b{s}' = \b{s}^\pi\) for
some permutation \(\pi\).

We can also describe a broader class of permutations that fixes the
center-of-mass vectors of a given sequence. Generally speaking, elements of this
class will not fix the corresponding bias matrices.

\begin{lemma}
If \(\b{s} = (s_1, \dots, s_\l)\) is a sequence and \(\pi = (k_1\ k_1 + 1) \cdots
(k_m\ k_m + 1) \in \perms{[l]}\) with \(k_{a + 1} > k_a + 1\) for all \(a \in
\set{1, \dots, m - 1}\), then \(\com(\b{s}) = \com(\b{s}^\pi)\) if \(\set{s_{k_1},
\dots, s_{k_m}} = \set{s_{k_1 + 1}, \dots, s_{k_m + 1}}\) and \(\abs{\set{s_{k_1},
\dots, s_{k_m}}} = m\).

\begin{proof}
Suppose that \(i \in \set{s_1, \dots, s_k}\). Then \(i = s_{k_b} = s_{k_d + 1}\) for
some \(b, d \in [m]\). Noting that \(L_i(\b{s}^\pi) = [L_i(\b{s}) \setminus
\set{k_b, k_d + 1}] \cup \set{k_b + 1, k_d}\), we see that
\begin{align*}
  \com_i(\b{s}) &= \com_i(\b{s}) + \frac{[k_b + (k_d + 1)] - [k_b + (k_d + 1)]}{c_i(\b{s})} \\
                &= \com_i(\b{s}) + \frac{[(k_b + 1) + k_d] - [k_b + (k_d + 1)]}{c_i(\b{s}^\pi)} \\
                &= \com(\b{s}^\pi).
\end{align*}
If \(i \notin \set{s_1, \dots, s_k}\), then \(\com_i(\b{s}) = \com_i(\b{s}^\pi)\)
since \(L_i(\b{s}^\pi) = L_i(\b{s})\). Hence, \(\com_i(\b{s}) = \com_i(\b{s}^\pi)\)
for all \(i \in \neurons\), thus completing the proof.
\end{proof}
\end{lemma}

\begin{example}
Let \(\b{s} = 123123\) and \(\pi = (1\ 2) (3\ 4) (5\ 6)\). Then \(\b{s}\) and \(\pi\)
meet the stipulations of the previous lemma since \(\set{s_1, s_3, s_5} =
\set{s_2, s_4, s_6} = \set{1, 2, 3}\). Noting that \(\b{s}^\pi = 211332\), we see
that
\[
\com(\b{s}) = \com(\b{s}^\pi) =\begin{bmatrix}
 \nicefrac{7}{2} \\
 \nicefrac{5}{2} \\
 \nicefrac{9}{2} \\
\end{bmatrix}.
\]\end{example}

\section{The fundamental polytope}
\label{sec:orgheadline12}
\label{sec:bias-space}

The bias matrix arose in an attempt to capture the tendency of pairs of neurons
to fire in a particular order. These tendencies can be computed from any
individual sequence, but the pairwise nature of the biases hints at an
underlying network of neurons with pairwise preferences for spiking in
particular orders. From this perspective, a sequence is really just a (noisy)
manifestation of the neural network's biases. As such, it may be useful to
understand the properties that characterize all bias matrices. By considering
the biases that arise from sequences, we aim to derive a bias-matrix formulation
that is independent of any individual sequence. In this section, we develop such
a formulation and conjecture that each collection of biases in this formulation
is realizable as the collection of biases of an actual sequence.

We begin by consideration of the bias-count matrix \(C(\b{s}) =
[c_{ij}(\b{s})]_{i, j \in \neurons}\). We already know a few properties that must
be satisfied:
\begin{itemize}
\item Its entries are natural numbers: \(c_{ij} \in \N_0\) for all \(i, j \in
  \neurons\).
\item Its diagonal entries are binomial coefficients: \(c_{ii} = \binom{c_i}{2}\) for
all \(i \in \neurons\).
\item For all neurons \(i, j \in \neurons\) with \(i \neq j\), \(c_{ij} + c_{ji} = c_i
  c_j\).
\end{itemize}
But we moved from the bias-count matrix to the skew-bias matrix because we want
the individual bias entries to have meaning independent of other entries. When
the above properties are translated into the skew-bias matrix, we find only two
properties:
\begin{itemize}
\item It is skew-symmetric: \(\beta_{ij} = -\beta_{ji}\) for all \(i, j \in \neurons\).
\item Its entries are bounded and rational: \(\beta_{ij} \in [-1, 1] \cap \Q\) for all
\(i, j \in \neurons\).
\end{itemize}
Though the properties listed here do not fully characterize the skew-biases in
general (as we will see below), they do provide a convenient starting place from
which we can further restrict as we look for additional properties. The skew
nature of the matrices \(\skbias(\b{s})\) allows us to describe the matrix using
only the above-diagonal entries of each matrix.

\begin{definition}
The \emph{bias space} \(\B(\neurons)\) of the neuron set \(\neurons = [n]\) is the
\(\binom{n}{2}\)-dimensional hypercube \([-1, 1]^{\binom{n}{2}}\) with coordinates
indexed by neuron pairs \((i, j) \in \neurons^2\) with \(i < j\). When \(\b{s}\) is a
sequence, the vector \([\beta_{ij}(\b{s})]_{i < j}\) consisting of the
upper-triangular portion of \(\skbias(\b{s})\) is the corresponding point in the
bias space. We will abuse terminology and refer to \(\skbias(\b{s})\) as being in
the bias space for notational/terminological convenience.
\end{definition}

\begin{definition}
A point \(x \in \B(\neurons)\) is \emph{obtainable} if \(x\) is in the closure of the set
\(\set{\skbias(\b{s}) \mid \b{s} \in \sequences}\).
\end{definition}

Which points in \(\B(\neurons)\) are obtainable? Let us first consider one of the
simplest classes of sequences: those with exactly two active neurons. In this
case---that is, when \(\neurons = [2]\)---the only bias that we consider is
\(\beta_{12}\); further, the bias space is just the interval \([-1, 1]\). Here, we
find that the biases are actually dense in the bias space.

\begin{lemma}
\label{lem:bias-dense-for-2-neurons}
For any \(x \in [-1, 1]\) and any \(\eps > 0\), there exists a sequence \(\b{s} \in
\sequences\) such that \(\supp(\b{s}) = [2]\) and \(\abs{\beta_{12}(\b{s}) - x} <
\eps\).

\begin{proof}
Let \(x \in [-1, 1]\) and \(\eps > 0\) be given. Choose \(\l \in \N\) such that
\(\frac{2}{\l - 1} < \eps\), and pick \(k \in \N\) as \(k = \floor{\half (1 +
x)(\l - 1) + 1}\). Note that \(0 \leq 1 + x \leq 2\) since \(-1 \leq x \leq 1\); so,
\(0 \leq \half (1 + x)(\l - 1) \leq \l - 1\) and, thus, \(k \in [\l]\). Moreover,
\(\abs{k - \half (1 -x) \l} \leq 1\) by choice of \(k\). Define the sequence \(\b{s}
= (s_1, \dots, s_\l)\) by \(s_k = 2\) and \(s_m = 1\) for \(m \in [\l] \setminus
\set{k}\). Now, \(c_{12}(\b{s}) = k - 1\), which implies that \(b_{12}(\b{s}) =
\frac{c_{12}(\b{s})}{c_1(\b{s}) c_2(\b{s})} = \frac{k - 1}{(\l - 1) \cdot 1} =
\frac{k -1}{\l - 1}\) and that \(\beta_{12}(\b{s}) = 2 b_{12}(\b{s}) - 1 =
\frac{2k - 2}{\l - 1} - 1\). Now,
\begin{align*}
  \abs{\beta_{12}(\b{s}) - x}
    &= \abs{\frac{2k - 2}{\l - 1} - 1 - x} \\
    &= \abs{\frac{2k - 2 - (1 + x)(\l - 1)}{\l - 1}} \\
    &= \frac{2}{\l - 1} \abs{k - \parens[\big]{\thalf (1 + x)(\l - 1) + 1}} \\
    &\leq \frac{2}{\l - 1} \\
    &< \eps,
\end{align*}
thus completing the proof.
\end{proof}
\end{lemma}

But it is not generally the case that every point in the bias space is
obtainable. To get some intuition about why some points are not obtainable,
consider the matrix
\[
A =\begin{bmatrix}
 0 & 1 & -1 \\
 -1 & 0 & 1 \\
 1 & -1 & 0 \\
\end{bmatrix}.
\]Suppose that \(A = \skbias(\b{s})\) for some sequence \(\b{s}\). Because all
off-diagonal entries are in \(\set{1, -1}\), we must have that \(\b{s}\) is a pure
sequence. Considering the entries more closely, we also find the following:
\begin{itemize}
\item (\(a_{12} = 1\)) All spikes from neuron 1 precede all spikes from neuron 2.
\item (\(a_{23} = 1\)) All spikes from neuron 2 precede all spikes from neuron 3.
\item (\(a_{13} = -1\)) All spikes from neuron 3 precede all spikes from neuron 1.
\end{itemize}
The first two points (\(a_{12} = a_{23} = 1\)) tell us that all spikes from neuron
1 must precede all spikes from neuron 3. But this is a contradiction to the
third observation! Hence, we cannot have that \(A = \skbias(\b{s})\) for any
sequence \(\b{s}\). Indeed, we will see in \Cref{cor:consistent-bias} that \(A\) is
not obtainable.

Although we have not yet shown that \(A\) is not obtainable, the point is
well-taken: There are non-trivial relationships among the entries \(\beta_{ij}\),
\(\beta_{jk}\), and \(\beta_{ik}\) (that is, among the collection of biases between
the neurons \(i\), \(j\), and \(k\)). In fact, the above-observed property holds in a
more general case: If \(\b{s} \in \sequences\) is pure, then we must have that
\(\beta_{ik} = 1\) if \(\beta_{ij} = \beta_{jk} = 1\) for distinct neurons \(i, j, k
\in \neurons\). The logic for seeing that this holds is exactly the same as in
the motivating example. As seen in the proof of
\Cref{lem:bias-dense-for-2-neurons}, only one spike from the second neuron was
necessary to obtain (in the sense of density) all possible biases between
exactly two neurons. As such, we begin our investigation of biases among three
neurons by considering sequences with exactly one spike from a third neuron.

\begin{lemma}
For any sequence \(\b{s} \in \sequences\) with distinct neurons \(i, j, k \in
\supp(\b{s})\) and with \(c_k(\b{s}) = 1\), \(b_{ik} (1 - b_{jk}) \leq b_{ij} \leq
1 - b_{jk} (1 - b_{ik})\).

\begin{proof}
Since neuron \(k\) spikes only once, it is easy to construct a sequence that
minimizes \(m_{ij}\) for fixed (rational) values of \(b_{ki}\) and \(b_{kj}\). Since
\(b_{ki} = \frac{c_{ki}}{c_k c_i} = \frac{c_{ki}}{c_i}\) and \(b_{kj} =
\frac{c_{kj}}{c_k c_j} = \frac{c_{kj}}{c_j}\), we see that \(c_{ki} = b_{ki} c_i\)
and \(c_{kj} = b_{kj} c_j\) are proportions of the total number of times that
neurons \(i\) and \(j\) spike, respectively. In particular, when considering the
restricted sequences containing \(k\) and only one other neuron, we find that
there is only one possibility for each restricted sequence:
\[
    \b{s}|_{\set{i, k}} = i^{c_{ik}} \cdot k \cdot i^{c_{ki}}
    \qquad \text{and} \qquad
    \b{s}|_{\set{j, k}} = j^{c_{jk}} \cdot k \cdot j^{c_{kj}} ,
\]
where we recall that the symbol \(\cdot\) represents concatenation. In joining
these restricted sequences to form the full sequence \(\b{s}\), we minimize
\(b_{ij}\) by placing \(j\) before \(i\) whenever possible, resulting in the
following: \[ \b{s} = j^{c_{jk}} \cdot i^{c_{ik}} \cdot k \cdot j^{c_{kj}} \cdot
i_{c_{ki}} .\] It is now straightforward to count \(c_{ij}\):
\[
    c_{ij} = c_{ik} c_{kj}
           = (b_{ik} c_i) (b_{kj} c_j)
           = b_{ik} (1 - b_{jk}) c_i c_j .
\]
Because \(\b{s}\) was chosen to minimize \(c_{ij}\), dividing by the product \(c_i
c_j\) yields one of the desired inequalities: \(b_{ij} \leq b_{ik} (1 - b_{jk})\).
Similar reasoning allows us to maximize \(c_{ij}\) with \(\b{s}'\): \[ \b{s}' =
i^{c_{ik}} \cdot j^{c_{jk}} \cdot k \cdot i^{c_{ki}} \cdot j_{c_{kj}} ,\] which
results in the count
\begin{align*}
  c_{ij} &= c_{ik} c_j + c_{ki} c_{kj} \\
         &= b_{ik} c_i c_j + (b_{ki} c_i) (b_{kj} c_j) \\
         &= (b_{ik} + b_{ki} b_{kj}) c_i c_j \\
         &= [b_{ik} + (1 - b_{ik}) (1 - b_{jk})] c_i c_j \\
         &= (1 - b_{jk} + b_{ik} b_{jk}) c_i c_j \\
         &= [1 - b_{jk} (1 - b_{ik})] c_i c_j ,
\end{align*}
thus yielding the other desired inequality: \(b_{ij} \geq 1 - b_{jk} (1 -
b_{ik})\). This completes the proof.
\end{proof}
\end{lemma}

As it turns out, however, not every point in the bias space \(\B([3])\) is
obtainable using only a single spike from the third neuron. Consider, for
example, the sequence \(\b{s} = 3123\) and its bias matrix
\[
\skbias(\b{s}) =\begin{bmatrix}
 0 & 1 & 0 \\
 -1 & 0 & 0 \\
 0 & 0 & 0 \\
\end{bmatrix}.
\]Suppose that there is some sequence \(\b{s}'\) such that \(c_3(\b{s}') = 1\) and
\(\skbias(\b{s}') = \skbias(\b{s})\). Since \(\beta_{12}(\b{s}') = 1\), the
restricted sequence \(\b{s}'|_{[2]}\) must be pure. But then the single spike from
neuron 3 cannot fall in the middle of the spikes for neuron 1 and in the middle
of the spikes for neuron 2. Hence, no such \(\b{s}'\) can exist. Still, not all
points in the bias space are obtainable even when we consider arbitrary
sequences on three neurons. But this should not come as too much of a surprise:
Our intuition for sequences says that if \(i\) follows \(j\) and \(k\) follows \(j\),
then we would expect for \(k\) to follow \(i\). The inequalities in the lemma below
are restrictions of such a form.

\begin{proposition}
For any sequence \(\b{s}\) with neurons \(i, j, k \in \supp(\b{s})\) such that \(i <
j < k\), the inequalities \(0 \leq b_{ij} + b_{jk} - b_{ik} \leq 1\) hold.

\begin{proof}
Let \(\b{s} = (s_1, \dots, s_\l)\) be a sequence. First, note that the following
equality holds because every triple \(\l_i, \l_j, \l_k \in [\l]\) with \(s_{\l_i} =
i\), \(s_{\l_j} = j\), and \(s_{\l_k} = k\) (of which there are \(c_i c_j c_k\)) will
produce exactly one of the six orderings \(ijk\), \(jki\), \(kij\), \(ikj\), \(jik\), and
\(kji\): \[ c_{ijk} + c_{ikj} + c_{jik} + c_{jki} + c_{kij} + c_{kji} = c_i c_j
c_k \]

We now show that the following relationships exist between pairwise-bias counts
and third-order bias counts:
\begin{align*}
  c_k c_{ij} &= c_{ijk} + c_{ikj} + c_{kij} \\
  c_i c_{jk} &= c_{jki} + c_{jik} + c_{ijk} \\
  c_j c_{ki} &= c_{kij} + c_{kji} + c_{jki}
\end{align*}
To see that these relationships hold, consider the first one. For each
subsequence of \(\b{s}\) that is equal to \$(i, j)\$---of which there are
\$c\(_{\text{ij}}\)\$---each of the \(c_k\) occurrences of \(k\) in \(\b{s}\) can produce only one
of the triples \((i, j, k)\), \((i, k, j)\), and \((k, i, j)\); moreover, each such
combination must produce one such triple. Hence, the equality holds. Similar
arguments work for each of the other two equations.

Summing the above equations, we get the following inequality:
\begin{align*}
  c_k c_{ij} + c_i c_{jk} + c_j c_{ki}
    &= 2 (c_{ijk} + c_{jki} + c_{kij}) + (c_{ikj} + c_{jik} + c_{kji}) \\
    &= 2 (c_{ijk} + c_{jki} + c_{kij} + c_{ikj} + c_{jik} + c_{kji}) \\
    &\phantom{=} - (c_{ikj} + c_{jik} + c_{kji}) \\
    &= 2 c_i c_j c_k - (c_{ikj} + c_{jik} + c_{kji}) \\
    &\leq 2 c_i c_j c_k.
\end{align*}
Dividing the resultant inequality by \(c_i c_j c_k\) yields the inequality
\(b_{ij} + b_{jk} + b_{ki} \leq 2\). By considering the sum of the same equations
again, we find another inequality:
\begin{align*}
  c_k c_{ij} + c_i c_{jk} + c_j c_{ki}
    &= 2 (c_{ijk} + c_{jki} + c_{kij}) + (c_{ikj} + c_{jik} + c_{kji}) \\
    &= (c_{ijk} + c_{jki} + c_{kij} + c_{ikj} + c_{jik} + c_{kji}) \\
    &\phantom{=} + (c_{ijk} + c_{jki} + c_{kij}) \\
    &= c_i c_j c_k + (c_{ikj} + c_{jik} + c_{kji}) \\
    &\geq c_i c_j c_k.
\end{align*}
Dividing this by \(c_i c_j c_k\) yields the inequality \(b_{ij} + b_{jk} + b_{ki}
\geq 1\). Putting these two inequalities together, we have that \(1 \leq b_{ij} +
b_{jk} + b_{ki} \leq 2\). Recalling that \(b_{ki} = 1 - b_{ik}\), this becomes \(0
\leq b_{ij} + b_{jk} - b_{ik} \leq 1\), thus completing the proof.
\end{proof}
\end{proposition}

\begin{corollary}
\label{cor:consistent-bias}
For any sequence \(\b{s}\) with neurons \(i, j, k \in \supp(\b{s})\) such that \(i <
j < k\), the inequalities \(-1 \leq \beta_{ji} + \beta_{kj} - \beta_{ik} \leq 1\)
hold.
\end{corollary}

As sequences become less pure, our intuition about one neuron "following"
another breaks down, and we run into tendencies that would be contradictory in
the case of a pure sequence.

\begin{example}
Consider the sequence \(\b{s} = 2331112123\). The skew-bias matrix of \(\b{s}\) is
\[
\skbias(\b{s}) =\begin{bmatrix}
 0 & 1/6 & -1/3 \\
 -1/6 & 0 & 1/9 \\
 1/3 & -1/9 & 0 \\
\end{bmatrix}.
\]Upon inspection, we see that (1) neuron 1 tends to precede neuron 2, (2) neuron
2 tends to precede neuron 3, and (3) neuron 3 tends to precede neuron 1.
\end{example}

Inequalities of the form in \Cref{cor:consistent-bias} essentially ensure that
the pairwise-biases among any three neurons are consistent with each other
(i.e., if \(j\) follows \(i\) and \(k\) follows \(j\), then \(k\) follows \(i\)), a sort of
pseudo-transitivity that is guaranteed when biases arise from an actual
sequence. Further, the collection of all such inequalities carves out a polytope
in the bias space, which we will call the \emph{consistent-bias polytope}. Because
three distinct neurons were required for this corollary, there are
\(\binom{\nneurons}{3}\) such inequalities, each of which is really two distinct
inequalities. Since the bias space itself is defined by inequalities of the form
\(-1 \leq \beta_{ij} \leq 1\) with \(i < j\), this polytope can be described by a
total of \(2 \binom{\nneurons}{3} + 2 \binom{\nneurons}{2}\) inequalities:
\begin{itemize}
\item \(\beta_{ij} + \beta_{jk} - \beta_{ik} \leq 1\) for all \(i < j < k\),
\item \(-\beta_{ij} - \beta_{jk} + \beta_{ik} \leq 1\) for all \(i < j < k\),
\item \(\beta_{ij} \leq 1\) for all \(i < j\),
\item \(-\beta_{ij} \leq 1\) for all \(i < j\).
\end{itemize}

The inequalities from \Cref{cor:consistent-bias} can be rearranged to bound the
bias of \(i\) to fire before \(k\) if the other biases are known: \(\beta_{ij} +
\beta_{jk} - 1 \leq \beta_{ik} \leq \beta_{ij} + \beta_{jk} + 1\). This
inequality is the direct equivalent to the intuition that \(k\) should follow \(i\)
if \(k\) follows \(j\) and \(j\) follows \(i\). But longer chains of this form do not
create inequalities that are more restrictive than those arising from such
chains of length three. For considering possible biases among more than three
neurons, we start by looking at the simplest of such biases: those arising from
pure sequences.

\begin{definition}
A \emph{pure-bias matrix} is a matrix \(A\) such that \(A = \skbias(\b{s})\) for some
pure sequence \(\b{s} \in \sequences\).
\end{definition}

For each pure-bias matrix \(A\), there is a sequence \(\b{s}\) with one spike per
neuron such that \(\skbias(\b{s}) = A\). Let us start by considering the sequence
\(\b{s} = (1, 2, \dots, n)\). Here, we see that \(i\) precedes \(j\) whenever \(i < j\);
as such, the corresponding entries of \(\skbias(\b{s})\) are \(\beta_{ij}(\b{s}) =
1\) and \(\beta_{ji}(\b{s}) = -1\). For example,
\[
\skbias(1234) =\begin{bmatrix}
 0 & 1 & 1 & 1 \\
 -1 & 0 & 1 & 1 \\
 -1 & -1 & 0 & 1 \\
 -1 & -1 & -1 & 0 \\
\end{bmatrix}.
\]Indeed, up to row and column permutations, the bias matrix of each pure sequence
is of this form, as the following lemma shows.

\begin{lemma}
Suppose that \(\b{s} \in \sequences\) is pure. Then \(\skbias(\b{s}) = P^{-1} A P\)
for \(A = \skbias(12 \cdots n)\) and some permutation matrix \(P\).

\begin{proof}
Let \(\b{s} \in \sequences\) be pure. Without loss of generality, we can assume
that \(\b{s} = (s_1, \dots, s_n)\) has one spike per neuron since, for a pure
sequence, it is clear that \(\skbias(\b{s}) = \skbias(\s(\b{s}))\). Note that
\(\beta_{s_i, s_j}(\b{s}) = \beta_{ij}(12 \cdots n)\) since \(s_i\) precedes \(s_j\)
in \(\b{s}\) if and only if \(i\) precedes \(j\) in \(12 \cdots n\). Define \(\pi \in
\perms{\neurons}\) by \(\pi(s_i) = i\), and let \(P = [p_{ij}]_{i, j \in \neurons}\)
be the permutation matrix with entries \[ p_{ij} = \begin{cases} 1, & \pi(j) =
i; \\ 0, & \text{otherwise}. \end{cases} \] Letting \(C = P^{-1} A P =
[c_{ij}]_{i, j \in \neurons}\) and \(D = AP = [d_{ij}]_{i, j \in \neurons}\), we
find that
\begin{align*}
  c_{s_i, s_j}
  &= \sum_{k = 1}^n p_{k, s_i} d_{k, s_j} \\
  &= \sum_{k = 1}^n p_{k, s_i} \parens{\sum_{\l = 1}^n a_{k, \l} p_{\l, s_j}} \\
  &= \sum_{k = 1}^n p_{k, s_i} a_{k, \pi(s_j)} \\
  &= a_{\pi(s_i), \pi(s_j)} \\
  &= a_{ij} \\
  &= \beta_{ij}(12 \cdots n) \\
  &= \beta_{s_i, s_j}(\b{s}).
\end{align*}
Therefore, \(\skbias(\b{s}) = P^{-1} AP\) as claimed, completing the proof.
\end{proof}
\end{lemma}

Pure sequences are extreme from the perspective of biases. Indeed, each bias
corresponding to a pure sequence is a vertex of the the bias space hypercube. As
we have already seen in this section, not all vertices of the bias space are
legitimate biases. We now know exactly which vertices correspond to legitimate
biases. Moreover, we have a simple count for how many vertices are obtainable
biases: There is one for each permutation of the neuron set \(\neurons = [n]\);
that is, \(n!\) of the \(2^{\binom{n}{2}}\) bias-space vertices are obtainable
biases.

But we can glean more from this investigation of pure biases. Consider two pure
sequences with one spike per neuron that are as close to each other as possible,
that is, sequences that differ by a single transposition of spikes. For the sake
of example, we will consider \(\b{s} = 1234 \cdots n\) and \(\b{s}' = \b{s}^{(1\
2)} = 2134 \cdots n\). From the perspective of the neurons in \(\neurons
\setminus [2]\), neurons \(1\) and \(2\) both behave exactly the same (since both \(1\)
and \(2\) precede all other spikes in both \(\b{s}\) and \(\b{s}'\)). Many other
sequences share this property with \(\b{s}\) and \(\b{s}'\); in fact, we can
completely characterize the collection of all such sequences as \(\set{(s_1,
\dots, s_\l, 3, 4, \dots, n) \mid (s_1, \dots, s_\l) \in \sequences([2])}\). That
is, we can prepend an arbitrary sequence from the neuron set \([2]\) to the
beginning of the sequence \(34 \cdots n\). But we know something about the biases
arising from the sequences in \(\sequences([2])\): Every bias is obtainable! This
tells us that the entire line joining \(\b{s}\) to \(\b{s}'\) in the bias space is
obtainable as a bias. This is similarly true for any such pair of "adjacent"
sequences with one spike per neuron. More surprisingly, each obtainable bias is
in the convex hull of the collection of pure biases. Before showing this claim
to be true, we name this polytope and show its relationship to the
consistent-bias polytope.

\begin{definition}
The \emph{fundamental polytope} of \(\sequences\) is the convex hull of the set of pure
biases arising from sequences in \(\sequences\).
\end{definition}

\begin{figure}
  \centering
  \includegraphics[height=2in]{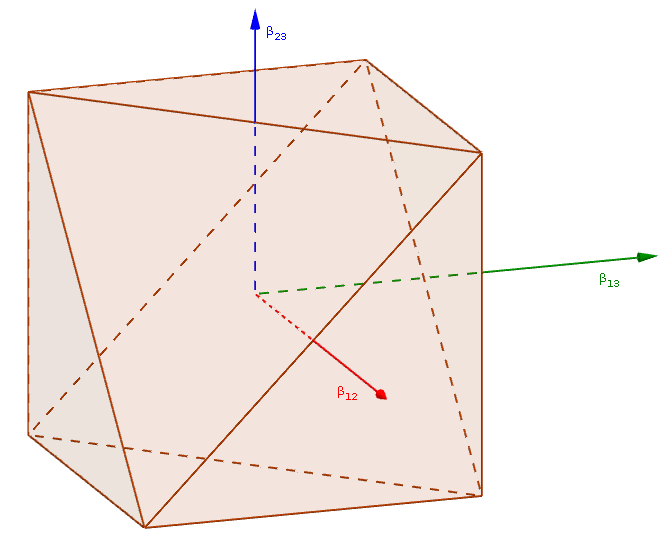}
  \includegraphics[height=2in]{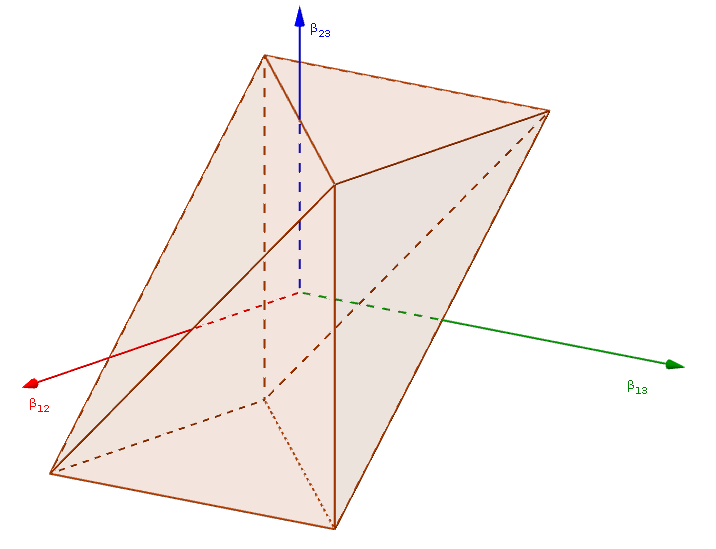}
  % ./images/polytope.png
  \caption{\label{fig:fundamental-polytope}When $\neurons = [3]$, the
    fundamental polytope of $\sequences$ is 3-dimensional. Above, this polytope
    is shown below from two angles. In this case of only three neurons, the
    fundamental polytope is equal to the consistent-bias polytope. Note that $n
    = 3$ is a special case: \textbf{These polytopes are not equal in general.}
    Since they are equal here, we can easily write down the hyperplanes that
    serve as its boundaries: $-1 \leq \beta_{12} + \beta_{23} - \beta_{13} \leq
    1$, $-1 \leq \beta_{12} \leq 1$, $-1 \leq \beta_{13} \leq 1$, and $-1 \leq
    \beta_{23} \leq 1$}
\end{figure}

\begin{lemma}
The consistent-bias polytope contains the fundamental polytope.

\begin{proof}
Since the vertices of the fundamental polytope are obtainable biases, they must
satisfy the constraints that specify the consistent-bias polytope; moreoever,
since both polytopes are convex, we must have that the fundamental polytope is
contained in the consistent-bias polytope.
\end{proof}
\end{lemma}

\begin{theorem}
\label{thm:fundamental-polytope}
If \(\b{s} \in \sequences\) with \(\supp(\b{s}) = \neurons\), then the bias
corresponding to \(\b{s}\) is contained in the fundamental polytope.

\begin{proof}
Let \(\b{t} = 12 \cdots n\), and let \(P_n = \{ \skbias(\b{t}^\pi) \mid \pi \in S_n
\}\) be the collection of pure-bias matrices. Also, let \(\b{s} = (s_1, \dots,
s_\l)\) be any sequence with \(\supp(\b{s}) = \neurons\). We need only to show that
\(\skbias(\b{s})\) is a convex combination of the elements of \(P_n\). For each
subsequence \(\b{s}'\) of \(\b{s}\) that is equal to \(\b{t}^\pi\) for some \(\pi \in
\perms{[n]}\), we have that \(\skbias(\b{s}') \in P_n\) by definition of \(P_n\). Let
\(\Sub(\b{s}) = \set{I \sset [\l] \mid \b{s}_I = \b{t}^\pi \text{ for some } \pi
\in \perms{[n]}}\), and define the sum \(D = [d_{ij}]\) to be \[ D =
\frac{1}{\prod_{i = 1}^n c_{i}(\b{s})} \sum_{I \in \Sub(\b{s})}
\skbias(\b{s}_I). \] Because \(\skbias(\b{s}_I) \in P_n\), the proof will be
finished if \(D = \skbias(\b{s})\) and if the coefficients of its summands sum
to 1.

To see that the latter of these requirements holds, first note that all
coefficients are equal. Moreover, there are exactly \(\prod_{i = 1}^n
c_{i}(\b{s})\) summands in the definition of \(D\): Choosing a subsequence of
\(\b{s}\) that is and ordering of \(\neurons\) is the same as choosing one index
from each of the sets \(L_i(\b{s})\) for \(i \in \neurons\), which can obviously be
done in \(\prod_{i = 1}^n c_{i}(\b{s})\) ways since \(c_i(\b{s}) = |L_i(\b{s})|\)
for each \(i \in \neurons\).

To see that \(D = \skbias(\b{s})\), consider an arbitrary entry \(d_{ij}\):
\begin{align*}
  d_{ij}
    &= \frac{1}{\prod_{k = 1}^n c_{k}(\b{s})}
       \sum_{I \in \Sub(\b{s})} \beta_{ij}(\b{s}_I) \\
    &= \frac{1}{\prod_{k = 1}^n c_{k}(\b{s})}
       \parens[\bigg]{\sum_{\substack{I \in \Sub(\b{s}) \\
                            c_{ij}(\b{s}_I) = 1}}
       1 +
       \sum_{\substack{I \in \Sub(\b{s}) \\
                       c_{ji}(\b{s}_I) = 1}}
       -1} \\
    &= \frac{1}{\prod_{k = 1}^n c_{k}(\b{s})}
       \bigg([c_{ij}(\b{s}) - c_{ji}(\b{s})] \cdot
             \prod_{\mathclap{k \notin \{i, j\}}} c_k(\b{s}) \bigg) \\
    &= \frac{c_{ij}(\b{s}) - c_{ji}(\b{s})}{c_i(\b{s}) c_j(\b{s})} \\
    &= \beta_{ij}(\b{s})
\end{align*}
Hence, we see that the fundamental polytope contains all biases.
\end{proof}
\end{theorem}

Though all biases are contained in the fundamental polytope, the question of
whether or not the biases are dense in the fundamental polytope is still an open
question.

\begin{question}
Is every point in the fundamental polytope obtainable?
\end{question}

\section{The bias network of a sequence}
\label{sec:orgheadline13}
\label{sec:bias-network}
\FloatBarrier

Neuronal templates are appealing because they provide a simple combinatorial
picture of neuronal sequences, but this representation is lacking in many
aspects. The bias matrix can be used to define a more-robust combinatorial
representation of these sequences.

\begin{definition}
Let \(A = [a_{ij}]_{i, j \in \neurons}\) be a bias matrix. We define the \emph{bias
network of \(A\)} to be the simple digraph \(G(A)\) with vertext set \(\neurons\) and
edge set \(E(A) := \set{(i, j) \mid a_{ij} > 0}\). The \emph{bias network of a sequence
\(\b{s}\)} is \(G(\b{s}) := G(\skbias(\b{s}))\) with edge set \(E(\b{s}) =
E(\skbias(\b{s}))\).
\end{definition}

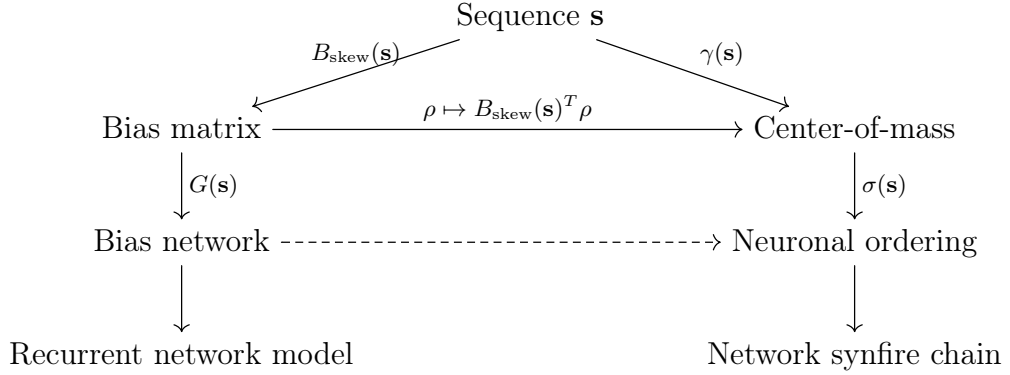
\begin{figure}
  \[
    \begin{tikzcd}
      \phantom{a} & \text{Sequence }\b{s} \ar{dl}[above]{\skbias(\b{s})} \ar{dr}{\gamma(\b{s})} & \\
      \text{Bias matrix} \ar{d}{G(\b{s})}
      \ar{rr}{\rho \mapsto \skbias(\b{s})^T \rho}
      & & \text{Center-of-mass} \ar{d}{\s(\b{s})} \\
      \text{Bias network} \ar{d} \ar[dashed]{rr} & & \text{Neuronal ordering} \ar{d} \\
      \text{Recurrent network model} & & \text{Network synfire chain}
    \end{tikzcd}
  \]
  \caption{\label{fig:bias-network}The bias network model. This diagram
    represents how the bias network serves as a generalization of neuronal
    templates. Starting with a sequence $\b{s}$, the center-of-mass vector and
    ordering can be computed directly (right side). Alternately, the same
    vectors can be computed via the bias matrix; this also allows for the
    generation of other orderings by varying the firing-rate vector $\rho$ used
    to generate the center-of-mass vector. The dashed line across the middle
    indicates that, in the case of pure sequences, the center-of-mass ordering
    is recoverable from the bias network, as shown in
    \Cref{lem:pure-transitive}. At the bottom level, these two combinatorial
    objects can be used to generate sequences using recurrent networks and
    network synfire chains, topics beyond the scope of this paper.}
\end{figure}

\begin{definition}
A \emph{tournament} is a complete oriented graph.
\end{definition}

\begin{definition}
A digraph \(D\) with edge-set \(E\) is \emph{transitive} if \((v_1, v_3) \in E\) whenever
\((v_1, v_2), (v_2, v_3) \in E\).
\end{definition}

The vertices of a transitive tournament can be totally ordered by reachability
(i.e., vertex \(v\) precedes vertex \(u\) in the ordering if \((v, u)\) is an edge).
The following lemma shows one way in which bias networks generalize neuronal
templates: by unambiguously inducing a total ordering in the case of a pure
sequence.

\begin{lemma}
\label{lem:pure-transitive}
If \(A\) is a pure-bias matrix, then \(G(A)\) is a transitive tournament.

\begin{proof}
Let \(\neurons = [n]\). If \(n = 1\), then the claim is trivially true. Let \(\b{s}\)
be a sequence with \(\skbias(\b{s}) = A\), and suppose that \(n > 1\). Since \(\b{s}\)
is pure, some neuron \(i \in \neurons\) must spike first. Then \(\b{s}|_{\neurons
\setminus \set{i}}\) is a pure sequence on \(n - 1\) neurons. By induction on \(n\),
we have that \(G(\skbias(\b{s}|_{\neurons \setminus \set{i}}))\) is a transitive
tournament. Because \(i\) precedes all other neurons, we have that
\(\beta_{ij}(\b{s}) = 1\) for each \(j \in \neurons \setminus \set{i}\). In
particular, this means that the graph \(G(\skbias(\b{s})) = A\) is transitive,
completing the proof.
\end{proof}
\end{lemma}

As a result of \Cref{lem:pure-transitive}, it is clear that all possible
transitive tournaments can be realized as the bias network of some sequence. But
it is not obvious which other digraphs are realizable, and it would be
surprising if a sequence could be constructed to match an arbitrary list of
pairwise orderings (as a simple digraph specifies).

\begin{question}
\label{qstn:possible-bias-networks}
Which digraphs are realizable as the bias network of some sequence?
\end{question}

The following examples explore this question.

\begin{example}
The sequence \(\b{s} = 12444433131124\) has skew-bias matrix
\[
\begin{bmatrix}
 0 & \nicefrac{1}{4} & -\nicefrac{1}{3} & -\nicefrac{1}{5} \\
 -\nicefrac{1}{4} & 0 & 0 & -\nicefrac{1}{5} \\
 \nicefrac{1}{3} & 0 & 0 & -\nicefrac{3}{5} \\
 \nicefrac{1}{5} & \nicefrac{1}{5} & \nicefrac{3}{5} & 0 \\
\end{bmatrix}.
\]This allows us to easily see that \(G(\b{s})\) has adjacency matrix
\[
\begin{bmatrix}
 0 & 1 & 0 & 0 \\
 0 & 0 & 0 & 0 \\
 1 & 0 & 0 & 0 \\
 1 & 1 & 1 & 0 \\
\end{bmatrix}.
\]Graphically, we can visualize \(G(\b{s})\) as below.
\begin{center}
  \begin{tikzpicture}[->,>=stealth',shorten >=1pt,auto,node distance=0.75in,
      thick,main node/.style={circle,draw,font=\sffamily\bfseries}]

    \node[main node] (1) {1};
    \node[main node] (2) [right of=1] {2};
    \node[main node] (3) [below of=1] {3};
    \node[main node] (4) [below of=2] {4};

    \path[every node/.style={font=\sffamily\small}]
    (1) edge node {} (2)
    (3) edge node {} (1)
    (4) edge node {} (1)
    edge node {} (2)
    edge node {} (3);
  \end{tikzpicture}
\end{center}
Notice that \(G(\b{s})\) is not a tournament because it does not contain an edge
between vertices \(2\) and \(3\), which happened because \(\beta_{23}(\b{s}) =
\beta_{32}(\b{s}) = 0\).
\end{example}

\begin{example}
The sequence \(\b{s} = 2331112123\) has bias network \(G(\b{s})\) with adjacency
matrix
\[
\begin{bmatrix}
 0 & 1 & 0 \\
 0 & 0 & 1 \\
 1 & 0 & 0 \\
\end{bmatrix}.
\]Graphically, \(G(\b{s})\) is shown below.
\begin{center}
\begin{tikzpicture}[->,>=stealth',shorten >=1pt,auto,node distance=0.75in,
thick,main node/.style={circle,draw,font=\sffamily\bfseries}]

\node[main node] (1) {1};
\node[main node] (2) [below left of=1] {2};
\node[main node] (3) [below right of=1] {3};

\path[every node/.style={font=\sffamily\small}]
(1) edge node {} (2)
(2) edge node {} (3)
(3) edge node {} (1);
\end{tikzpicture}
\end{center}
So \(G(\b{s})\) is clearly not transitive since \((1, 2), (2, 3) \in E(\b{s})\) but
\((1, 3) \notin E(\b{s})\).
\end{example}

The answer to \Cref{qstn:possible-bias-networks} is surprising: We have control
over each individual tendency in a bias network.

\begin{theorem}
\label{thm:bias-network}
Every bias network is possible.
\end{theorem}

To prove this theorem, we first need some other results and definitions.

\begin{definition}
The \emph{reverse} of a sequence \(\b{s} = (s_1, \dots, s_\l)\) is the sequence
\(\b{s}^T = (s_\l, s_{\l - 1}, \dots, s_1)\).
\end{definition}

Note that the reverse of the reverse of a sequence is the original sequence;
that is \((\b{s}^T)^T = \b{s}\) for any sequence \(\b{s} \in \sequences\). The
reason for the notation \(\b{s}^T\) is seen in the following lemma.

\begin{lemma}
\label{lem:reverse-transpose}
For any sequence \(\b{s}\), \(\skbias(\b{s}^T) = \skbias(\b{s})^T\).

\begin{proof}
Consider neurons \(i, j \in \neurons\). For each index pair \((k, k') \in
L_{ij}(\b{s})\), we must have that \((k', k) \in L_{ji}(\b{s}^T)\). Similarly, for
each index pair \((m, m') \in L_{ij}(\b{s}^T)\), we must have that \((m', m) \in
L_{ji}(\b{s})\). Hence, \(c_{ij}(\b{s}) = \abs{L_{ij}(\b{s})} =
\abs{L_{ji}(\b{s}^T)} = c_{ji}(\b{s}^T)\) for arbitrary neuron pairs. Now, \[
\beta_{ji}(\b{s}^T) = \frac{c_{ji}(\b{s}^T) - c_{ij}(\b{s}^T)}{c_{ji}(\b{s}^T) +
c_{ij}(\b{s}^T)} = \frac{c_{ij}(\b{s}) - c_{ji}(\b{s})}{c_{ij}(\b{s}) +
c_{ji}(\b{s})} = \beta_{ij}(\b{s}) = -\beta_{ji}(\b{s}) . \] In particular,
\(\skbias(\b{s}^T) = -\skbias(\b{s}) = \skbias(\b{s})^T\) since \(\skbias\) is
skew-symmetric, thus finishing the proof.
\end{proof}
\end{lemma}

\begin{lemma}
\label{lem:change-bias-entry}
Given a sequence \(\b{s}\), a sign value \(a \in \set{-1, 1}\), and distinct neurons
\(i, j \in \neurons\), there is a sequence \(\b{s}'\) such that
\begin{enumerate}
\item \(\sign(\beta_{ij}(\b{s}')) = a\) and
\item \(\sign(\beta_{qr}(\b{s}')) = \sign(\beta_{qr}(\b{s}))\) for \(\set{q, r} \neq
   \set{i, j}\).
\end{enumerate}

\begin{proof}
Recall that, for any distinct neurons \(i, j \in \neurons\), \(\beta_{ij} =
\frac{c_{ij} - c_{ji}}{c_{ij} + c_{ji}}\). Since \(c_{ij} + c_{ji}\) is a count of
neuron pairs, it is positive; in particular, this tells us that
\(\sign(\beta_{ij}) = \sign(c_{ij} - c_{ji})\). Now, let \(\b{s} \in \sequences\) be
given and fix \(i, j \in \neurons\) as distinct neurons.

Since \(\skbias\) is skew-symmetric, assume without loss of generality that \(a =
1\) (by exchanging \(i\) and \(j\) if \(a = -1\)). Pick \(k \in \N\) such that \(2k^2 >
c_{ji}(\b{s}) - c_{ij}(\b{s})\), and define \(\b{s}' := i^k \cdot j^k \cdot \b{s}
\cdot i^k \cdot j^k\). We will first show that \(\beta_{ij}(\b{s}') > 0\), implying
that \(\sign(\beta_{ij}(\b{s}')) = 1 = a\). Note that \(c_{ij}(\b{s}') = 3k^2 + k
\cdot c_j(\b{s}) + c_i(\b{s}) \cdot k + c_{ij}(\b{s})\) and that \(c_{ji}(\b{s}')
= k^2 + k \cdot c_i(\b{s}) + c_j(\b{s}) \cdot k + c_{ji}(\b{s})\). Now,
\(c_{ij}(\b{s}) - c_{ji}(\b{s}) = 2k^2 + c_{ij}(\b{s}) - c_{ji}(\b{s}) > 0\), as
claimed.

Let \(q, r \in \neurons\) be distinct with \(\set{q, r} \neq \set{i, j}\). Suppose
that \(\set{q, r} \cap \set{i, j} = \emptyset\). Then \(\beta_{qr}(\b{s}') =
\beta_{qr}(\b{s}'|_{\neurons \setminus \set{i, j}}) =
\beta_{qr}(\b{s}|_{\neurons \setminus \set{i, j}}) = \beta_{qr}(\b{s})\). In
particular, \(\sign(\beta_{qr}(\b{s}')) = \sign(\beta_{qr}(\b{s}))\). Now, suppose
that \(q \in \set{i, j}\) but \(r \notin \set{i, j}\). Then \(c_{qr}(\b{s}') = k
\cdot c_r(\b{s}) + c_{qr}(\b{s})\) and \(c_{rq}(\b{s}') = c_{rq}(\b{s}) +
c_r(\b{s}) \cdot k\); hence, \(c_{qr}(\b{s}') - c_{rq}(\b{s}') = c_{qr}(\b{s}) -
c_{rq}(\b{s})\). In particular, \(\sign(\beta_{qr}(\b{s}')) =
\sign(c_{qr}(\b{s}') - c_{rq}(\b{s}')) = \sign(c_{qr}(\b{s}) - c_{rq}(\b{s})) =
\sign(\beta_{qr}(\b{s}))\). Therefore, the lemma holds.
\end{proof}
\end{lemma}

\begin{definition}
A sequence \(\b{s}\) is a \emph{palindrome} if \(\b{s}^T = \b{s}\).
\end{definition}

\begin{lemma}
\label{lem:palindrome}
If \(\b{s} \in \sequences([n])\) is a palindrome, then \(\skbias(\b{s}) = \b{0}_n\),
where \(\b{0}_n\) is the \(n \x n\) zero matrix.

\begin{proof}
Since \(\b{s}\) is a palindrome, we have that \(\b{s} = \b{s}^T\). Using
\Cref{lem:reverse-transpose} and the fact that the skew-bias matrix is
skew-symmetric, we have that \(\skbias(\b{s}) = \skbias(\b{s}^T) =
\skbias(\b{s})^T = -\skbias(\b{s})\). In particular, \(\beta_{ij}(\b{s}) =
-\beta_{ij}\) for each \(i, j \in \neurons\), implying that \(\beta_{ij} = 0\). This
completes the proof.
\end{proof}
\end{lemma}

We are now ready to show that every simple digraph is realizable as a bias
network.

\begin{proof}[Proof of \Cref{thm:bias-network}]
Let \(D\) be an arbitrary digraph. Then \(D\) has skew-symmetric adjacency matrix \(A
\in \set{-1, 0, 1}^{n \x n}\). We want to show that there exists a sequence
\(\b{s}\) on neuron set \(\neurons = [n]\) such that \(\skbias(\b{s}) = A\).

Let \(\b{s}_0\) be a palindrome with \(\supp(\b{s}_0) = \neurons\). By
\Cref{lem:palindrome}, we know that \(\skbias(\b{s}_0) = \b{0}_n\) is the \(n \x n\)
zero matrix. Define \(I := \set{(i, j) \in \neurons^s \mid a_{ij} > 0}\) to be the
set of index pairs at which \(A\) is positive, and let \(I = \set{(i_1, j_1),
\dots, (i_m, j_m)}\) be an enumeration of \(I\). We proceed by induction on \(m\). If
\(m = 1\), then \(A\) has only one positive entry \(a_{i_1 j_1}\). Since \(\b{s}_0\) is
the zero matrix, \Cref{lem:change-bias-entry} provides a sequence \(\b{s}_1\) such
that \(\beta_{i_1 j_1}(\b{s}_1) = 1\) and \(\beta_{qr}(\b{s}_1) = 0\) for all \(q, r
\in \neurons\) with \(\set{q, r} \neq \set{i, j}\); that is \(\skbias(\b{s}_1) = A\).

Suppose now that \(m > 1\), and assume for induction that there is a sequence
\(\b{s}_{m - 1}\) such that \(\beta_{i_m j_m}(\b{s}_{m - 1}) = \beta_{j_m,
i_m}(\b{s}_{m - 1}) = 0\) and \(\beta_{qr}(\b{s}_{m - 1}) = a_{qr}\) for all other
\(q, r \in \neurons\). \Cref{lem:change-bias-entry} now provides a sequence
\(\b{s}_m\) such that \(\beta_{i_m j_m}(\b{s}_m) = 1\) and \(\beta_{qr}(\b{s}_m) =
\beta_{qr}(\b{s}_{m - 1})\) for all other \(q, r \in \neurons\). Then
\(\skbias(\b{s}_m) = A\). Hence, by induction, the lemma holds for any \(m > 0\).
\end{proof}

\chapter{Comparing sequences}
\label{sec:orgheadline19}
\label{ch:comparing}

\section{Correlation between sequences}
\label{sec:orgheadline15}

In the brain, it is often the case that signals are very noisy (due to neuronal
misfires, improper spike sorting, etc.); as a result, comparing sequences
directly to each other may not be so effective. Additionally, complications
arise when one wants, for instance, to compare relatively long neuronal
sequences (e.g., place-field sequences) to relatively short neuronal sequences
(e.g., SWR sequences): How does one meaningfully (and generically) compare a
sequence with 10 spikes to a sequence with 100 spikes? Even if a meaningful
comparison were developed to directly compare the sequences, this comparison
would likely be computationally expensive since these techniques usually involve
modifying one sequence until it becomes the other sequence.
As we attempt to compare sequences in a
less direct way, we must ask a natural question: What does it mean for two
sequences to be similar (or even "the same")? As we will hopefully show, the
bias matrix provides a somewhat natural framework for answering this question
and, thus, for comparing sequences.

Recall that the bias matrix arose out of an attempt to generalize our intuitive
notion of a pure neuronal sequence. Even with this representation of sequences
as biases (which are vectors), careful consideration should be given to what
method is employed to compare these vectors. It is tempting to want to compare
biases using the standard Euclidean distance, but this may not be the most
meaningful of comparisons. To see this, once again consider pure sequences.
Recalling \Cref{lem:recover-pure-order}, the bias matrix of a pure sequence
contains sufficient information to determine the original neuronal ordering.
Pure sequences, though, are not the only sequences for which the bias matrix
allows us to unambiguously reconstruct an ordering in the same manner. Indeed,
some of the sequences corresponding to these other bias matrices even match up
well with our original intuition.

\begin{example}
\label{ex:scaled-pure-bias}
Consider the spike-raster plot in \Cref{fig:impure-sequence}.
\begin{figure}
  \centering
  \includegraphics[width=1in]{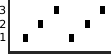}
  \caption{\label{fig:impure-sequence}Spike raster plot of an impure sequence
    that induces an unambiguous neuronal ordering}
\end{figure}
Though the spikes of the neurons are interlaced, it seems obvious to the casual
observer that there is only one neuronal ordering that makes sense: neuron \(1\)
fires, then neuron \(2\), and finally neuron \(3\). This casual observation seems
slightly less obvious when considering the sequence representation \(\b{s} =
123123\), but it emerges again upon computing the bias matrix:
\[
\skbias(\b{s}) =\begin{bmatrix}
 0 & \nicefrac{1}{2} & \nicefrac{1}{2} \\
 -\nicefrac{1}{2} & 0 & \nicefrac{1}{2} \\
 -\nicefrac{1}{2} & -\nicefrac{1}{2} & 0 \\
\end{bmatrix}= \half \skbias(123).
\]The skew-bias matrix of \(\b{s}\) is a scaling of that of the pure sequence 123.
As such, \(\skbias(\b{s})\) induces the same ordering as \(\skbias(123)\), namely
the ordering \((1, 2, 3)\).
\end{example}

From this perspective, it seems that the underlying "ordering," if you will, of
a bias matrix is characterized simply by the direction in which the bias matrix
points (when considered as a vector). This further suggests that the magnitude
of a bias vector is not so much an indication of an underlying order as it is an
indication of the purity of a sequence. It is with this perspective that we
define a correlation between sequences.

\begin{definition}
We define the \emph{product} of sequences \(\b{s}\) and \(\b{s}'\) to be \(\inprod{\b{s},
\b{s}'} := \sum_{i, j \in \omega} \beta_{ij}(\b{s}) \beta_{ij}(\b{s}')\) for
\(\omega = \supp(\b{s}) \cap \supp(\b{s}')\). Note that \(\inprod{\b{s}, \b{s}'} =
\inprod{\b{s}|_\omega, \b{s}'|_\omega} = \inprod{\skbias(\b{s}|_\omega),
\skbias(\b{s}'|_\omega)}\), where the last expression is the standard Euclidean
inner product of \(\skbias(\b{s}|_\omega)\) and \(\skbias(\b{s}'|_\omega)\). Also,
we define the \emph{magnitude} of a sequence \(\b{s}\) to be \(\norm{\b{s}} :=
\sqrt{\inprod{\b{s}, \b{s}}}\). With these expressions in hand, we can finally
define the \emph{correlation of sequences \(\b{s}\) and \(\b{s}'\)} to be
\[
\corr(\b{s}, \b{s}') :=
\frac{\inprod{\b{s}, \b{s}'}}
{\norm{\b{s}|_{\supp(\b{s}')}} \cdot \norm{\b{s}'|_{\supp(\b{s})}}}
\]
\end{definition}

In terms of geometry, \(\corr(\b{s}, \b{s}')\) is the cosine of the angle between
\(\skbias(\b{s}|_\omega)\) and \(\skbias(\b{s}'|_\omega)\), where again \(\omega =
\supp(\b{s}) \cap \supp(\b{s}')\). In these definitions, we restrict to the
common set of neurons in part because noise is a major consideration in real
data. A real sequence might contain a spike from a neuron that should not have
been active but not contain a spike from a neuron that should have been active.
Though we cannot know for sure which spikes "should" and "should not" be in a
sequence, we attempt to mitigate this problem slightly by comparing sequences
based only on the collection of neurons that are active in both sequences.

\begin{example}
Let \(\b{s} = 12444433131124\) and \(\b{s}' = 2431\). Then
\[
\skbias(\b{s}) =\begin{bmatrix}
 0 & \nicefrac{1}{4} & -\nicefrac{1}{3} & -\nicefrac{1}{5} \\
 -\nicefrac{1}{4} & 0 & 0 & -\nicefrac{1}{5} \\
 \nicefrac{1}{3} & 0 & 0 & -\nicefrac{3}{5} \\
 \nicefrac{1}{5} & \nicefrac{1}{5} & \nicefrac{3}{5} & 0 \\
\end{bmatrix}\qquad \text{and} \qquad
\skbias(\b{s}') =\begin{bmatrix}
 0 & -1 & -1 & -1 \\
 1 & 0 & 1 & 1 \\
 1 & -1 & 0 & -1 \\
 1 & -1 & 1 & 0 \\
\end{bmatrix}.
\]Noting that \(\supp(\b{s}) = \supp(\b{s}') = [4]\), the correlation between
\(\b{s}\) and \(\b{s}'\) is \(\corr(\b{s}, \b{s}') = \frac{\inprod{\b{s},
\b{s}'}}{\norm{\b{s}} \cdot \norm{\b{s}'}}\). Now, each component of this can be
computed as follows:
\begin{align*}
\inprod{\b{s}, \b{s}'}
&= \sum_{i, j \in \neurons} \beta_{ij}(\b{s}) \beta_{ij}(\b{s}') \\
&= 2 \sum_{i = 1}^{4 - 1} \sum_{j = i + 1}^4 \beta_{ij}(\b{s}) \beta_{ij}(\b{s}') \\
&= 2 \parens{-\tfrac{1}{4} + \tfrac{1}{3} + \tfrac{1}{5} + 0 - \tfrac{1}{5} + \tfrac{3}{5}} \\
&= \frac{41}{60} \\
\norm{\b{s}}
&= \sqrt{2 \parens{\tfrac{1}{16} + \tfrac{1}{9} + \tfrac{1}{25} + 0
- \tfrac{1}{25} + \tfrac{9}{25}}}
= \frac{47 \sqrt{2}}{60} \\
\norm{\b{s}'} &= \sqrt{12} = 2 \sqrt{3}
\end{align*}
Putting this all together, we find that \(\corr(\b{s}, \b{s}') = \frac{41}{47
\sqrt{6}} \approx 0.6168\).
\end{example}

\section{Significance of correlation}
\label{sec:orgheadline16}
\label{sec:significance}

While we do now have the ability to compare arbitrary sequences using our
correlation measure, there is still an issue to be addressed: How do we
interpret a correlation value? Despite coming to the bias-matrix representation
of a sequence and the correlation measure by fairly intuitive means, a
correlation value does not seem to have an intuitive interpretation beyond (1)
its sign indicating correlation or anti-correlation and (2) larger magnitude
indicating strength of (anti-)correlation.

\begin{figure}
  \centering
  \begin{subfigure}[b]{4.07in}
    \includegraphics[height=1in]{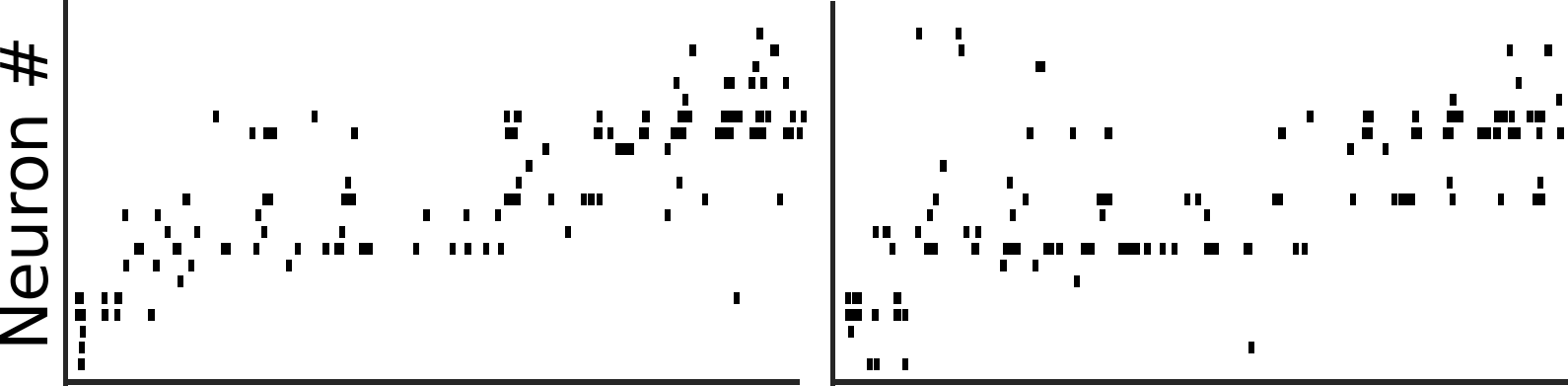}
    \caption{\label{fig:correlation-not-enough-arm}Arm-run}
  \end{subfigure}
  \hspace{1.5em}
  \begin{subfigure}[b]{0.69in}
    \includegraphics[height=1in]{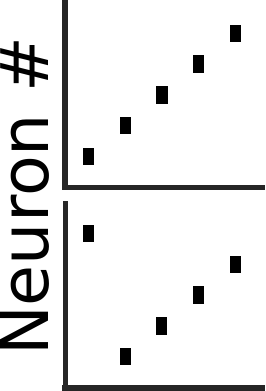}
    \caption{\label{fig:correlation-not-enough-pure}Pure}
  \end{subfigure}
  \caption{\label{fig:correlation-not-enough}Both of these pairs of sequences
    seem to be similar (to the eye), so we might expect them to have relatively
    high correlation values. However, neither pair does: The arm-run sequences
    have a correlation value of 0.52572, and the pure sequences have a
    correlation value of 0.2. These low values are somewhat unsatisfactory since
    maximally uncorrelated sequences have a correlation value of zero.}
\end{figure}

In \Cref{fig:correlation-not-enough}, we see an example of how a correlation
value alone does not necessarily say so much about the similarity of two
sequences. Why does something like this happen? This is partly an issue of
angles and dimensionality. Consider the pure sequences in
\Cref{fig:correlation-not-enough-pure}. By moving just a single spike, we
modified a large proportion of the entries in the pairwise-bias matrix. If these
short sequences had a common subsequences prepended to them, then the
correlation value would be much higher even though the same number of spikes are
"out of order"; this is because a much larger percentage of the pairwise-bias
matrix entries are unmodified in this case. When considering the arm-run
sequences in \Cref{fig:correlation-not-enough-arm}, we notice that each of the
neurons has spikes that are interlaced with spikes from many of the other
neurons. When spikes are interlaced in a consistent way like in Example
\ref{ex:scaled-pure-bias}, bias matrices end up being scalings of pure biases. These
spikes, however, are not interlaced in such consistent ways; and the result is
that the bias matrices end up pointing in fairly different directions. Still, to
the eye, these long arm-run sequences could be much more different from each
other if we just shifted some of their spikes around.

And here we stumble onto some key intuition: We evaluate the similarity of one
sequence \(\hat{\b{s}}\) to another \(\b{s}\) not just on how close \(\hat{\b{s}}\) is
to \(\b{s}\) but also on how much more different they could be from each other. In
a sense, we ask the question of where their similarity falls in relation to the
similarities of other pairs of sequences.

\begin{definition}
Given sequences \(\b{s}\) and \(\hat{\b{s}}\), the \emph{significance of correlation} of
\(\hat{\b{s}}\) to \(\b{s}\) is the proportion of shufflings of \(\hat{\b{s}}\) that
are more strongly correlated to \(\b{s}\) than \(\hat{\b{s}}\) is to \(\b{s}\). More
precisely, letting \(\Pi = \set{\pi \in \perms{[\l']} : \abs{\corr(\b{s},
\hat{\b{s}})} \leq \abs{\corr(\b{s}, \hat{\b{s}}^\pi)}}\), then the significance
of the correlation of \(\hat{\b{s}}\) to \(\b{s}\) is \(\sig_{\b{s}}(\hat{\b{s}}) :=
\frac{|\Pi|}{\len(\hat{\b{s}})!}\).
\end{definition}

\begin{example}
We used computational techniques (as will be described in
\Cref{sec:monte-carlo}) to approximate the significance values corresponding to
the correlations of \Cref{fig:correlation-not-enough}. In our simulations, the
significance of the correlation values of the arm-run sequences and of the pure
sequences were approximately 0.00 and 0.40, respectively. Using a standard
\$p\$-value of 0.05, This indicates that the correlation value between the arm-run
sequences is significant but that the correlation value between the pure
sequences is not significant.
\end{example}

\section{Global significance}
\label{sec:orgheadline17}
\label{sec:global-significance}

The previous section allows us to compare collections of sequences while putting
each correlation on more even footing; i.e., we can set a threshold for
significance instead of for (absolute value of) correlation. Now, we are brought
to the question of similarity of one collection of sequences to another. Let \(U
= (\b{u}_1, \dots, \b{u}_a)\) and \(V = (\b{v}_1, \dots, \b{v}_b)\) be lists of
sequences in \(\sequences\). We want to ask how similar is \(U\) to \(V\). One natural
place to start answering this question is to consider all of the correlation
values between a sequence in \(U\) and a sequence in \(V\).

\begin{definition}
We define the \emph{correlation matrix} between \(U\) and \(V\) to be \(\Corr(U, V) :=
[\corr(\b{u}, \b{v})]_{\b{u} \in U,\ \b{v} \in V}\).
\end{definition}

Though we can look for commonalities and statistics among the entries of this
matrix (e.g., that it has all positive entries or by considering the
distribution of values), we face a similar problem as when looking at individual
correlation values: Aside from basic properties (such as having only positive
entries), we do not have a good interpretation of the magnitude of correlation
(as demonstrated in the examples of the previous section). As such, we will
approach the answer to this qeustion in the same way as for individual
correlations: by considering significance values. But we will need to iterate
the process, in a sense, because we are comparing collections instead of just
individual sequences.

\begin{definition}
Define the \emph{significance matrix} of correlation of \(U\) to \(V\) to be \(\Sig_V(U)
:= [\sig_\b{v}(\b{u})]_{\b{u} \in U,\ \b{v} \in V}\).
\end{definition}

The significance matrix allows us to identify how significantly correlated each
sequence in \(U\) is to each sequence in \(V\). Still, though, the question remains:
How do we convert this information to some sort of meaningful value to tell us
how similar \(U\) is to \(V\)? Given a significance parameter \(p \in (0, 1)\), we can
count how many pairs \((\b{u}, \b{v}) \in U \x V\) satisfy \(\sig_\b{v}(\b{u}) <
p\). Though this gives us, for instance, the percentage of pairs that are
significantly correlated (as determined by \(p\)), we have still not quite reached
a satisfactory measure. What percentage of pairs is a significant percentage? It
really depends on the collections \(U\) and \(V\). As we compared \(\corr(\b{s},
\hat{\b{s}})\) to a distribution in the previous section, we must find an
appropriate distribution of values to which we can compare the count \(\sig_V(U,
p) := \abs{\set{(\b{u}, \b{v}) \in U \x V \mid \sig_\b{v}(\b{u}) < p}}\).
Naturally, as before, we want to compare this value to the counts gotten by
comparing a random shuffling of the sequenes in \(U\) to \(V\). If \(\vec{\pi} =
(\pi_1, \dots, \pi_a) \in \Pi(U) := \perms{[\len(\b{u}_1)]} \x \cdots \x
\perms{[\len(\b{u}_a)]}\), define \(U^{\vec{\pi}} := (\b{u}_1^{\pi_1}, \dots,
\b{u}_a^{\pi_a})\). Now, the \emph{global significance of correlation} of \(U\) to \(V\)
is \(g_V(U, p) = \frac{|A|}{|\Pi(U)|}\) where \(A = \set{\vec{\pi} \in \Pi(U) \mid
\sig_V(U^{\vec{\pi}}, p) \geq \sig_V(U, p)}\). Hence, we have a measure telling
us how significantly correlated the list \(U\) is to the list \(V\).

\section{Monte-Carlo computation of significance}
\label{sec:orgheadline18}
\label{sec:monte-carlo}

Mathematically, the significance of correlation and global significance of
correlation values described in the previous two sections allow us to have
confidence that two sequences or two collections of sequences share (or do not
share) similarities. In practice, however, the distributions necessary for
computation of these values are often prohibitively large. For instance, if
\(\hat{\b{s}}\) is a sequence with only ten spikes, calculation of
\(\sig_\b{s}(\hat{\b{s}})\) requires computation of over \(3.6\) million correlation
values. And the situation gets a lot worse as the length of \(\hat{\b{s}}\) grows,
not to mention the added complexities when trying to compute a global
significance value.

Because of this difficulty, we employ a Monte-Carlo method to approximate the
two types of significance values. In the analysis that follows in
\Cref{ch:analysis}, each sequence is a part of some collection of sequences. We
leverage this fact by simultaneously approximating the distribution for both
types of significance values. Let \(T\) be some (large) natural number
representing the number of trials in our Monte-Carlo approximations. Further,
let \(U = (\b{u}_1, \dots, \b{u}_a)\) and \(V\) be lists of sequences; and let \(p
\in (0, 1)\) be a significance parameter. For each \(\b{u} \in U\) and each \(\b{v}
\in V\), we will approximate \(\sig_\b{v}(\b{u})\) in such a way that the resultant
distribution approximations of all such pairings can also be used to approximate
the distributions necessary for computing \(g_V(U, p)\).

Let \(\vec{\pi}_1, \dots, \vec{\pi}_T \in \Pi(U)\) be \(T\) randomly selected lists
of permutations. For each \(t \in [T]\), we compute and store the correlation
matrix \(\Corr(V, U^{\vec{\pi}_k})\). This list of correlation matrices will allow
us to approximate all significance values. First, to compute
\(\sig_\b{v}(\b{u}_m)\), note that the list \(\corr(\b{v}, \b{u}_m^{\pi_{1, m}}),
\dots, \corr(\b{v}, \b{u}_m^{\pi_{T, m}})\) can be extracted from the list of
correlation matrices; hence, we approximate \(\sig_\b{v}(\b{u}_m)\) by
\(\widehat{\sig}_\b{v}(\b{u}_m) = \frac{|A|}{T}\) where \(A = \{t \in
[T] : |\corr(\b{v}, \b{u}_m)| < |\corr(\b{v}, \b{u}_m^{\pi_{t, m}})|\}\). Indeed,
this process produces an approximation \(\widehat{\Sig}_V(U)\) of the significance
matrix \(\Sig_V(U)\).

To approximate \(g_V(U)\) using the list of computed correlation matrices, we note
first that the list \(U^{\vec{\pi}_k}\) is a shuffling not only of the list \(U\)
but also of the list \(U^{\vec{\pi}_{t'}}\) for each \(t' \in [T] \setminus
\set{t}\). As such, we can also approximate the significance matrices
\(\Sig_V(U^{\vec{\pi}_t})\) by \(\widehat{\Sig}_V(U^{\vec{\pi}_t}) =
[\widehat{\sig}_\b{v}(\b{u}_m^{\pi_{t, m}})]_{\b{v} \in V, m \in [a]}\) with
\(\widehat{\sig}_\b{v}(\b{u}_m^{\pi_{t, m}}) = \frac{|A|}{T}\) where \(A = \{t'
\in [T] : |\corr(\b{v}, \b{u}_m^{\pi_{t, m}})| < |\corr(\b{v}, \b{u}_m^{\pi_{t',
m}})|\}\). With these matrices in hand, the approximation of \(g_V(U, p)\) is
straightforward: Let \(\widehat{\sig}_V(U', p) = |\{(\b{u}, \b{v}) \in U
\x V : \widehat{\sig}_V(U', p)\}|\), from which we define \(\hat{g}_V(U, p) =
\frac{|A|}{T}\) where \(A = \{t \in [T] : \widehat{\sig}_V(U^{\vec{\pi}_t},
p) \geq \widehat{\sig}_V(U, p)\}\).

\chapter{Experimental set-up and data analysis}
\label{sec:orgheadline23}
\label{ch:experiment}

The tools introduced in the preceding chapters were developed for use in the
analysis of real-world data. The lab of Eva Pastalkova at the Howard Hughes
Medical Institute's Janelia Research Campus graciously provided all of the data
used in this document. The following section describes the experimental setup
from which this data was collected and the various types of neuronal sequences
that are considered in our analysis. \Cref{sec:swr-detection} then describes
details of our sequence detection techniques. Finally, \Cref{ch:analysis}
uses bias matrices to perform some basic analyses of the data.

\section{The experimental setup and event types}
\label{sec:orgheadline20}
\label{sec:experiment}

\begin{figure}
  \centering
  \includegraphics[width=2.5in]{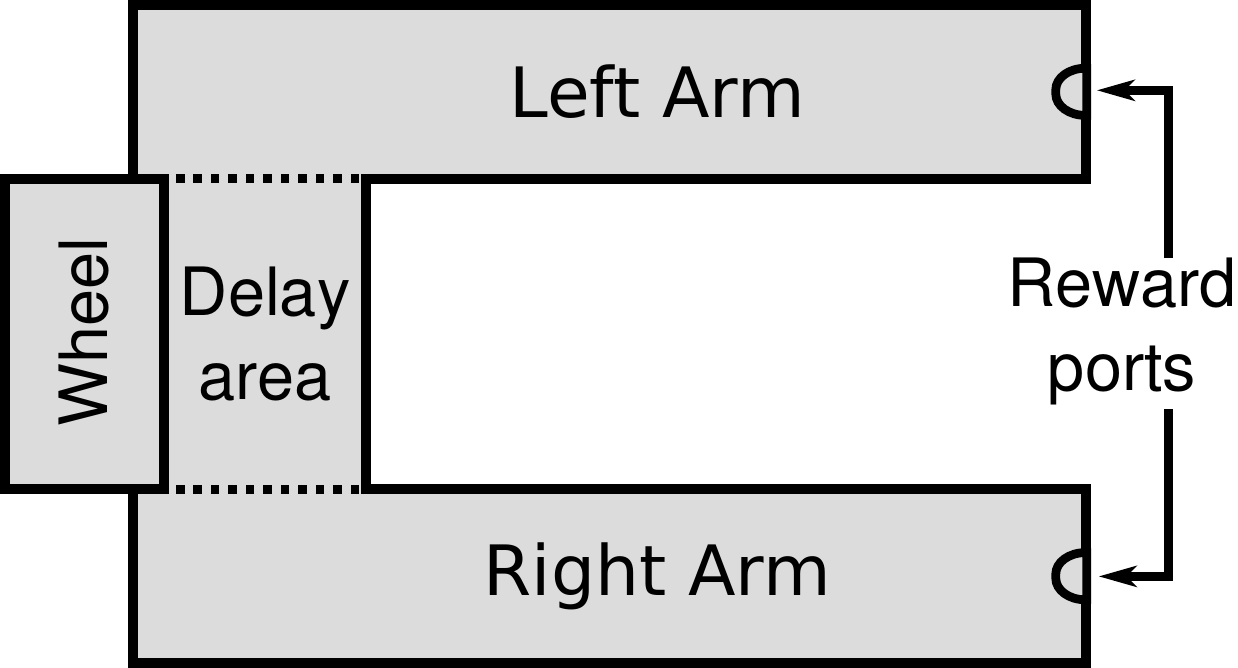}
  \caption{\label{fig:experiment-setup}Alternation memory task. The three main
    areas of interest for this task are (1) the delay area and running wheel,
    (2) the left and right arms/corridors, and (3) the reward area at the end of
    each arm. The dotted lines surrounding the delay area represent mechanically
    movable doors that are used to hold the animals in the delay area until the
    wheel-run portion of each trial has been completed.}
\end{figure}

Rats were trained to perform a basic memory task in the U-shaped track pictured
in \Cref{fig:experiment-setup}. In short, the task of the animal can be
described as follows: To receive a reward, the animal must alternate between
running to the left reward port and the right reward port. The task is
complicated for the animal by the delay area, which forces the animal to run in
the running wheel for a specified amount of time before the doors of the delay
area open and allow the animal to proceed toward a reward port. Each trial of
the experiment consists of a run in the wheel, a visit to a reward port, and a
return to the delay area. By requiring the animal to spend time running in the
wheel before choosing to run to the left or right, the animal is forced to
remember the arm to which it should next run instead of being allowed to run
immediately from one arm to the other.

In addition to performing this task under normal conditions, each animal
performed the task after the drug muscimol had been injected into the animal's
medial septum. This injection has the effect of shutting down the medial septum,
which is the area of the brain that serves as the primary drive of the theta
oscillations that are thought to be an integral part of the generation of
place-field sequences in the hippocampus during locomotion.

During performance of this task, the rat's location within the track was
recorded. Further, local-field potentials (LFPs) were recorded at various depths
across the pyramidal layer of the CA1 region of the rat's hippocampus. Neuronal
spiking information was then extracted from these LFPs via a process called
spike sorting. For our purposes, we will say that an \emph{event} is all of the
information (LFPs, spike trains, animal location, etc.) within some time window.
The sequence corresponding to an event is the list of all spikes that occur
within that event. The various event/sequence types that we will be
investigating are described below.

\begin{description}
\item[{Wheel-run events}] are behavioral events recorded while the animal is running
in the wheel to satisfy the delay requirement. As such, each wheel-run
event is recorded over a period of approximately 10 seconds. These event
types are generally the longest in duration of all of the event types that
we consider; thus, the corresponding wheel-run sequences are usually the
longest of our recorded sequence types. Because the animals are running
during these sequences, the hippocampus exhibits a prominent theta
oscillation under normal conditions.
\item[{Arm-run events}] are behavioral events recorded while an animal is running
through either arm of the track in either direction between the delay area
and a reward area. The duration of each arm-run event is the length of time
that it takes the animal to get where it is going, usually around 2
seconds. Though shorter than wheel-run sequences, arm-run sequences are
still usually quite long. The locomotion of the animal is again accompanied
by a prominent theta oscillation. Because the animals are actually moving
through the track during these sequences, the sequences exhibit place
fields under normal conditions. We divide these sequences into four
subtypes based on the trajectory of the animal; specifically, the animal
can run to (outbound) or from (inbound) either the left or right reward
area.
\item[{Sharp-wave ripple (SWR) events}] usually occur while an animal is not in
motion; hence, they primarily occur in the reward areas and in the delay
area outside of the wheel, though we placed no explicit restrictions on the
location or movement of the animal in our detection of these events. As
described in Section \ref{sec:orgheadline1}, these events are characterized by a
hyper-polarization in the LFP across the pyramidal layer in the CA1 region
of the hippocampus, which generally lasts from around 50 to 250
milliseconds. Given the shorter timescale than the behavioral events, these
sequences are generally much shorter than wheel-run and arm-run sequences.
\item[{Refined SWR events}] are the bursts of neuronal activity that accompany SWRs.
These bursts of activity are not always perfectly aligned with the
corresponding sharp wave, which (as we will see in the next section) is the
feature we use to detect the SWR events. Given that our analysis is
performed on sequences of neuronal spikes, it is important that we strive
to find the sequences of spikes that most accurately represent the neuronal
activity that accompanies the SWR events, regardless of whether or not
those sequences are aligned with the accompanying LFP envelope. This
refinement process results in the discarding of many events that do not
contain sufficient spiking information. Thus, there are fewer refined SWR
sequences than (unrefined) SWR sequences; but the sequences that remain are
often longer.
\end{description}

\section{SWR detection and refinement}
\label{sec:orgheadline21}
\label{sec:swr-detection}

Unlike arm-run and wheel-run events, SWR events are not behavioral in nature. In
particular, no action or activity of an animal will tell us when a SWR event is
occurring. As such, we must find a way to determine when these events are
occurring. To do this, we consider one prominent feature of SWR events: the
sharp wave. During a sharp wave, the LFPs across the pyramidal layer of CA1
become hyperpolarized, resulting in the characteristic shape that has already
been observed (see \Cref{fig:ripple-probe-envelope}). Our method for detecting
SWR events can be summarized as follows:
\begin{enumerate}
\item Select two LFP channels, a lower channel (i.e., one that drops in value
during SWRs) and an upper channel (i.e., one that increases in value during
SWRs). In lieu of an automated method for performing this, all of the LFP
channels were selected by hand. The selection process was done by gaining an
inuitive understanding of what SWR events look like and then by selecting the
channels that most-faithfully represented the desired LFP separation.
\item Center each LFP at zero by computing a local average for each. This is
necessary to account for changes in the LFPs that occur over a longer time
scale than SWRs occur on. To do this, we computed the average output for each
channel over the interval of one second preceding each point in time; we then
subtracted the resultant local averages from the original signals.
\item Create a single sharp-wave signal by subtracting the lower centered-LFP from
the upper centered-LFP, subtracting the overall mean of the resultant signal,
normalizing by the standard deviation of the resultant signal, and smoothing
the resultant signal using a gaussian filter.
\item Threshold the sharp-wave signal to determine initial estimates for SWR event
windows. We set the base-line threshold at 1.25 standard deviations of the
sharp-wave signal.
\item Filter, combine, and modify the initial event estimates to ensure that each
event satisfies properties characteristic of SWRs.
\begin{itemize}
\item Split single events into multiple events. Sometimes, SWRs happen close enough
together that the separation in the LFP from the first event has not ended
before the separation from the second event has begun. As a result, the
detection process up to this point joins those multiple events into a single
event. This process of splitting is done by finding the peaks of the second
derivative of the smoothed sharp-wave signal; if a large enough peak occurs
within an event, then the event is split at that point in time.
\item Remove events that do not last for some minimum amount of time (25 ms).
\item Shorten events that last for too long. It is sometimes the case that the
minimum-required LFP separation of a detected event can persist for longer
than the underlying SWR event. To account for this while not discarding
legitimate SWRs, we enforce a maximum SWR duration (250 ms) by taking the
event in a 250-ms time window surrounding the highest point of LFP
separation within the event.
\item Remove events in which the peak of the sharp-wave signal is not large
enough (4 standard deviations of the sharp-wave signal).
\item Remove events that do not contain a sudden LFP separation. Without this
characteristic feature, the event is not a SWR. We enforce this by ensuring
that the absolute value of the first derivative of the sharp-wave signal
passes above a minimum value.
\item Ensure that enough spikes happen close enough to each other within each
event. This is done by summing up gaussian functions centered at the spikes
within each event and ensuring that the resultant signal passes above a
minimum value.
\end{itemize}
\end{enumerate}

The parameters used in this detection process were fine-tuned by hand until the
resultant events were largely undeniably SWRs but also while not being so
restrictive as to eliminate many good candidates for SWR events.

\begin{figure}
  \centering
  \begin{subfigure}[b]{2.5in}
    \includegraphics[width=2.5in]{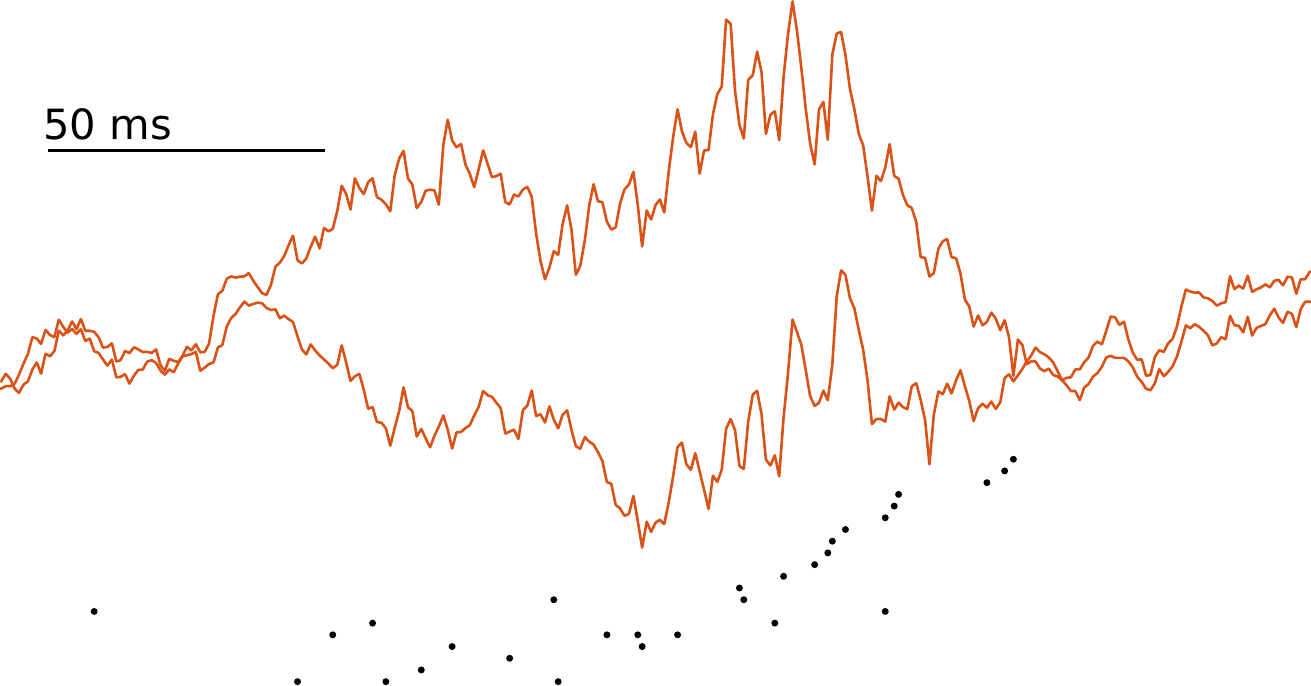}
    \caption{Normal condition}
  \end{subfigure}
  \hspace{2em}
  \begin{subfigure}[b]{2.5in}
    \includegraphics[width=2.5in]{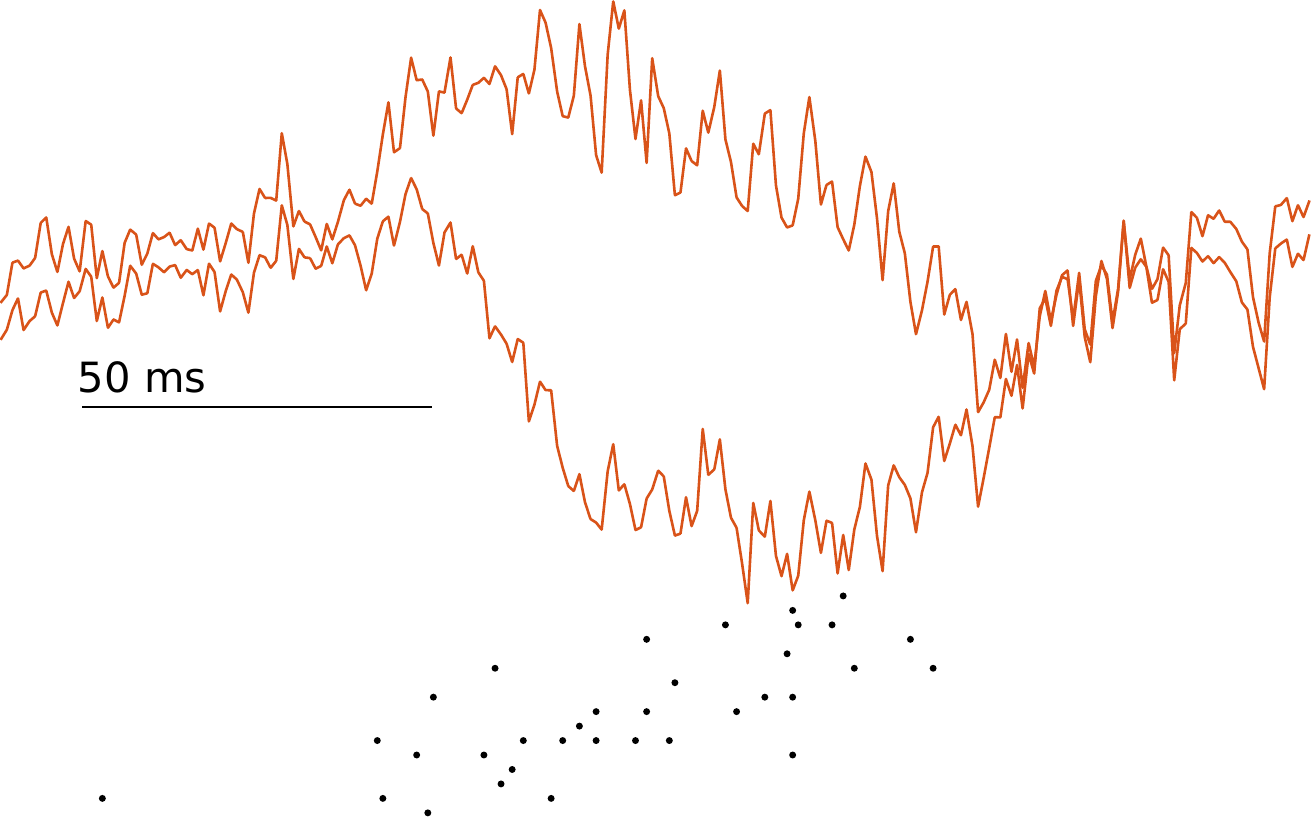}
    \caption{Muscimol condition}
  \end{subfigure}
  \newline
  \begin{subfigure}[b]{2.5in}
    \includegraphics[width=2.5in]{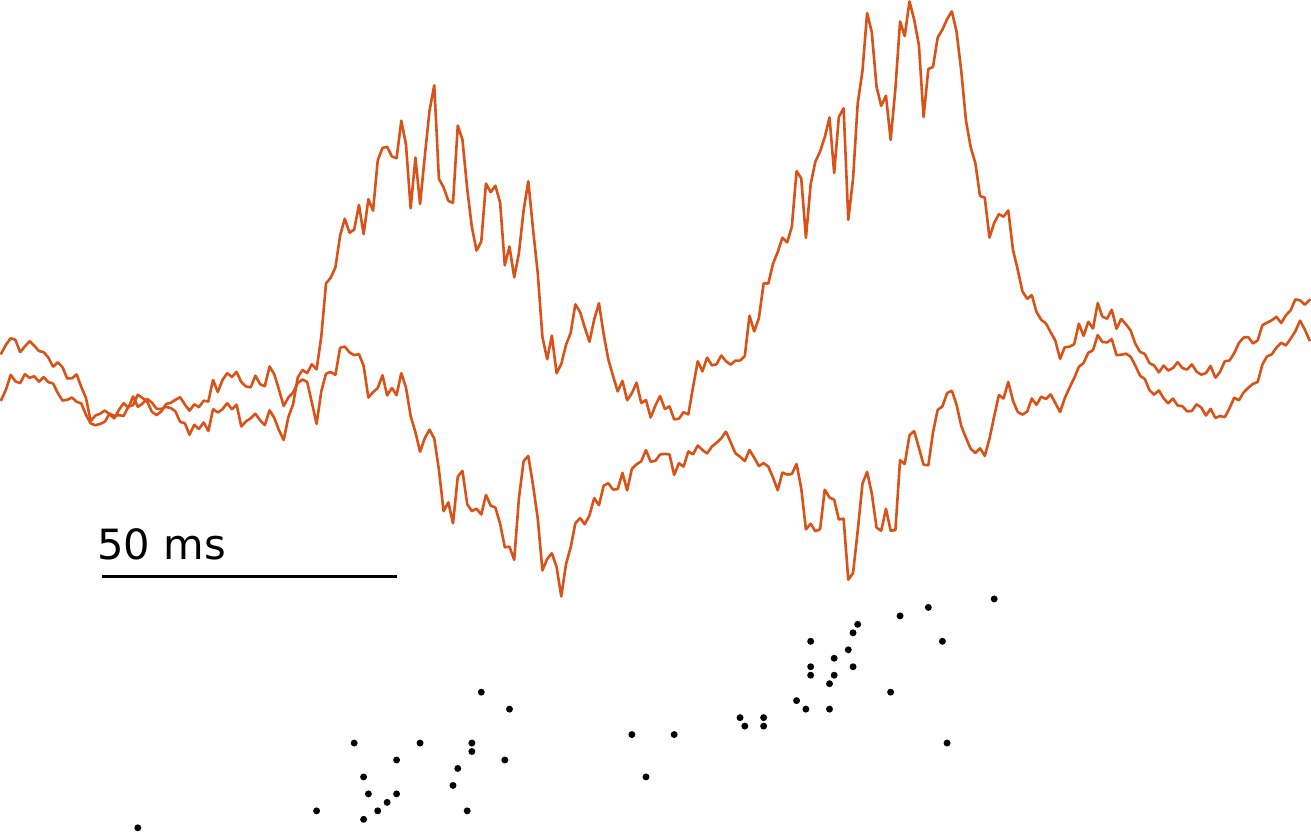}
    \caption{Doublet}
  \end{subfigure}
  \hspace{2em}
  \begin{subfigure}[b]{2.5in}
    \includegraphics[width=2.5in]{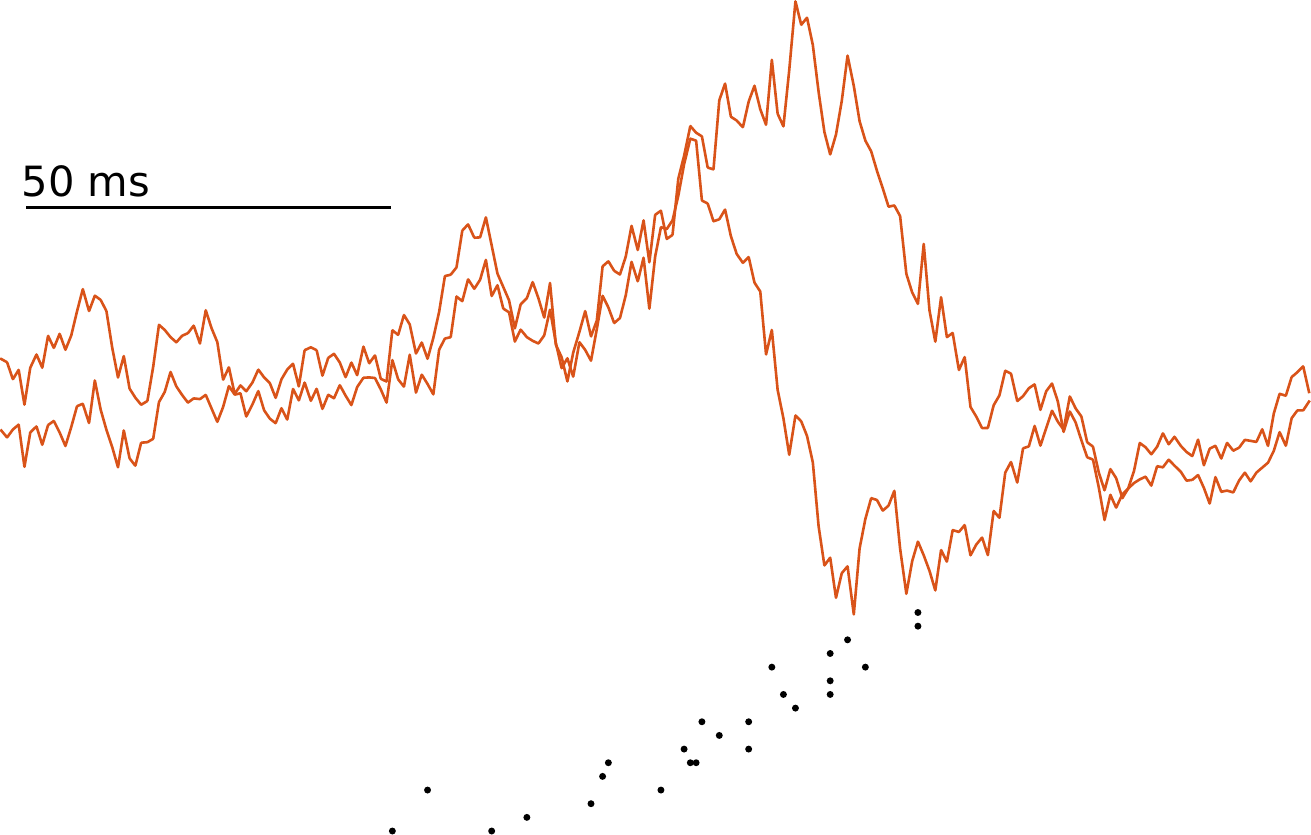}
    \caption{\label{fig:swr-needs-refinement}Needs refinement}
  \end{subfigure}
  \caption{\label{fig:swr-examples}SWR events. Panels~(a) and~(b) show standard
    examples of SWRs. Panel~(c) shows an example of a doublet, two SWR events
    that occur so close in time as to be joined together. Panel~(d) shows an
    example of a SWR for which the burst of neural activity significantly
    precedes the accompanying sharp wave. Such an event requires refinement.}
\end{figure}

Although detection based on the sharp wave gives us a good starting place for
SWR detection, a detection algorithm based solely on the sharp wave is
insufficient for our needs. While the bursts of neural activity in a SWR are (by
definition) accompanied by a sharp wave, it is not the case that the two types
of activity are perfectly synchronized in time. See
\Cref{fig:swr-needs-refinement} for an example of this. Since our analysis is of
neuronal sequences, it is crucial that we have as accurate a set of sequences as
possible. Our attempt to extract the events corresponding to the
neuronal-activity bursts of SWR events consists of four basic steps:
\begin{enumerate}
\item Detect the intrinsic sequences of a recording. These are formed by splitting
the spike train of a recording wherever the gap between successive spikes
exceeds some threshold and then by extracting the sequences from each
resultant spike train.
\item Discard all intrinsic sequences that do not overlap with any SWR event.
\item Join together those intrinsic sequences that overlap with a common SWR event.
The idea is that the SWR event should represent a single pice of sequential
information. Any gap that occurs within such a joining of intrinsic sequences
is justified as being within a single sequence by arguing that other
neurons---neurons that were not detected---were likely spiking within that
interval.
\item Of the resulting collection of sequences, keep only those that meet certain
requirements. First, we impose a requirement on the minimum number of active
neurons; a sequence does not contain much useful information to us if, for
instance, it contains spikes from only one neuron. Next, we impose a duration
condition; there is reason to believe that SWRs have minimum and maximum
duration. Lastly, we require that the sequence is isolated in time from its
nearest intrinsic sequences; this is an attempt to ensure that the activity
captured by the sequence is a burst of activity accompanying a SWR and not
due to increased neural activity in the area for some other reason.
\end{enumerate}
This process generally results in significantly fewer events than the sharp-wave
detection process.

\section{Sequence comparison in \emph{in-vivo} hippocampus}
\label{sec:orgheadline22}
\label{ch:analysis}

It is believed that theta oscillations are necessary for the generation of
hippocampal sequences. Analyses suggesting this result \cite{Wang2014} compare
observed hippocampal sequences under normal and theta-inhibited (i.e., muscimol)
conditions to neuronal templates generated, for instance, by considering the
order of place fields in a track under normal conditions. The introduction of
the bias matrix allows for the direct comparison of observed sequences to each
other. In this \namecref{ch:analysis}, we use the techniques of
\Cref{ch:comparing} to perform similarity analyses on the hippocampal sequences
described in \Cref{sec:experiment}. In particular, we will attempt to determine
whether the sequences generated under the influence of muscimol are different
from those generated under normal conditions. For reference, brief descriptions
of the sequence types are given in \Cref{tbl:sequence-types}.

\begin{table}[htb]
\caption{\label{tbl:sequence-types}
Hippocampal sequence types.}
\centering
\begin{tabular}{lll}
\textbf{Type} & \textbf{Subtypes} & \textbf{Description}\\
\hline
Wheel-run &  & Behavioral sequences generated during the\\
 &  & enforced delay period while running in\\
 &  & the wheel\\
\hline
Arm-run & Left / outbound & Behavioral sequences generated as an animal\\
 & Left / inbound & traverses a path between the delay area\\
 & Right / outbound & and a reward area\\
 & Right / inbound & \\
\hline
Refined SWR &  & Non-behavioral sequences from SWRs\\
\hline
\end{tabular}
\end{table}

\begin{question}
\label{qstn:sequences-change}
Are hippocampal sequences the same after elimation of theta oscillation?
\end{question}

\begin{figure}[h!]
  \centering
  \begin{subfigure}[b]{1.5in}
    \includegraphics[width=1.5in]{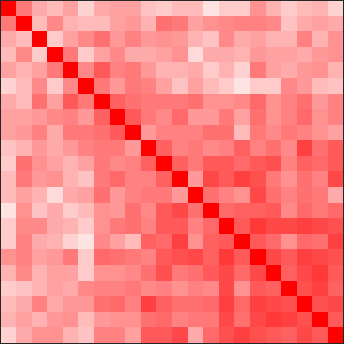}
    \caption{Wheel correlation}
  \end{subfigure}
  \hspace{2em}
  \begin{subfigure}[b]{1.5in}
    \includegraphics[width=1.5in]{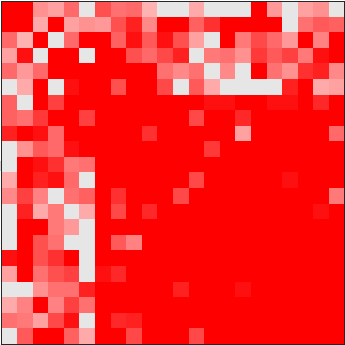}
    \caption{Wheel significance}
  \end{subfigure}
  \\[1em]

  \begin{subfigure}[b]{1.5in}
    \includegraphics[width=1.5in]{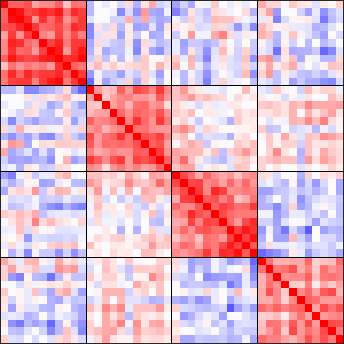}
    \caption{Arm correlation}
  \end{subfigure}
  \hspace{2em}
  \begin{subfigure}[b]{1.5in}
    \includegraphics[width=1.5in]{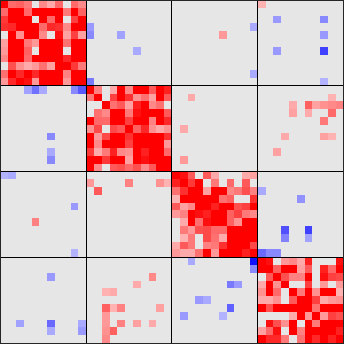}
    \caption{Arm significance}
  \end{subfigure}
  \\[1em]

  \begin{subfigure}[b]{1.5in}
    \includegraphics[width=1.5in]{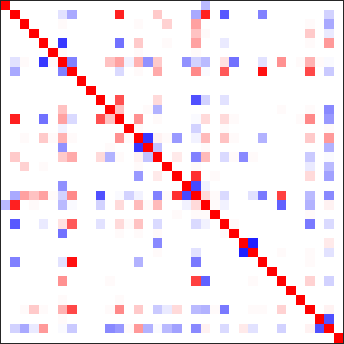}
    \caption{SWR correlation}
  \end{subfigure}
  \hspace{2em}
  \begin{subfigure}[b]{1.5in}
    \includegraphics[width=1.5in]{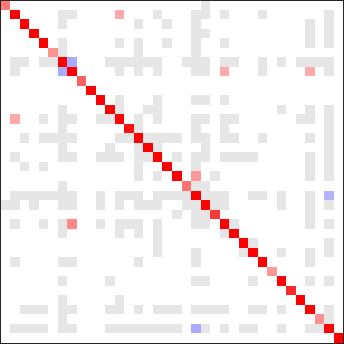}
    \caption{SWR significance}
  \end{subfigure}

  \caption{\label{fig:matrix-examples}Correlation and significance matrices. Red
    entries indicate positive correlation, blue entries indicate negative
    correlation, and white entries indicate incomparable sequences; in
    significance matrices, gray entries indicate insignificant correlation
    values. Panels~(a) and~(b) show that all wheel sequences are positively
    correlated to each other and that a majority of the correlations are
    significant. Panels~(c) and~(d) show matrices for arm-run sequences; events
    are sorted by their subtypes: left/outbound, left/inbound, right/outbound,
    left/inbound. Lines delineate subtypes. The checkering pattern indicates
    postitive correlation within a subtype and negative correlations between
    inbound and outbound events. Few correlations between subtypes are
    significant. Panels~(e) and~(f) show the sparsity of correlation matrices
    for SWR sequences; additionally, few of the comparable SWR events are
    significantly correlated. The all-red diagonals indicate that each event is
    positively and significantly correlated to itself. All matrices were
    generated from the same recording.}
\end{figure}

In the end, our analysis suggests that eliminating theta oscillation eliminates
(at least some of) the structure of behavioral hippocampal sequences. We
approach this \namecref{qstn:sequences-change} by setting \emph{reference sets} of
sequences to which we can compare normal-condition sequences and
muscimol-condition sequences. Since the question we are trying to answer is
about whether there is a change from the normal behavior, these reference sets
should consist of normal-condition sequences. For instance, when comparing
normal-condition wheel-run sequences to muscimol-condition wheel-run sequences,
the reference set will be the collection of normal-condition wheel-run
sequences. The basic tools used in this section are the correlation and
significance matrices introduced in \Cref{sec:global-significance}.
\Cref{fig:matrix-examples} provides an example of the correlation and
significance matrices for each sequence type. The examples in this figure
illustrate how the correlation matrix can be a useful tool for observing
structure in collections sequences. In particular, notice the difference in the
matrix structure between wheel, arm, and SWR correlations. Where the wheel-run
correlation matrix displays that all correlations are positive, the arm-run
correlation matrix shows positive and negative correlations existing in a highly
structured manner (switching sign as we go between comparing subtypes of these
sequences); the SWR correlation matrix, on the other hand, shows that positive
and negative correlations exist in a much less structured way.

\begin{figure}
  \centering
  \begin{subfigure}[b]{2.8in}
    \includegraphics[width=2.8in]{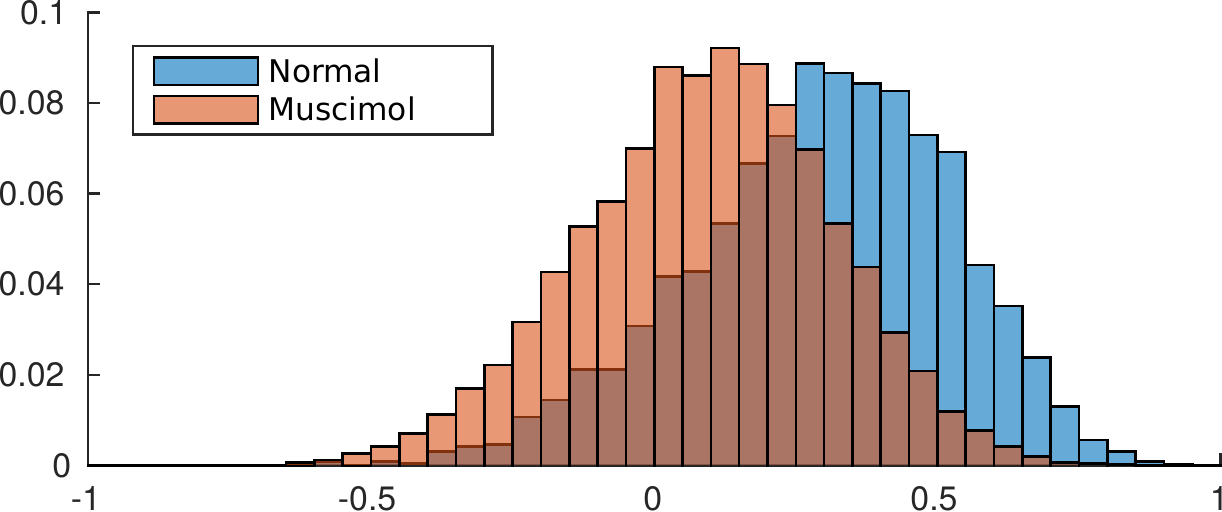}
    \caption{Wheel correlation}
  \end{subfigure}
  \hspace{1em}
  \begin{subfigure}[b]{2.8in}
    \includegraphics[width=2.8in]{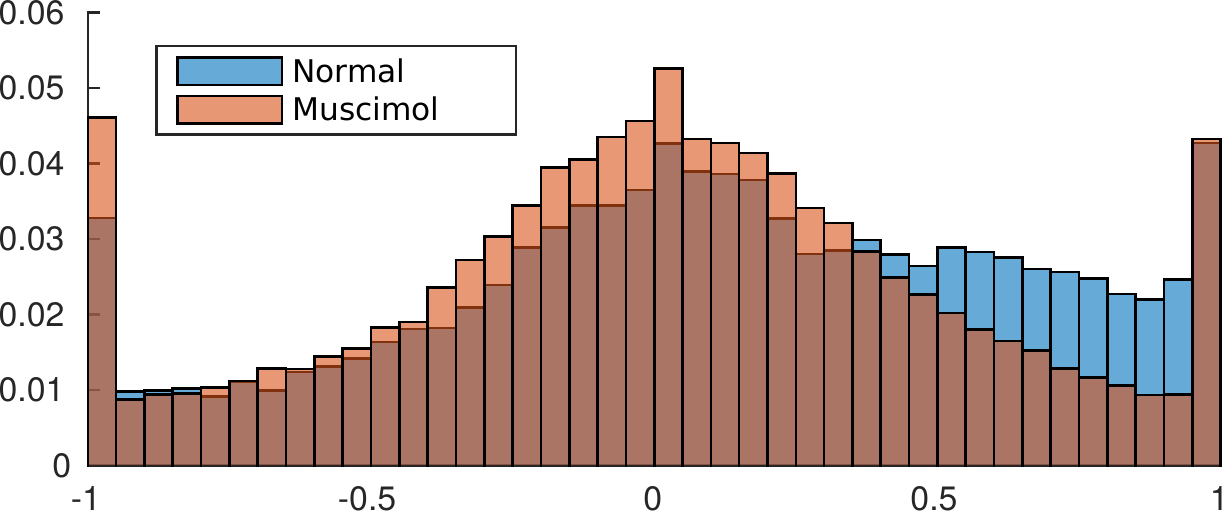}
    \caption{Arm correlation}
  \end{subfigure}
  \\[1em]
  \begin{subfigure}[b]{2.8in}
    \includegraphics[width=2.8in]{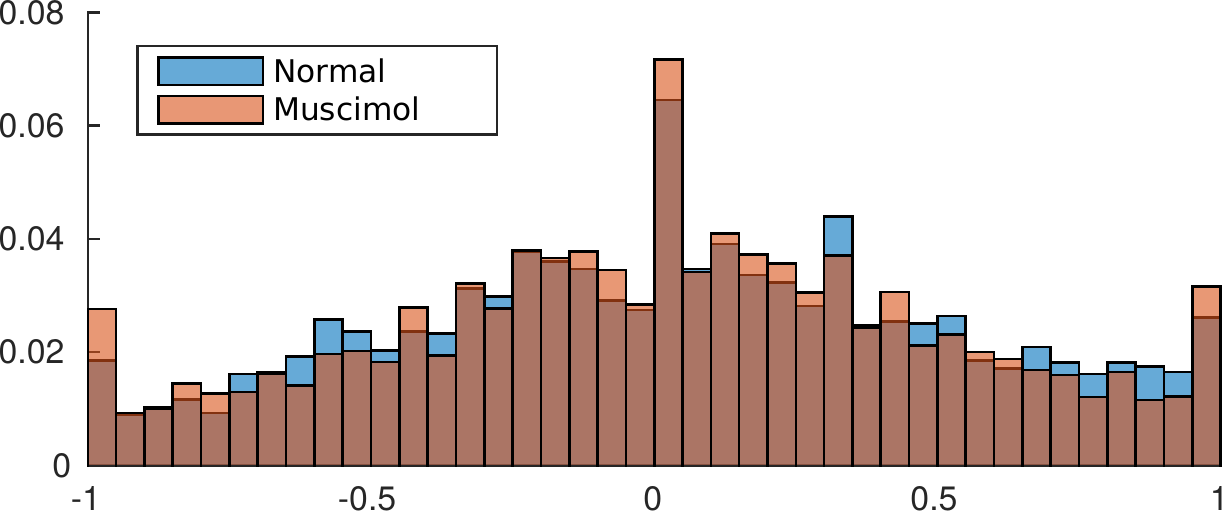}
    \caption{SWR correlation}
  \end{subfigure}

  \caption{\label{fig:corr-histograms}Histograms of correlation values. For
    each type of sequence, the corresponding plot above shows the distribution
    of correlation values for normal-condition and muscimol-condition
    correlation values across all data sets. In panel~(a), the wheel-run
    histograms show the greater tendency of normal-condition sequences to be
    positively correlated (88.75\% positive) than their muscimol counterparts
    (67.86\% positive). Panel~(b) illustrates a similar phenomenon for
    normal-condition arm-run sequences to exhibit stonger positive
    correlations than their muscimol counterparts. Panel~(c) shows a much less
    pronounced difference when considering SWR sequences, though the
    two-sample Kolmogorov-Smirnov test indicates that the distributions are
    indeed different ($p$-value of $0.0097$).}
\end{figure}

The differences in structure between the correlation matrices shown in
\Cref{fig:matrix-examples} can be observed over all data sets in
\Cref{fig:corr-histograms}, which shows histograms of the correlation values for
the different types. For instance, looking at the distributions of correlation
values for arm-run sequences, the normal-condition correlations tend to be more
positive than the muscimol-condition correlations. This observation is in line
with the intended purpose of the muscimol injections to eliminate the theta
oscillation that is believed to aid in the formation of place-field sequences.

\begin{figure}
  \centering
  \begin{subfigure}[b]{1.7in}
    \includegraphics[width=1.7in]{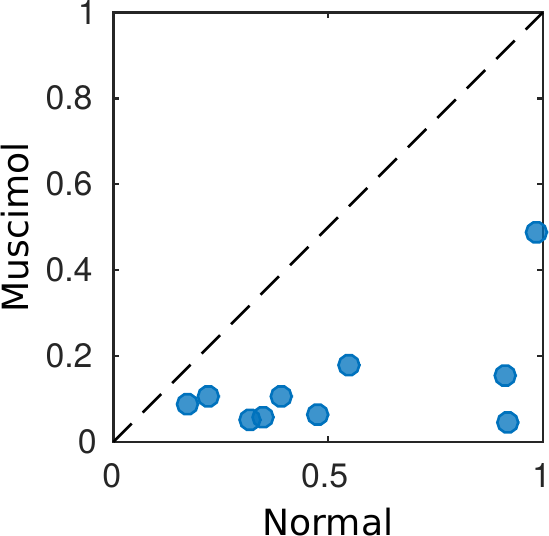}
    \caption{Wheel}
  \end{subfigure}
  \begin{subfigure}[b]{1.7in}
    \includegraphics[width=1.7in]{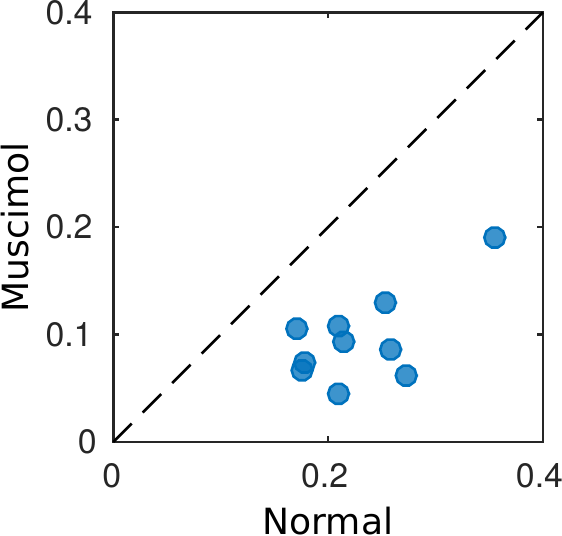}
    \caption{Arm}
  \end{subfigure}
  \begin{subfigure}[b]{1.7in}
    \includegraphics[width=1.7in]{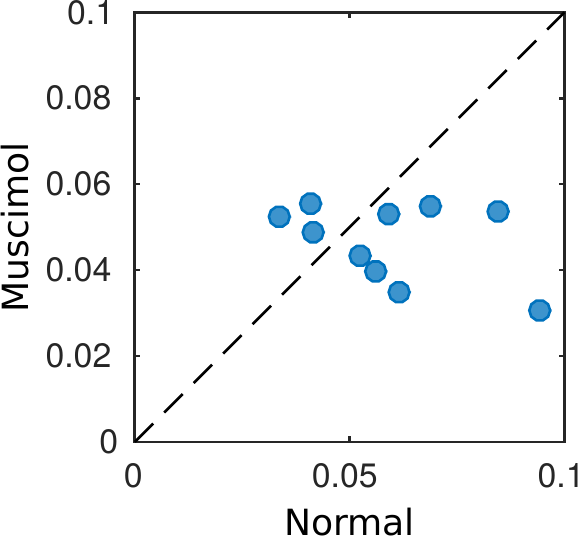}
    \caption{SWR}
  \end{subfigure}
  \caption{\label{fig:sig-proportion}Proportion of correlations that are
    significant. For each recording and each correlation/significance matrix, we
    consider the proportion of significance values that are below some threshold
    (0.05); this process results in two values, one for the normal-condition
    sequences and one for the muscimol-condition sequences, the horizontal and
    vertical axes, respectively, of the plots above. Panels~(a) and~(b) show
    that injection of muscimol is accompanied by a systematic drop in the
    proportion significant correlations for both sets of behavioral sequences.
    This universal drop in significance is not observed when considering SWR
    sequences in panel~(c).}
\end{figure}

In addition to looking for changes in the distributions of correlation values
after injection of muscimol, we also investigate the change in proportion of
correlations that are significant after injection of muscimol, as shown in
\Cref{fig:sig-proportion}. For behavioral events, muscimol is accompanied by a
reduction in the proportion of correlations that are significant. The reduction
is somewhat modest for arm-run sequences when compared to the reduction for
wheel-run sequences. This suggests that theta oscillation is more critical for
the formation of normal-condition wheel-run sequences than for the formation of
normal-condition arm-run sequences.

In conclusion, our analysis demonstrates via bias-matrix correlations that
behavioral hippocampal sequences lose some of their structure when theta
oscillation is removed by the injection of musimol into an animal's medial
septum.

\backmatter
\appendix
\addtocontents{toc}{\protect\setcounter{tocdepth}{0}}

\chapter{Data analysis codebase}
\label{sec:orgheadline27}
\label{ch:readme}

Much of the code written for the data analysis performed in this thesis is
stored in a Git repository at \url{https://github.com/zjroth/neural-events}
along with a corresponding README file.

\chapter{Supplementary figures}
\label{sec:orgheadline28}

\begin{figure}[H]
  \centering
  \begin{subfigure}[b]{1.7in}
    \includegraphics[width=1.7in]{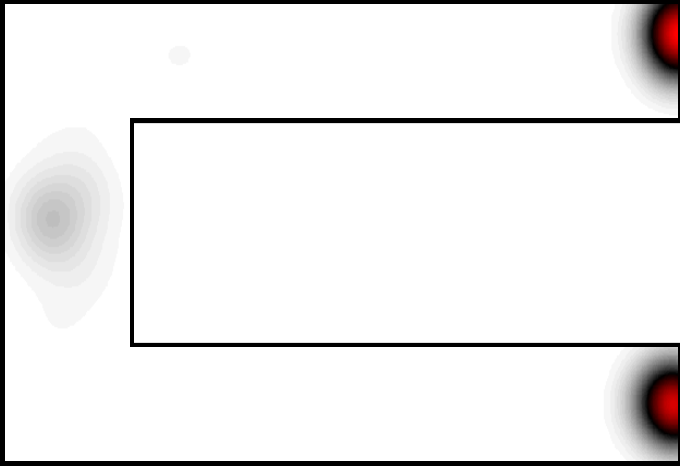}
    \caption{Normal condition}
  \end{subfigure}
  \hspace{1.7em}
  \begin{subfigure}[b]{1.7in}
    \includegraphics[width=1.7in]{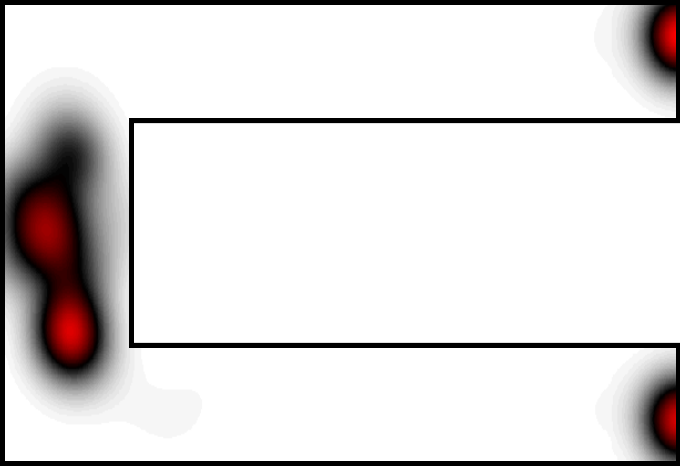}
    \caption{Muscimol condition}
  \end{subfigure}
  \hspace{1.7em}
  \begin{subfigure}[b]{1.7in}
    \includegraphics[width=1.7in]{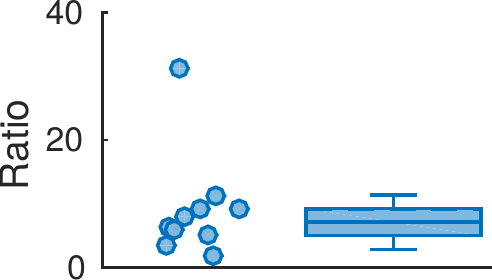}
    \caption{Increase of occurrence}
  \end{subfigure}
  \caption{\label{fig:swr-location}SWR occurrence. Panel~(a) shows that under
    normal conditions, SWR events occur almost exclusively at the reward areas,
    though some small number occur within the delay area. Panel~(b) shows that
    injection of muscimol is accompanied by an increased number of SWRs
    occurring within the delay area. Panel~(c) shows the ratio of the number of
    muscimol events to the number of normal-condition events on a per-recording
    basis.}
\end{figure}

\begin{figure}
  \centering
  \begin{subfigure}[b]{2.5in}
    \includegraphics[width=2.5in]{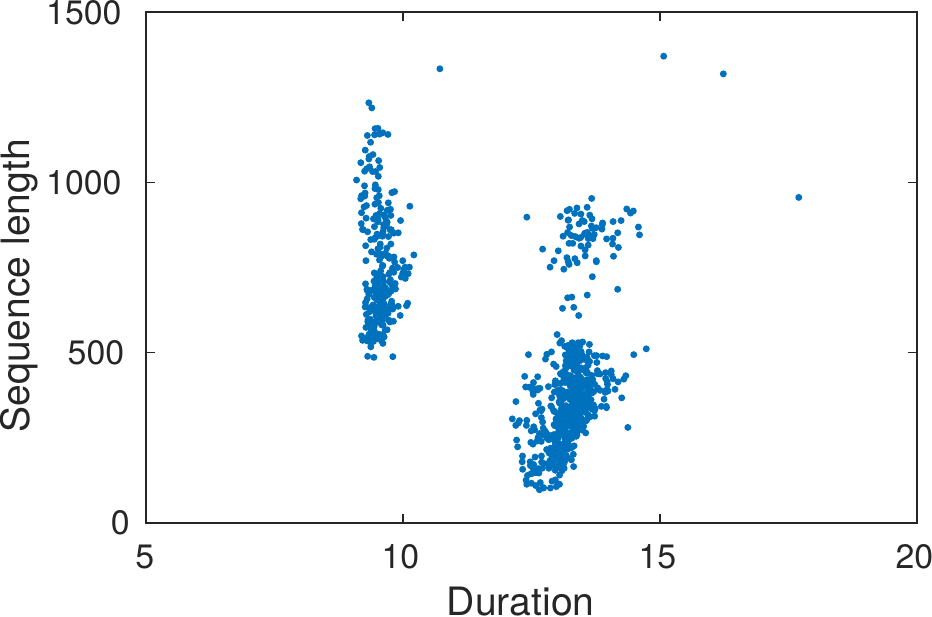}
    \caption{Wheel}
  \end{subfigure}
  \hspace{1em}
  \begin{subfigure}[b]{2.5in}
    \includegraphics[width=2.5in]{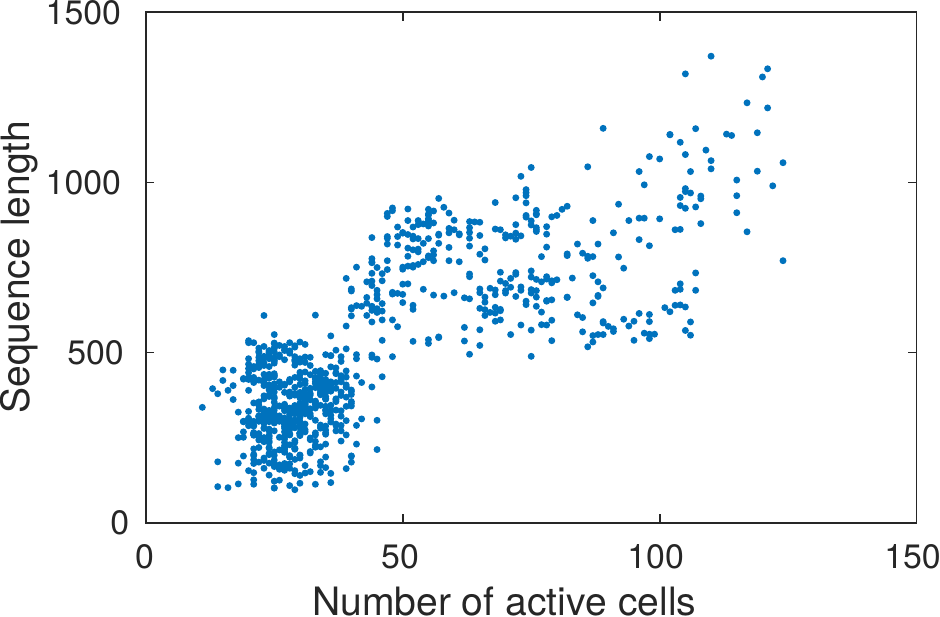}
    \caption{Wheel}
  \end{subfigure}
  \\[1.5em]

  \begin{subfigure}[b]{2.5in}
    \includegraphics[width=2.5in]{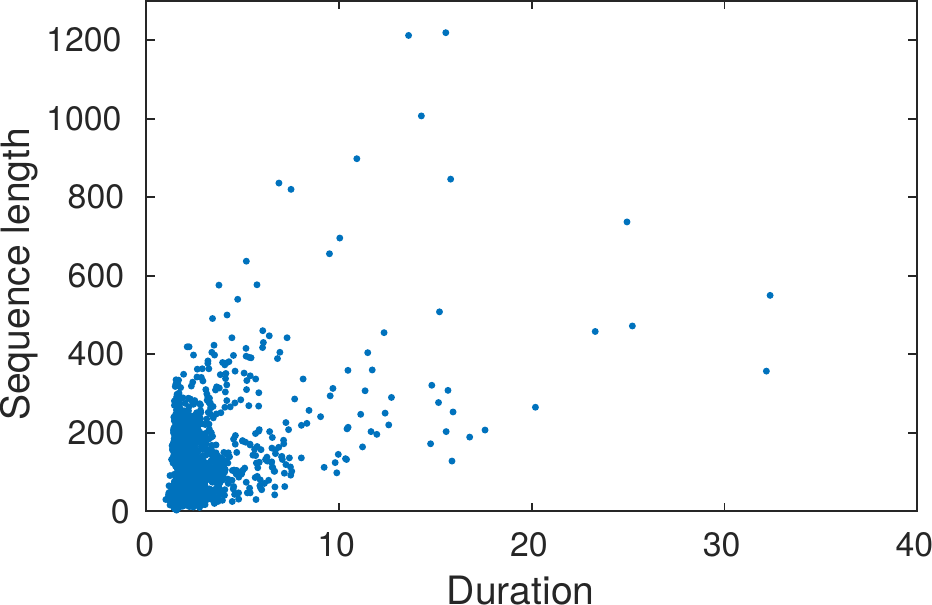}
    \caption{Arm}
  \end{subfigure}
  \hspace{1em}
  \begin{subfigure}[b]{2.5in}
    \includegraphics[width=2.5in]{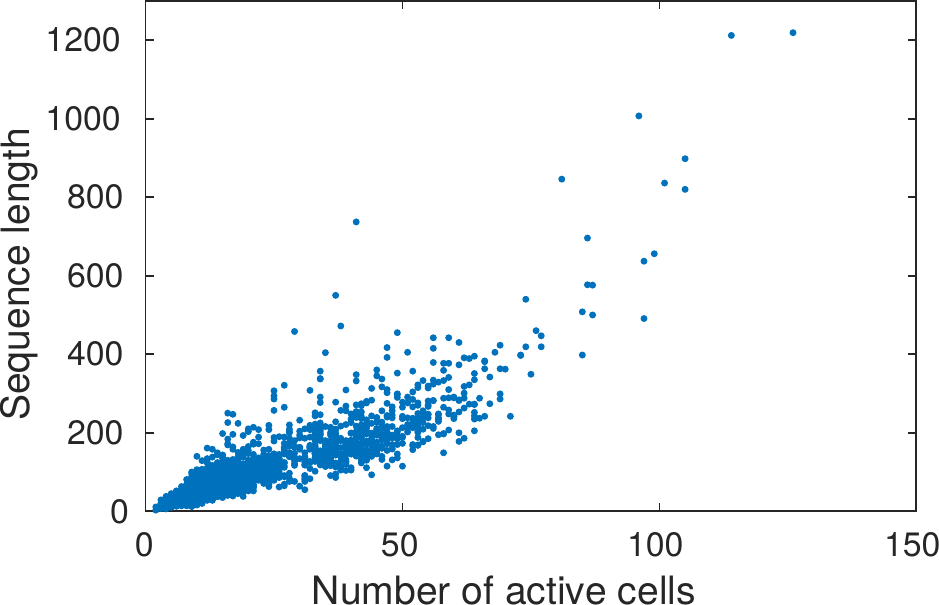}
    \caption{Arm}
  \end{subfigure}
  \\[1.5em]

  \begin{subfigure}[b]{2.5in}
    \includegraphics[width=2.5in]{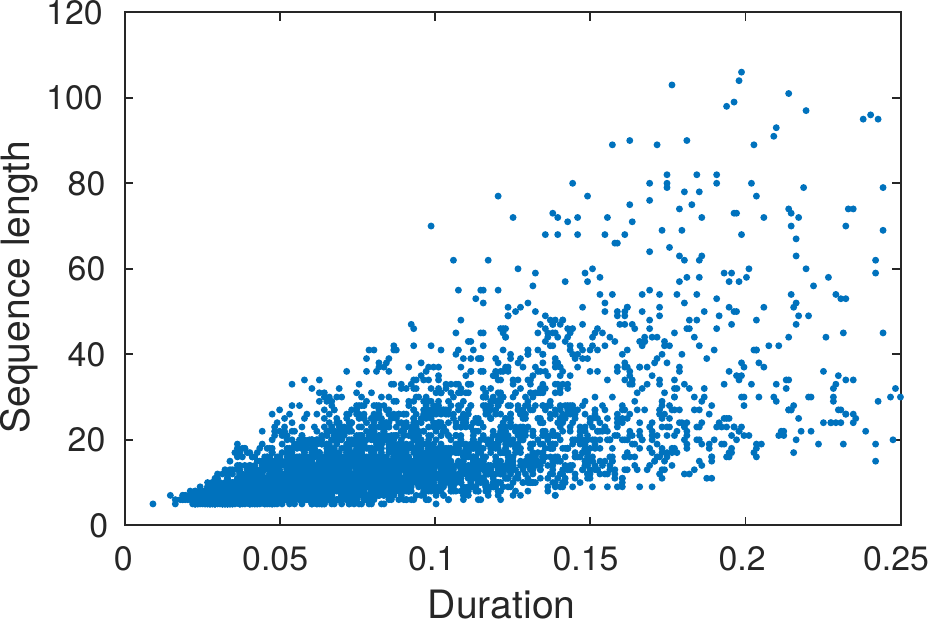}
    \caption{SWR}
  \end{subfigure}
  \hspace{1em}
  \begin{subfigure}[b]{2.5in}
    \includegraphics[width=2.5in]{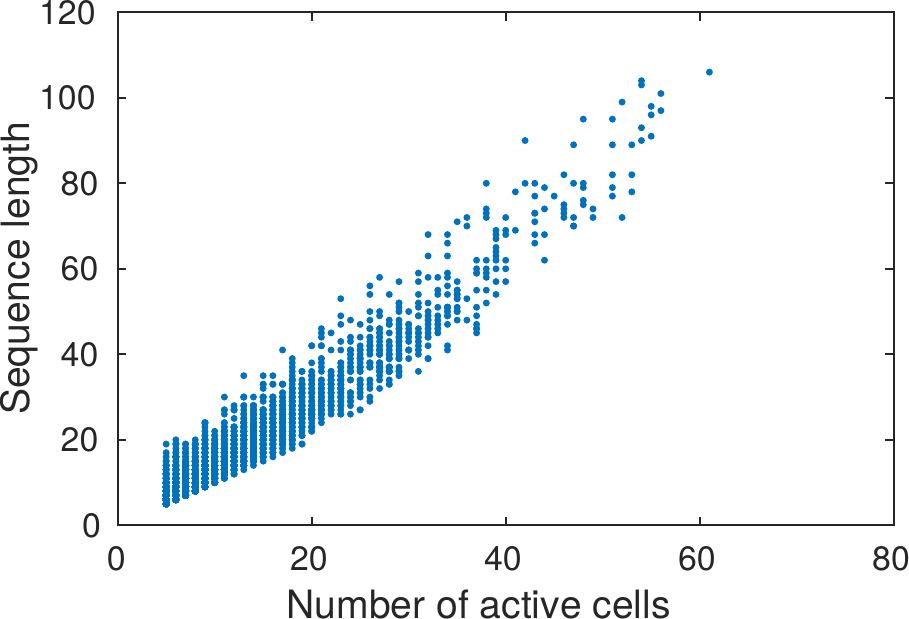}
    \caption{SWR}
  \end{subfigure}
  \caption{Sequence length. The left column shows the relationship between the
    duration of an event and the corresponding sequence's length. The right
    column shows the relationship between the total number of neurons that are
    active within an event and the corresponding sequence's length. Each plot in
    the right column shows that a linear dependence exists between the number of
    active cells in an event and the total length of the sequence.}
\end{figure}

\begin{figure}
  \centering
  \begin{subfigure}[b]{2.5in}
    \includegraphics[width=2.5in]{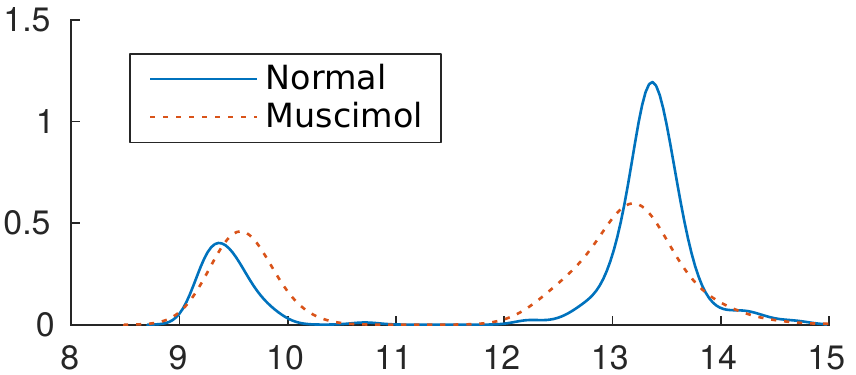}
    \caption{\textbf{Wheel-run.} Median: 13.30 / 12.94; KS: $0.0046$}
  \end{subfigure}
  \hspace{2em}
  \begin{subfigure}[b]{2.5in}
    \includegraphics[width=2.5in]{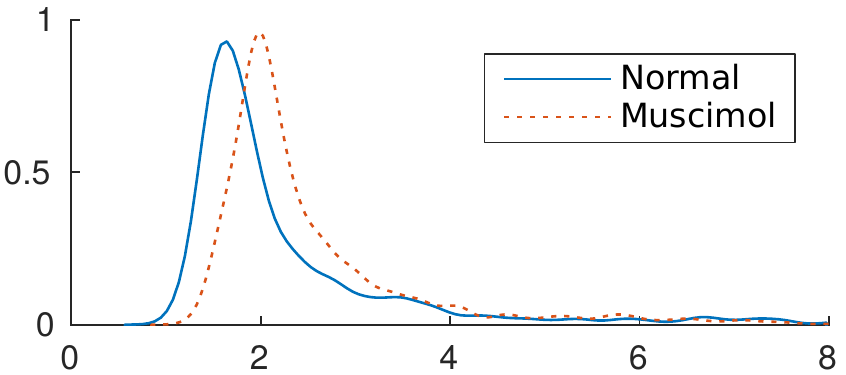}
    \caption{\textbf{Arm-run.} Median: 1.84 / 2.16; KS: $1$.}
  \end{subfigure}
  \\[1.5em]
  \begin{subfigure}[b]{2.5in}
    \includegraphics[width=2.5in]{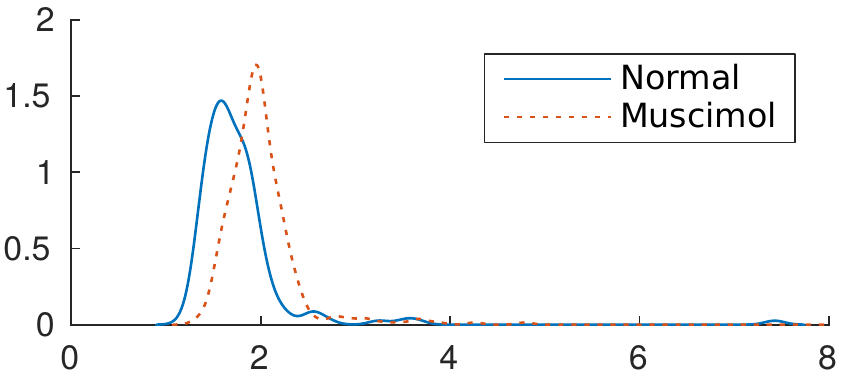}
    \caption{\textbf{Left / outbound.} Median: 1.66 / 1.96; KS: $0$.}
  \end{subfigure}
  \hspace{2em}
  \begin{subfigure}[b]{2.5in}
    \includegraphics[width=2.5in]{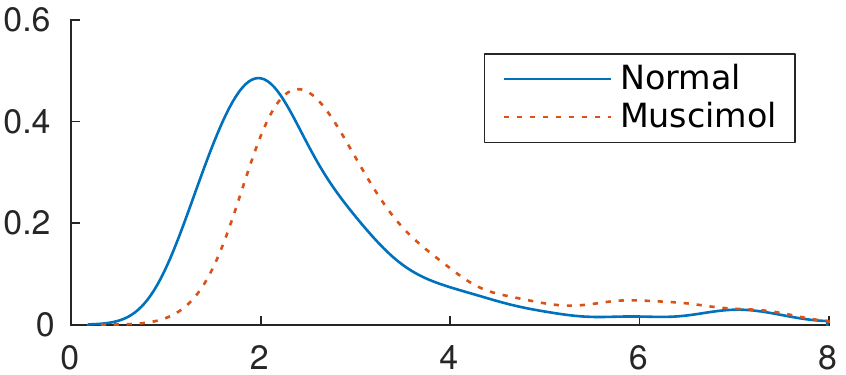}
    \caption{\textbf{Left / inbound.} Median: 2.25 / 2.74; KS: $1$.}
  \end{subfigure}
  \\[1.5em]
  \begin{subfigure}[b]{2.5in}
    \includegraphics[width=2.5in]{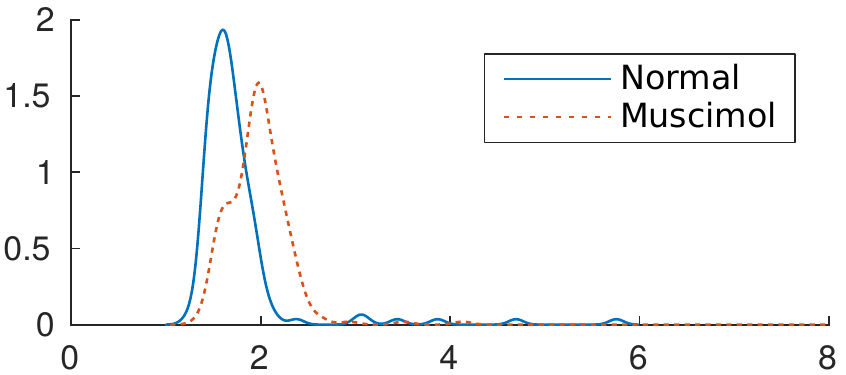}
    \caption{\textbf{Right / outbound.} Median: 1.64 / 1.97; KS: $8$.}
  \end{subfigure}
  \hspace{2em}
  \begin{subfigure}[b]{2.5in}
    \includegraphics[width=2.5in]{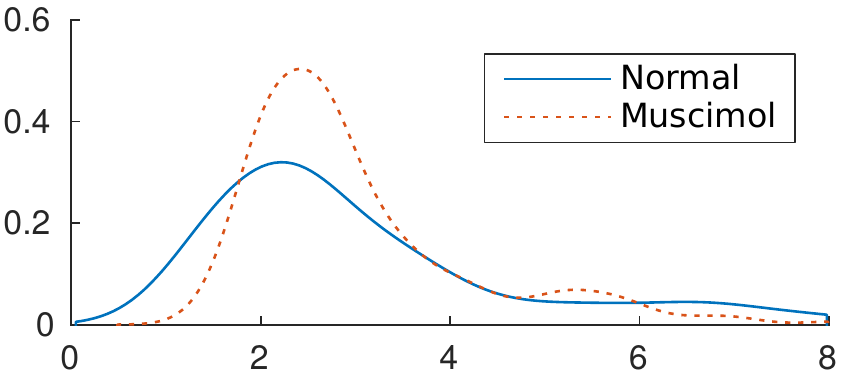}
    \caption{\textbf{Right / inbound.} Median: 2.65 / 2.70; KS: $0$.}
  \end{subfigure}
  \\[1.5em]
  \begin{subfigure}[b]{2.5in}
    \includegraphics[width=2.5in]{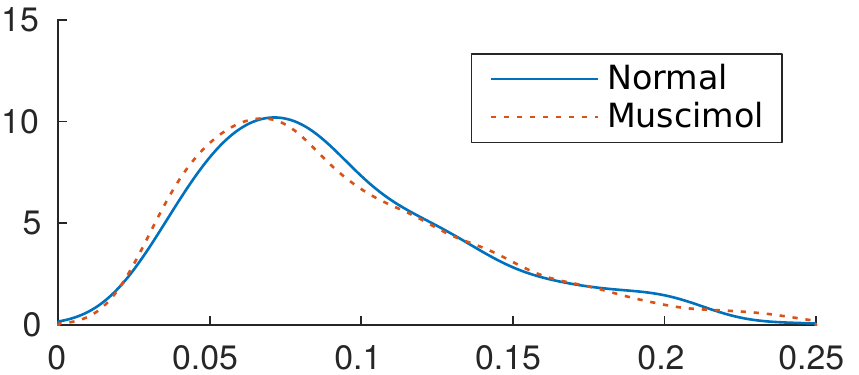}
    \caption{\textbf{Refined SWR.} Median: 0.084 / 0.082; KS: $2.99$}
  \end{subfigure}
  \caption{\label{tbl:event-duration}Event duration. For each event type, the
    distribution of event durations is shown under normal (solid line) and
    muscimol (dashed line) conditions. The median of each distribution is shown
    as normal / median. In each case, the two-sample Kolmogorov-Smirnov test
    (KS) tells us the $p$-value with with we reject the null hypothesis that the
    two distributions are the same. For each behavioral event type (all but
    SWR), the injection of muscimol tends to create slightly longer events. In
    contrast, muscimol tends to create slightly shorter SWR events.}
\end{figure}

\begin{figure}[H]
  \centering
  \includegraphics[width=4in]{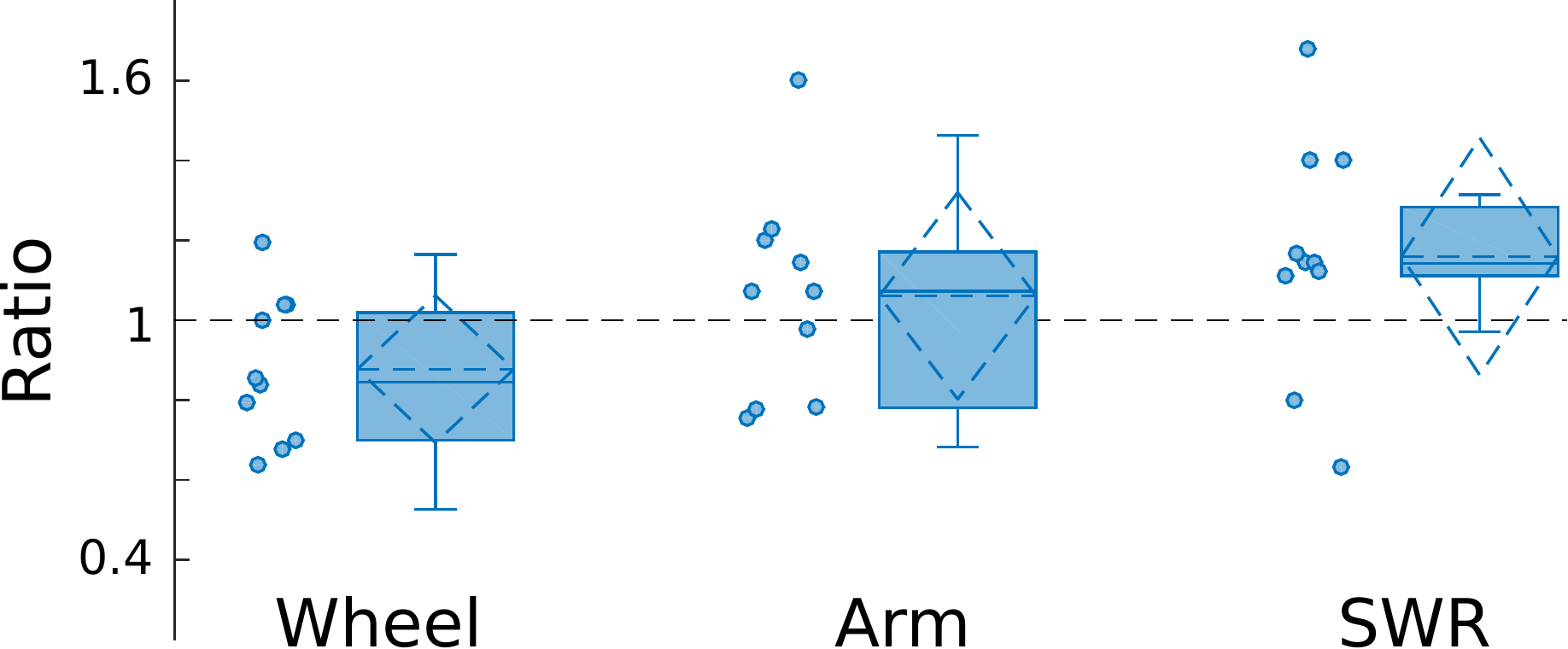}
  \caption{\label{fig:active-neurons}Number of active neurons. The vertical axis
    shows the ratio of the median number of active neurons in a muscimol event
    to the median number of active neurons in an event under normal conditions.
    After injection of muscimol, the median number of active cells in wheel-run
    sequences tends to decrease slightly, whereas the median number of The
    median number of active neurons changes after. No change is observed in the
    number of active neurons in arm-run sequences.}
\end{figure}

\nocite{FmaToolbox,Varshney2011,Tatsuno2015,Wang2014}
\printbibliography
\end{document}

%% file: header.tex
\usepackage{amsmath, amsfonts, amsthm, hyperref, mathtools}
\usepackage{titlesec, listings, multicol, nicefrac, tikz, tikz-cd}
\usepackage{float, lipsum, cleveref, array, minibox}

\usepackage{caption,subcaption}
\usepackage[section]{placeins}
\usepackage{tocbibind}
\usepackage{color}

\usetikzlibrary{arrows}

\newcommand{\brackets}[4]{
  \if\relax\detokenize{#3}\relax
    \left#1\relax#4\relax\right#2\relax
  \else
    #3#1\relax#4\relax#3#2\relax
  \fi
}
\newcommand{\set}[2][]{\brackets{\{}{\}}{#1}{#2}}
\newcommand{\brkts}[2][]{\brackets{[}{]}{#1}{#2}}
\newcommand{\parens}[2][]{\brackets{(}{)}{#1}{#2}}

\newcommand{\inprod}[2][]{\brackets{\langle}{\rangle}{#1}{#2}}
\newcommand{\abs}[2][]{\brackets{|}{|}{#1}{#2}}
\newcommand{\norm}[2][]{\brackets{\|}{\|}{#1}{#2}}
\newcommand{\floor}[2][]{\brackets{\lfloor}{\rfloor}{#1}{#2}}

\renewcommand{\l}{\ell}

\newcommand{\nneurons}{{|\neurons|}}

\newcommand{\neurons}{{\mathcal{N}}}
\newcommand{\sequences}{{\mathcal{S}}}
\newcommand{\B}{{\mathfrak{B}}}

\newcommand{\bias}{B_{\text{prob}}}
\newcommand{\skbias}{B_{\text{skew}}}
\DeclareMathOperator{\corr}{corr}
\DeclareMathOperator{\Corr}{Corr}
\DeclareMathOperator{\sig}{sig}
\DeclareMathOperator{\Sig}{Sig}
\DeclareMathOperator{\com}{center}
\DeclareMathOperator{\len}{len}
\DeclareMathOperator{\sign}{sign}

\newcommand{\x}{\times}
\newcommand{\sset}{\subseteq}
\newcommand{\R}{\mathbb{R}}
\newcommand{\N}{\mathbb{N}}
\newcommand{\Q}{\mathbb{Q}}
\newcommand{\eps}{\epsilon}

\newcommand{\g}{\gamma}
\newcommand{\s}{\sigma}
\newcommand{\w}{\omega}

\renewcommand{\b}[1]{\mathbf{#1}}

\newcommand{\half}[1][1]{\frac{#1}{2}}
\newcommand{\thalf}[1][1]{\tfrac{#1}{2}}

\DeclareMathOperator{\supp}{supp}
\DeclareMathOperator{\Sub}{Sub}
\newcommand{\perms}[1]{{\mathrm{Perms}(#1)}}

\newtheorem{theorem}{Theorem}
\newtheorem{proposition}[theorem]{Proposition}
\newtheorem{lemma}[theorem]{Lemma}
\newtheorem{corollary}[theorem]{Corollary}

\theoremstyle{definition}
\newtheorem{definition}[theorem]{Definition}
\newtheorem{example}[theorem]{Example}
\newtheorem{question}[theorem]{Question}
\newtheorem{notation}[theorem]{Notation}

\newcommand{\bmtx}[1]{\begin{bmatrix}#1\end{bmatrix}}
\newcommand{\sbmtx}[1]{\left[\begin{smallmatrix}#1\end{smallmatrix}\right]}

\titleformat{\paragraph}[hang]{\normalfont\normalsize\bfseries}{\theparagraph}{0em}{}